\documentclass[notitlepage,nofootinbib,12pt]{revtex4-2}
\usepackage{amssymb}
\usepackage{amsfonts}
\usepackage{graphicx}
\usepackage{amsfonts}
\usepackage{amsmath}
\usepackage{amssymb}
\usepackage{amsthm}
\usepackage{bm}
\usepackage{braket}
\usepackage{calc}
\usepackage{cancel}
\usepackage{color}
\usepackage{dsfont}
\usepackage[inline, shortlabels]{enumitem}
\usepackage{fancybox}
\usepackage{fancyhdr}
\usepackage{fixmath}
\usepackage{float}
\usepackage{graphicx}
\usepackage{hyperref}
\usepackage{mathrsfs}
\usepackage{mathtools}
\usepackage{natbib}
\usepackage{slashed}
\usepackage{subcaption}
\usepackage{tcolorbox}										% http://ctan.org/pkg/tcolorbox
\usepackage{textcomp}
\usepackage{transparent}
\usepackage{verbatim}
\usepackage{xcolor}
\usepackage{xparse}
\usepackage{tikz}
\usepackage{tikz-feynman,contour}
\numberwithin{equation}{section}
\newcommand{\PD}{\partial}
\newcommand{\dbar}{d\hspace*{-0.08em}\bar{}\hspace*{0.1em}}
\usepackage{tikz}
\usetikzlibrary{calc,positioning,shadows.blur,decorations.pathreplacing}
\usepackage{etoolbox}

\tikzset{%
brace/.style = { decorate, decoration={brace, amplitude=5pt} },
mbrace/.style = { decorate, decoration={brace, amplitude=5pt, mirror} },
label/.style = { black, midway, scale=0.5, align=center },
toplabel/.style = { label, above=.5em, anchor=south },
leftlabel/.style = { label,rotate=-90,left=.5em,anchor=north },   
bottomlabel/.style = { label, below=.5em, anchor=north },
force/.style = { rotate=-90,scale=0.4 },
round/.style = { rounded corners=2mm },
legend/.style = { right,scale=0.4 },
nosep/.style = { inner sep=0pt },
generation/.style = { anchor=base }
}

\newcommand\particl[7][white]{%
\begin{tikzpicture}[x=1cm, y=1cm]
\path[fill=#1,blur shadow={shadow blur steps=5}] (0.1,0) -- (0.9,0)
arc (90:0:1mm) -- (1.0,-0.9) arc (0:-90:1mm) -- (0.1,-1.0)
arc (-90:-180:1mm) -- (0,-0.1) arc(180:90:1mm) -- cycle;
\ifstrempty{#7}{}{\path%[fill=purple!50!white]
(0.6,0) --(0.7,0) -- (1.0,-0.3) -- (1.0,-0.4);}
\ifstrempty{#6}{}{\path%[fill=green!50!black!50] (0.7,0) -- (0.9,0)
arc (90:0:1mm) -- (1.0,-0.3);}
\ifstrempty{#5}{}{\path%[fill=orange!50!white] (1.0,-0.7) -- (1.0,-0.9)
arc (0:-90:1mm) -- (0.7,-1.0);}
\draw[\ifstrempty{#2}{dashed}{black}] (0.1,0) -- (0.9,0)
arc (90:0:1mm) -- (1.0,-0.9) arc (0:-90:1mm) -- (0.1,-1.0)
arc (-90:-180:1mm) -- (0,-0.1) arc(180:90:1mm) -- cycle;
\ifstrempty{#7}{}{\node at(0.825,-0.175) [rotate=-45,scale=0.2] {#7};}
\ifstrempty{#6}{}{\node at(0.9,-0.1)  [nosep,scale=0.17] {#6};}
\ifstrempty{#5}{}{\node at(0.8,-0.8)  [nosep,scale=0.6] {#5};}
\ifstrempty{#4}{}{\node at(0.1,-0.1)  [nosep,anchor=west,scale=0.25]{#4};}
\ifstrempty{#3}{}{\node at(0.1,-0.85) [nosep,anchor=west,scale=0.3] {#3};}
\ifstrempty{#2}{}{\node at(0.1,-0.5)  [nosep,anchor=west,scale=1.5] {#2};}
\end{tikzpicture}
}

\begin{document}
%-------------------------------------------------------------------------------------------------------------------------------------------------------% 
\title{Quark mass generation due to  scalar fields with zero dimension}
%-------------------------------------------------------------------------------------------------------------------------------------------------------% 
\author{J. Miller}
\affiliation{Ariel University, Ariel, 40700, Israel}

\author{ M.A.Zubkov }
\affiliation{Ariel University, Ariel, 40700, Israel}

%-------------------------------------------------------------------------------------------------------------------------------------------------------% 
\begin{abstract}\noindent 
We propose a model of dynamical symmetry breaking, in which a new type of 
fundamental scalar fields of zero mass-dimension
mediate the couplings of fermions to the gravitational field, 
represented here as a tetrad field in the same manner as
Riemann - Cartan gravity.
In our model, the tetrad couples to the standard model fermions  non-minimally,
and the very coupling coefficients are the fundamental scalar fields.
There are exactly 36 scalar fields  
in the model, which are distinguishable by flavor indices on the fields.
This is the precise number of zero dimension scalar fields  
that leads to
a vanishing Weyl anomaly and a vanishing vacuum energy.
Precisely the same number of these very same scalar fields
is required for the coupling of all of the different standard model fermions
to the vielbein field.
At the same time their interaction with fermions gives rise to 
fermion mass terms, without the need to introduce a fundamental Higgs field. 
Within the proposed theory we construct a toy model that deals solely with the top and bottom quarks, and 
we 
demonstrate that their 
observable masses can appear in the action dynamically.
Moreover, this mechanism allows for the top and bottom quarks to acquire distinctly different masses,
as opposed to our previous, even simpler toy model 
that contained only the top quark. 
\end{abstract}
\date{\today}
%-------------------------------------------------------------------------------------------------------------------------------------------------------% 
\maketitle
\tableofcontents
%-------------------------------------------------------------------------------------------------------------------------------------------------------% 
\section{Introduction}
%-------------------------------------------------------------------------------------------------------------------------------------------------------%
In this article 
we put forward a new model of dynamical symmetry breaking.
We assume that the standard model (SM) fermions  couple to a background gravitational field,
which we represent as a tetrad field  in the same formalism as the Palatini action \cite{Perez:2005pm}.
The coupling of the fermions to the tetrad is the same non-minimal coupling 
as the Holst action \cite{Alexandrov:2008iy} (where the coupling coefficients are constants)
but with the coupling coefficients being 
scalar fields.
The SM has 36 different fermion particles.
Assuming that each one of the 36 fermion fields has a different coupling 
to the tetrad, then 36 different scalar fields 
make up the required coupling terms.
Separately, 
the following recently discovered observation was published in 
\cite{Boyle:2021jaz}. 
The current SM includes 
the leptons, quarks, photon, $W^\pm$ and $Z$ bosons, gluons, 
and the Higgs boson.
With this configuration of particles 
the Weyl anomaly and the vacuum energy are non-zero.
By removing the Higgs boson (which has mass dimension one)
and inserting 36 scalar fields with mass dimension zero,
then not only does the Weyl anomaly vanish, but so does the vacuum energy.
This is in striking agreement with the number of couplings required 
for the coupling of the tetrad field to the SM fermions.
These  observations  motivated our original hypothesis,
that adding 36 zero dimension scalar fields to the action 
could lead to dynamical symmetry breaking,
and in turn the  emergence of fermion mass terms.
We have shown that this mechanism does work in a simplified toy model %version of this model,
containing just one quark \cite{Miller:2022qil}.
Specifically, we showed that for the sector of the SM action 
containing just the top quark,
dynamical symmetry breaking occurs, and the predicted top quark mass 
is very close to its experimentally observed mass.
  In this article we 
generalize this model 
from the case of 
identical masses for both quarks in the third generation, to the case of distinct  mass terms for the 
top and bottom  quarks.
\par 
%-------------------------------------------------------------------------------------------------------------------------------------------------------% 
The Higgs mechanism is 
how  mass terms are ascribed to the $W^\pm$ and $Z$ bosons,
and to the quarks  and leptons. It was first conjectured between 1964 and 1967 in numerous papers,
of which the most seminal are refs. 
\cite{Higgs:1964ia,Higgs:1966ev,Kibble:1967sv,Englert:1964et,Guralnik:1964eu,Weinberg:1967kj},
 at the time when particle physicists were attempting 
to resolve the problem that the SM action  
had no mass terms for those particles.
The Higgs mechanism is  
a $\phi^4$ theory  containing  
an $SU(2)$ doublet of scalar fields %(with mass dimension one),
with a non-zero vacuum expectation value (VEV),
which later became  known as the Higgs Boson.
The $SU(2)\times U(1)$ gauge covariant derivative 
of the Higgs leads to mass terms for the $W^\pm$ and   $Z$,
but breaks the $SU(2)\times U(1)$ symmetry.
In addition a Yuakawa term comprising the Higgs coupled 
to the  fermion fields leads to fermion mass terms, but breaks chiral symmetry. 
These masses  are proportional to the VEV of the Higgs.
If  the VEV is non-zero  these masses will be non-zero,
and as a result the $SU(2)\times U(1)$ symmetry of the theory gets spontaneously broken down to 
$U(1)$, as originally observed by \cite{Nakada:1958zz,Nambu:1960tm,Schwinger:1962tn,Higgs:1964ia,Englert:1964et,Guralnik:1964eu}. 
The current standard electroweak theory was 
formulated \cite{Glashow:1961tr,Weinberg:1967tq,salam68,Politzer:1973fx,Cahn:1989by,Fritzsch:1973pi} 
based on the Higgs mechanism, in which spontaneous  $SU(2)\times U(1)$ symmetry breaking
leads to non-zero masses for the 
$W^\pm$, $Z$, 
quarks and leptons.
The Higgs Boson 
has been detected as a scalar excitation with a mass found at 125 GeV \cite{CMS:2012bfw,Mariotti:2015psa,CMS:2012qbp,CMS:2012dun,ATLAS:2012yve,Mountricha:2012cja},
consistent with theoretical predictions.\par 
%-------------------------------------------------------------------------------------------------------------------------------------------------------% 
Open questions remain about 
the Higgs sector of the SM
as eloquently summarized in \cite{Lane:2002wv},
of which fall into the following categories. 
The Higgs Boson is merely inserted into the 
action by hand with no consideration of how it emerges in nature, except  that without it
particles would be devoid of mass.
Quarks and leptons form matter.  Bosons are the force carriers 
that govern interactions between matter. Is there an analogous  role played by the Higgs Boson?
It's origins remain an enigma. 
The appearance of the Higgs in the action as a $\phi^4$ theory gives rise to self-interactions between the Higgs and itself.
What is the nature of this interaction? 
The SM inclusive of the Higgs 
bears a hierarchy problem with vastly disparate energy scales in the theory.
On the structure of the Higgs itself, is it a fundamental particle or composed of
known fundamental particles itself.
These questions and more   prompted  the pursuit of  composite Higgs models.
\par 
%-------------------------------------------------------------------------------------------------------------------------------------------------------%
One of the earliest composite Higgs model
is based on 
two papers 
\cite{Nambu:1961tp, Nambu:1961fr}
by Nambu Jona and Lasinio (NJL)
in 1961, inspired by the theory of superconductivity
by Bardeen, Cooper and Schrieffer (BCS) first published in \cite{Bardeen:1957mv}
and subsequently given a mathematically elegant form in
\cite{Bogolyubov:1958se,Bogolyubov:1958km,Bogolyubov:1958kj}.
Given that the idea behind the NJL model is replicated in so many  
 later composite Higgs models
based on four-fermion interactions, including our model,  
it is informative to give a paragraph here with an overview of its origins.
In BCS  theory  a pair of fermions join to form a condensate
 forming a type of superconductor.
An energy gap emerges  between the ground state and excited states
of the condensate, due to a self-interaction between the two fermions.
In \cite{Nambu:1961tp, Nambu:1961fr}
an analogy is drawn between this energy gap of condensates, and the energy gap
between the two energies of the  two pairs of solutions in a four-component Dirac spinor.
The analogy is as follows: just like the energy gap in a condensate is created by a self-interaction between fermions,
so too the mass of a fermion arises due to an interaction between massless fermions.
The NJL model inclusive of this interaction 
is based on a Lagrangian 
of the form \cite{Nambu:1961tp}
${\cal L}_{\rm NJL}=-\bar\psi\slashed\partial\psi+g_0 [ (\bar\psi\psi )^2- (\bar\psi\gamma_5\psi )^2 ]$
where $\psi$ is a Dirac fermion and $g_0$ a positive constant with dimensions $[{\rm mass}]^{-2}$.
The four-fermion terms $(\bar\psi\psi)^2$ are precisely analogous to similar four-fermion terms
in BCS theory.
Note that there is no bare mass to violate chiral symmetry.
The NJL model assumes an unknown fermion self-energy function, 
$\Sigma(m,g,\Lambda)$
where $m$ is the observed fermion mass, $g$ a coupling related to the bare coupling
$g_0$, and $\Lambda$ a  cut-off.
% A Dirac particle satisifes the full Dirac equation
Accordingly the full inverse propagator has the general form
\cite{Nambu:1961tp}
$i\gamma\cdot p+m=i\gamma\cdot p+m_0+\Sigma( m,g,\Lambda)=0$
where $m_0$ is the bare mass, giving rise to the 
 \emph{mass gap} equation
 $m-m_0=\Sigma(m,g,\Lambda)$.
From the Lagrangian ${\cal L}_{\rm NJL}$,
an expression for $\Sigma$ can be  derived,
leading to a gap equation of the form
\cite{Nambu:1961tp}
$m=m\, g_0\int \dbar  p(p^2+m^2-i\epsilon)^{-1}F(p,\Lambda)$
where $F(p,\Lambda)$ is a cut-off factor and $\int \dbar p(\dots)=(2\pi)^{-4}\int d^4p(\dots)$.
In this way the chiral condensate ($\bar\psi\psi$) in ${\cal L}_{\rm NJL}$
leads to an effective mass term and hence spontaneously broken chiral symmetry.
In the same year BCS theory was applied to the problem of dynamical symmetry breaking in the standard model
\cite{larkin:61} using a similar approach.
A more pedagogical description of the NJL model can be found for example in refs.
\cite{Bijnens:1992uz,Prades:1993ys,Scherer:2002tk}.
\par 
%-------------------------------------------------------------------------------------------------------------------------------------------------------%
Despite   its elegance, the NJL model falls short of constituting a viable model.
Firstly, the formation of chiral condensates at a certain energy means there is no confinement.
Secondly, the NJL model is non-renormalizable in four dimensions, so  at best 
it can only be an effective field theory, which needs UV completion.
Still, it remains a highly useful starting point for informing 
subsequent composite Higgs models based on the notion of a four-fermion interaction.
And importantly,  a precise relation for the mass of the Higgs boson 
can be extracted from the NJL model, described below.
\par 
%-------------------------------------------------------------------------------------------------------------------------------------------------------%
% In  an article in 1985 \cite{Nambu:1984ph} 
Nambu  \cite{Nambu:1984ph}
discovered a precise relation
between the masses of the boson excitation states 
of the theory, namely the Higgs bosons that emerge,
and the mass of the fermion that forms the condensate,
known as the Nambu sum rule. It has the form
$m_1^2+m_2^2+\dots=4m_f^2$
where $m_i$ is the mass of each Higgs boson and $m_f$ is the mass of the condensed fermion.
A similar relation was also discussed earlier in the original article on the  NJL approximation to QCD in 
\cite{Nambu:1961tp}.
It was pointed out in \cite{Volovik:2013bjq},
based on the Nambu sum rule  the  possibility that
 the 125 GeV excitation 
discovered 
\cite{CMS:2012bfw,Mariotti:2015psa,CMS:2012qbp,CMS:2012dun,ATLAS:2012yve,Mountricha:2012cja},
is not the only Higgs boson.
Higgs bosons appear in top-quark condensation models, where the masses of the
bosonic excitations are related to the top quark mass by a sum rule \cite{Volovik:2013bjq}  similar to the Nambu sum rule of the
NJL model \cite{Nambu:1961tp}. 
This sum rule suggests the existence of so called 
\emph{Nambu partners} for the 125 GeV Higgs boson,
whose masses can be predicted by the Nambu sum rule. For example, if there are only two states in the given channel, the mass of the Nambu
partner is $\sim$ 325 GeV.
They together satisfy the Nambu sum rule $m_1^2+m_2^2 = 4m_t^2$ , where $m_t\sim 175$ GeV is
the mass of the top quark.
Based on this idea it has been suggested by Volovik and Zubkov \cite{Volovik:2013bjq,Volovik:2012qq,Volovik:2013lta}
that other scalar excitations corresponding to the Higgs Boson may exist, as well as the 125 GeV excitation discovered in 
\cite{CMS:2012bfw,Mariotti:2015psa,CMS:2012qbp,CMS:2012dun,ATLAS:2012yve,Mountricha:2012cja}.
\par 
%-------------------------------------------------------------------------------------------------------------------------------------------------------%
A specific model based on the 
Higgs scalar being a fermion condensate
was 
put forward by Terazawa et. al. in \cite{Terazawa:1976eq},
which was later  made more precise  
\cite{Terazawa:1980nck} with the hypothesis that the Higgs is 
composed of  a top-quark condensate ($t\bar t$).
This was guided by the result that the mass of the Higgs scalar, 
as predicted by the Nambu sum rule,
indicates that the Higgs scalar is close to the   
$t\bar t$ threshold, and could behave like a $t\bar t$ bound state.
Later on 
in 1989, the original $t\bar t$
model of Terazawa et. al. was revived by Miransky, Tanabashi, Yamawaki \cite{Miransky:1989ds,Miransky:1988xi},
and later by Bardeen, Hill and Lindner \cite{Bardeen:1989ds,Marciano:1989xd,Hill:1991at,Hill:2002ap},
who proposed the top-quark condensate as a means of electroweak symmetry breaking.
In the $t\bar t$ model the 
Higgs is composite at short distance scales.
However 
in such $t\bar t$ models 
the energy scale of  new dynamics is assumed to be $\sim 10^{15}$ GeV, 
corresponding to a Higgs mass of the order $2 m_t \sim 350$ GeV 
\cite{Terazawa:1976eq,Miransky:1989ds,Bardeen:1989ds,Marciano:1989xd,Hill:1991at,Hill:2002ap}, 
which is starkly different from experimental observations.
%A large degree of fine-tuning of the parameters in the model is needed to obtain a smaller value for the Higgs mass.
In response to this problem the 
\emph{top seesaw} model was 
put forward by Chivukula, Dobrescu, Georgi and Hill \cite{Chivukula:1998wd}.
but 
in addition to the  top quark, an additional heavy fermion, $\chi$ \cite{Dobrescu:1997nm,Chivukula:1998wd}
is required by the model.
\par 
%-------------------------------------------------------------------------------------------------------------------------------------------------------% 
In the 1970s the technicolor (TC) model was formulated 
\cite{Susskind:1978ms,Weinberg:1975gm} to resolve issues arising from the SM with a fundamental Higgs boson
(listed in the opening remarks),
while at the same time 
predicting a Higgs mass in agreement with experiments.
The TC  model is
a gauge theory of fermions with no elementary scalars,
which also has 
dynamical electroweak symmetry breaking  and flavor symmetry breaking.
The Higgs, instead of being a fundamental scalar is composed of \emph{technifermions}, a new class of fermions 
interacting via technicolor gauge bosons. This interaction is attractive, and hence,
by analogy with BCS theory, generates fermion condensates. 
By itself the TC model
doesn't have mass terms for the quarks and leptons, so it 
was extended to the \emph{Extended Technicolor} (ETC) model \cite{Dimopoulos:1979es,Eichten:1979ah,Chivukula:2013,Sayre:2011ed},
but the ETC model is inconsistent with experimental constraints on flavor changing neutral currents,
and precision electroweak measurements.
Later on   the \emph{walking technicolor model}
\cite{Appelquist:1998rb,Gudnason:2006ug} 
was composed
that resolves the above issues, but it yields the wrong value for the top-quark mass.
For an in-depth review of the original TC model and its various derivatives 
the reader should consult Refs. \cite{Weinberg:1975gm,Weinberg:1979bn,Susskind:1978ms,Hill:2002ap,Chivukula:1990bc},
while for a more pedagogical introduction to the TC and ETC models see Refs. \cite{Lane:2002wv,Lane:1993wz}.
A more  recent alternative to the TC model can also be found in 
\cite{Simonov:2009hv,Simonov:2010wd,Simonov:2010mr}.
\par 
%-------------------------------------------------------------------------------------------------------------------------------------------------------%
Subsequently the partial compositeness model was conjectured,   by Kaplan
\cite{Kaplan:1991dc} initially. Therein
each SM particle has a heavy partner that can mix with it. 
Like this the SM particles are linear combinations of elementary and composite states
via 
a mixing angle.
The elegance of the partial compositeness model is its simplicity,
without the need for deviations beyond the standard model, thus  relying 
on the known fundamental particles only \cite{Redi:2011zi}.
% \par 
%-------------------------------------------------------------------------------------------------------------------------------------------------------% 
More recently, among a variety of theoretical papers that appeared between December 2015 and August 2016,  there are several that consider both
the $125$ GeV Higgs boson, and the hypothetical new heavier Higgs boson as composite due to the new strong interaction, 
as noticed  for example in Refs. 
\cite{Hong:2016uou,Matsuzaki:2016joz}. 
In a different approach there are other papers devoted to the description of the composite nature of the heavier ($750$ GeV) Higgs boson only 
\cite{Harigaya:2016pnu,
Nakai:2015ptz,
Franceschini:2015kwy,
Molinaro:2015cwg,
Bian:2015kjt,
Bai:2015nbs,
Cline:2015msi,
Ko:2016wce,
Harigaya:2016pnu,
Redi:2016kip,Harigaya:2016eol,
Foot:2016llc,
Iwamoto:2016ral,
Bai:2016vca,Kobakhidze:2015ldh,
Foot:2016llc}.
% \par 
%-------------------------------------------------------------------------------------------------------------------------------------------------------% 
Other  more modern composite Higgs  models consist of  
the walking model,
\cite{Holdom:1988gs,Holdom:1989aa}
the ideal walking  model
\cite{Yamawaki:1996vr,Fukano:2010yv,Rantaharju:2017eej,Rantaharju:2019nmh},
the technicolor scalar model
\cite{Foadi:2012bb}, 
Sannino's model of the generalized orientafold gauge theory approach to electroweak symmetry breaking
\cite{Sannino:2004qp},
and 
the walking model in higher-dimensional $SU(N)$ gauge theories
of Dietrich and Sannino
\cite{Dietrich:2006cm,Dietrich:2005jn}.
A more comprehensive review of modern composite Higgs models can be found in 
\cite{Cacciapaglia:2020kgq}.
\par 
%-------------------------------------------------------------------------------------------------------------------------------------------------------% 
We have discussed a number of 
composite Higgs models
based on fermion condensates.
It
is reasonable to suppose that
the $125$ GeV Higgs, as well as 
the extra 
scalars suggested in \cite{Volovik:2013bjq,Volovik:2012qq,Volovik:2013lta},
are composed of known SM fermions,
due to some unknown %strong 
interaction between them with a scale $\Lambda$ above $1$ TeV. 
If it exists,  such an interaction would  have  specific properties that make it distinctly different from the conventional TC interactions
\cite{Hill:2002ap,Lane:2002wv,Chivukula:2000mb}.
% \par 
%-------------------------------------------------------------------------------------------------------------------------------------------------------%
Firstly, these interactions must occur at appropriate energy scales such there is no confinement,
since otherwise they would confine quarks and leptons to extremely small regions of space $\sim 1/\Lambda$ to the point that
strong and weak interactions would take place on length scales unobservable. 
At the same time, these interactions must produce spontaneous symmetry breaking needed to make $W$ and $Z$ bosons massive,
and spontaneous chiral symmetry breaking to yield massive fermions.
A number of models with these properties have been 
thought of 
within the framework of 
topcolor models
\cite{Bardeen:1989ds,Simmons:2011aa,Marciano:1989xd,Cvetic:1997eb,Miransky:1988xi,Chivukula:2012cp,Hill:2002ap,Hill:1996te,Buchalla:1995dp}. 
Other candidates considered are topcolor assisted technicolor models
\cite{Hill:1994hp, Lane:1995gw, Popovic:1998vb, Braam:2007pm,Chivukula:2011ag},
which  combines both technicolor and topcolor ingredients.
Indeed in refs. \cite{Simonov:2009hv,Simonov:2010wd,Simonov:2010mr,Volovik:2013bjq,Volovik:2012qq} 
such a model was suggested in which chiral condensates 
 appear 
 comprising SM fermions,  not necessarily just the $t\bar t$ condensate. 
 This particular configuration 
provides masses for both the $W$, $Z$ bosons, and for the fermions.\par 
\par 
%-------------------------------------------------------------------------------------------------------------------------------------------------------%
Quantum gravity in the first order formalism with
either the Palatini action \cite{palatini1919,RovelliPalatini} or the Holst action \cite{Holst:1995pc},
shares a property with the above composite Higgs models: 
it
leads to a four - fermion interaction.
This four fermion interaction 
is now between spinor fields
coupled in a minimal way to the torsion field \cite{Perez:2005pm,Rovelli4fermion}, 
but importantly 
this
four - fermion interaction leads to fermion condensates 
of the type that could constitute the composite Higgs. 
%-------------------------------------------------------------------------------------------------------------------------------------------------------% 
A similar idea has already been 
suggested in Ref. \cite{Zubkov:2010sx}, namely that the torsion field coupled in a non-minimal way
to fermion fields \cite{Belyaev:1998ax,Shapiro:1994vs,Shapiro:2001rz}  could be a source of
dynamical electroweak symmetry breaking.
\par 
%-------------------------------------------------------------------------------------------------------------------------------------------------------%
Starting with 
a  theory of the  SM  coupled to gravity,
the fermion and gauge fields  are coupled to a background classical gravitational 
field, with
a right-handed neutrino for each generation also coupled to the background field.
This representation is a QFT on a classical background spacetime.
The Weyl anomaly
is a measure of the failure of the classically Weyl-invariant theory
to define a Weyl-invariant quantum theory,
and can be expressed in terms of the number of different fields present in the theory.
In the SM there are $n_{0}=4$ ordinary real scalars in the usual complex Higgs doublet, $n_{1/2}=3\times16=48$ Weyl spinors (16 per generation), $n_{1}=8+3+1=12$  gauge fields of
$SU(3)\times SU(2)\times U(1)$, and a single 
gravitational field ($n_{2}=1$).
With these combinations the Weyl anomaly is non-zero.
However suppose that the Higgs and graviton fields are not fundamental fields but rather composite, such that their contributions to the vacuum energy and Weyl anomaly 
can be dropped, and  $n_{0}=n_{2}=0$.  The implication is that by introducing $n_{0}'=36$  scalars with zero mass-dimension, 
then not only does the  Weyl anomaly vanish, but also   
the vacuum energy vanishes as well.  
\par 
%-------------------------------------------------------------------------------------------------------------------------------------------------------% 
This embodies our motivation for the starting point of 
this work: a model comprising
Dirac fermions of the SM 
with a non-minimal coupling to
gravity, that includes 
36 fundamental scalar fields.
The details of the construction of the theory is
described in  full in 
\S\ref{sec_nonminimal_coupling_of_fermions_to_gravity}.
We find that in this model,
the theory admits  mass terms for the fermions.
By using the Schwinger-Dyson equation for the fermion self-energy function,
a relationship is derived between the top-quark mass, the Planck mass (the ultraviolet cut-off point of the loop integral) and the strength of the inverse coupling between the fermion fields and the vielbein.
From this relation 
an estimation for the mass of the top-quark can be extracted.\par 
%-------------------------------------------------------------------------------------------------------------------------------------------------------%
This paper is organized in the following way.
In \S\ref{sec_vielbein_formalism} we introduce the tetrad formalism,
the Palatini action. We explain the origins of the form of the action 
of fermions coupled non-minimally to gravity and sketch the derivation.
In \S\ref{sec_weyl_anomaly} we expand on the definition of the Weyl anomaly, and 
show that  it vanishes with zero dimension scalars added to the SM.
We also summarize our previous toy model with just one quark.
In \S\ref{sec_fundamental_scalars_in_models_of_3_generations_of_SM_fermions}
we construct the action of our model: zero dimension scalars coupled to gravity. We then
focus in on the sector of the action containing the third generation of quarks coupled to
the gravitational (tetrad) field through zero dimension scalar fields.
In \S\ref{sec_SD_for_calculating_mf}
we apply the Schwinger-Dyson approach to derive the mass gap equations for the top and bottom quarks
and we present our numerical results for the masses and the couplings coefficients.
In \S\ref{sec_conclusions} we summarize our results and lay out our plans for the next piece of this research.
Appendix \S\ref{sec_dirac_matrices} contains a list of definitions for the Dirac matrices in the conventions used throughout this paper,
together with a number of useful identities.
In Appendix \S\ref{sec_wick_rotatns} we give details of the Wick rotations used for simplifying
the calculation of the one-loop contribution to the 
self-energy of the quarks (see Fig.~\ref{fig_1}).
Finally in Appendix \S\ref{sec_angular_integral}
gives the of how to evaluate the angular  integrals 
in the one-loop diagram in Fig.~\ref{fig_1}.\par 
%-------------------------------------------------------------------------------------------------------------------------------------------------------% 
Throughout this article
lower case Latin letters
$a,b,c\dots=0,1,2,3$
label internal Lorentz indices,
Greek letters 
$\mu,\nu,\dots=0,1,2,3$
label spacetime indices, and 
$\epsilon_{abcd}$ is the completely antisymmetric Levi-Civita symbol.
Lorentz indices are raised and lowered by the Minkowski metric $\eta_{ab}$ and 
spacetime indices are raised and lowered by a spacetime metric $g_{\mu\nu}$.
Metrics are assumed to be  `mostly negative', 
and specifically the Minkowski metric is
$\eta_{ab}={\rm diag}(1,-1,-1,-1)$.
%-------------------------------------------------------------------------------------------------------------------------------------------------------% 
\section{The vielbein formalism}
\label{sec_vielbein_formalism}\noindent
%-------------------------------------------------------------------------------------------------------------------------------------------------------% 
The aim of this section is to offer a brief introduction to the notations and conventions of the tetrad formalism,
and some insight into why it is chosen as a starting point for our model.
The majority of this overview is based on \cite{Perez:2005pm}.
Only the main results are given here. For more details of how these results are derived the interested reader should consult that reference.
\par 
In the vielbein formalism the metric $g_{\mu\nu}(x)$
is expressed in terms 
of a \emph{tetrad field } or \emph{vielbein  field} ${e^a}_\mu(x)$ through the completeness relation
%-------------------------------------------------------------------------------------------------------------------------------------------------------% 
\begin{equation} g_{\mu\nu}(x)={e^a}_{\mu} (x){e^b}_\nu (x)\eta_{ab}\label{completeness_relation}\end{equation}
%-------------------------------------------------------------------------------------------------------------------------------------------------------% 
where $\eta_{ab}={\rm diag}\left(-1,1,1,1\right)$
is the flat space Minkowski metric.
The  orthogonality relation,
%-------------------------------------------------------------------------------------------------------------------------------------------------------% 
\begin{equation} \eta_{ab}={e^\mu}_{a}(x){e^\nu}_b (x)g_{\mu\nu}(x)\label{orthogonality_relation}\end{equation}
%-------------------------------------------------------------------------------------------------------------------------------------------------------% 
where ${e^\mu}_{a}$ is the inverse tetrad, can be obtained directly from the completeness relation %in (\ref{completeness_relation})
via the inverse conditions 
%-------------------------------------------------------------------------------------------------------------------------------------------------------% 
\begin{equation} {e^\mu}_{a}{e^a}_\nu={\delta^\mu}_\nu\ ,\qquad {e^\mu}_a {e^b}_\mu={\delta^b}_a\ ,\label{inverse_conditions}\end{equation}
%-------------------------------------------------------------------------------------------------------------------------------------------------------% 
where ${\delta^b}_a$ and ${\delta^\mu}_\nu$ are kronecker delta functions.
% The dual vielbein fields $e^\mu_a$ that are dual to $e^a_\mu$ are defined through the inverse conditions in (\ref{inverse_conditions}).
\par 
%-------------------------------------------------------------------------------------------------------------------------------------------------------% 
The vielbein ${e^a}_{\mu} (x)$ acts as a map from tangent space at a given point $x$ to flat Minkowski space.
In this regard
the vielbein captures Einstein's intuition that spacetime is locally flat.
The metric is invariant under local $SO(3,1)$ gauge transformations
in the sense that (\ref{completeness_relation}) is invariant under 
%-------------------------------------------------------------------------------------------------------------------------------------------------------% 
\begin{equation}{e^a}_\mu(x)\to{\Lambda^a}_b {e^b}_\mu(x)\ .\label{vielbein_SO3_invariance}\end{equation}
\par 
%-------------------------------------------------------------------------------------------------------------------------------------------------------%
The action for the gravitational field itself is given by the Einstein-Hilbert (EH) action
$\int dx\,\sqrt{-g}R$, where $R$ is the Ricci scalar and $g $ is the determinant of the metric.
Its variation with respect to the metric yields the Einstein equations.
The action itself contains up to second order derivatives in the metric.
As an attempt to simplify the field equations Palatini \cite{palatini1919} 
proposed 
an action in which the metric and connection are independent variables, in order that the action will have up to first order derivatives only, thus simplifying
the equations of motion.
The Palatini action is given in terms of the tetrad instead of the metric,
and has the form
%-------------------------------------------------------------------------------------------------------------------------------------------------------% 
\begin{equation} S[e,\omega]= \int dx\, e\,{e^\mu}_a {e^\nu}_b {F_{\mu\nu}}^{ab} \ .\label{palatini_action} \end{equation}
%-------------------------------------------------------------------------------------------------------------------------------------------------------%
Here 
$e=\sqrt{-g}$ is the determinant of
${e^a}_\mu$ and  
${\omega_\mu}^{ab}$ is a Lorentz connection 
that defines a \emph{covariant partial derivative}
on fields with Lorentz ($a$) indices 
as $D_\mu V^a=\partial_\mu V^a+{\omega_\mu}^{ab} v_b$,
and in general a \emph{gauge-covariant exterior derivative} $D$
on $p$-forms. For example, for a one-form ${T^a}_\mu$
with a Lorentz index,
${D_{[\mu}T_{\nu]}}^a={\partial_{[\mu}T_{\nu]}}^a+{\omega_{[\mu}}^{ab}T_{\nu]b} $.
The latter is often written with spacetime indices omitted as
$D T^a={\rm d} T^a+{\omega^a}_b\wedge T^b$, where ${\rm d}$ is an exterior derivative.

$F$ is the curvature of $\omega$
given by
${F_{\mu\nu}}^{ab} v_b=[D_\mu,D_\nu] v^a=\partial_{[\mu}{\omega_{\nu]}}^{ab}+{{\omega_{[\mu }}^{ac}} {\omega_{\nu]c}}^{ b}$,
or equally
%-------------------------------------------------------------------------------------------------------------------------------------------------------%
\begin{equation}F^{ab}={\rm d}\omega^{ab}+{\omega^{a}}_c \wedge \omega^{cb}\ ,\label{2nd_Cartan_eqtn}\end{equation}
%-------------------------------------------------------------------------------------------------------------------------------------------------------%
where the first term is an exterior derivative, and spacetime indices have been omitted in (\ref{2nd_Cartan_eqtn}).
%-------------------------------------------------------------------------------------------------------------------------------------------------------% 
The invariance of the action (\ref{palatini_action})
with respect to variations 
of
the connection $\omega$ yields 
%-------------------------------------------------------------------------------------------------------------------------------------------------------%
\begin{equation}D_{[\mu}{e^a}_{\nu]}=0\ .\label{1st_Cartan_eqtn}\end{equation}
%-------------------------------------------------------------------------------------------------------------------------------------------------------%
Eq.~(\ref{1st_Cartan_eqtn}) with spacetime indices omitted has the form
$De^a\equiv{\rm d}e^a+{\omega^a}_be^b=0$.
The left-hand side is the definition of the \emph{torsion two-form}.
The equation of motion in 
(\ref{1st_Cartan_eqtn}) is the statement that the torsion vanishes.
%-------------------------------------------------------------------------------------------------------------------------------------------------------%
\par 
%-------------------------------------------------------------------------------------------------------------------------------------------------------%
Eq.~(\ref{1st_Cartan_eqtn}) has a unique solution, denoted by $\omega  = \omega[e]$,
with the form
%-------------------------------------------------------------------------------------------------------------------------------------------------------%
\begin{equation}
{\omega^{ab}}_{ \mu}[e] =   {e^a}_{ \alpha}\nabla_\mu\, {e^b}_{ \beta}\ \eta^{\alpha\beta}\ . \label{spin_connection}
\end{equation}
%-------------------------------------------------------------------------------------------------------------------------------------------------------% 
$\omega[e]$ is the torsion-free spin connection of the tetrad field $e$.
%-------------------------------------------------------------------------------------------------------------------------------------------------------% 
% \footnote{
It is analogous to
to the metric compatibility condition $\nabla_\mu g_{\alpha\beta}=0$ where $\nabla$ is the 
covariant spacetime derivative in GR,
whose unique solution is a torsion-free metric.%}
\par 
%-------------------------------------------------------------------------------------------------------------------------------------------------------% 
Substitution of $\omega=\omega[e]$ in the curvature  leads to
${F_{\alpha\beta}}^{ab}\left(\omega[e]\right)={R_{\alpha\beta}}^{ab}$
where 
${R_{\alpha\beta}}^{ab}={R_{\alpha\beta}}^{\mu\nu}{e^a}_\mu {e^b}_\nu$
and ${R_{\alpha\beta}}^{\mu\nu}$ is the ordinary Riemann tensor.
The invariance of the action (\ref{palatini_action}) 
under  variations 
of
$e$ leads to the vacuum Einstein equation.
%-------------------------------------------------------------------------------------------------------------------------------------------------------% 
Eq.~(\ref{1st_Cartan_eqtn}) is called the \emph{first Cartan equation}
while (\ref{2nd_Cartan_eqtn}) is called the \emph{second Cartan equation}.  
\par 
%-------------------------------------------------------------------------------------------------------------------------------------------------------% 
An extension of the Palatini action called the Holst action
bears the same equations of motion as (\ref{palatini_action}).
The Holst action 
\cite{Holst:1995pc} has the form
%-------------------------------------------------------------------------------------------------------------------------------------------------------% 
\begin{equation}
S_{\rm Holst}[e,\omega]=
\int dx\, e\,{e^\mu}_a {e^\nu}_b {F_{\mu\nu}}^{ab}
-\dfrac1{\gamma}
\int dx\, e\,{e^\mu}_a {e^\nu}_b \star {F_{\mu\nu}}^{ab}
\ ,\label{holst_action}\end{equation}
%-------------------------------------------------------------------------------------------------------------------------------------------------------% 
where 
${}^\star  {F_{\mu\nu}}^{ab}=\frac{1}{2}\epsilon^{ab}{}_{cd}{F_{\mu\nu}}^{cd}$
is the Hodge dual, 
and the parameter $\gamma$   is  the \emph{Immirzi parameter} \cite{Immirzi:1996dr}. 
The first term in (\ref{holst_action}) is identical to (\ref{palatini_action}), while 
the second distinguishes the Holst action  from the Palatini action.
The functional dependence of (\ref{holst_action}) on the tetrad is identical to (\ref{palatini_action}),
thus it's variation with $e$ yields the Einstein equations in the same manner.
Variation of (\ref{holst_action}) 
with  $\omega$ 
leads to the same equation of motion as (\ref{1st_Cartan_eqtn}), with the solution $\omega=\omega[e]$.
Substitution of $\omega=\omega[e]$  in (\ref{holst_action}) forces the second term to vanish 
due to the Bianchi identity 
$R_{[\alpha\beta\mu]\nu}=0$ .
\par 
%-------------------------------------------------------------------------------------------------------------------------------------------------------% 
The equation of motion (\ref{1st_Cartan_eqtn}) changes when fermions are introduced into the action,
resulting in a solution to to $\omega[e]$.
In the presence of a fermion field, $\psi$ the action (\ref{palatini_action})
becomes %\cite{Perez:2005pm}
%-------------------------------------------------------------------------------------------------------------------------------------------------------% 
\begin{equation} S[e,\omega,\psi]=S[e,\omega]+\dfrac{i}2\int dx\,\left(\bar\psi\gamma^a {e^\mu}_a D_\mu \psi 
-\overline{D_\mu\psi}\gamma^a{e^\mu}_a\psi\right)\ ,\label{S_f}\end{equation}
%-------------------------------------------------------------------------------------------------------------------------------------------------------% 
where $S[e,\omega]$ is given in (\ref{palatini_action}), $\gamma^a$ are the Dirac matrices, and 
%-------------------------------------------------------------------------------------------------------------------------------------------------------% 
\begin{equation}  D_\mu\psi=  \partial_\mu\psi-\frac{1}4{\omega_\mu}^{ab}\gamma_a\gamma_b \psi\ .   \label{covariant_derivative}\end{equation}
%-------------------------------------------------------------------------------------------------------------------------------------------------------% 
By varying $\psi$ in (\ref{S_f}) the $SO(3,1)$ gauge-covariant Dirac equation is implied:
%-------------------------------------------------------------------------------------------------------------------------------------------------------% 
\begin{equation} (i\gamma^a {e^\mu}_a D_\mu-m)\psi= 0\ .\label{cov_dirac_eqn}\end{equation}
\par 
%-------------------------------------------------------------------------------------------------------------------------------------------------------% 
Upon varying  (\ref{S_f}) with  $\omega$,  
the following equation of motion is obtained %\cite{Perez:2005pm}:
%-------------------------------------------------------------------------------------------------------------------------------------------------------% 
\begin{equation}\label{tff} 
D_\mu\left(e\, {e^{[\mu}}_a {e^{\nu]}}_b\right)=e\,{J_{ab}}^\nu
\  \end{equation}
%-------------------------------------------------------------------------------------------------------------------------------------------------------% 
where the \emph{fermion current} is given by
%-------------------------------------------------------------------------------------------------------------------------------------------------------% 
\begin{equation}{J_{ab }}^\nu=\dfrac1{4}{e^\nu}_d{\epsilon_{abc}}^d\,J^c\ ,\qquad J^c\equiv  \bar\psi\gamma_5\gamma^c\psi \ .  \end{equation}
%-------------------------------------------------------------------------------------------------------------------------------------------------------% 
It is immediately clear from (\ref{tff})  that the presence
of a fermion field in the  action introduces a torsion component in the connection arising from the fermion
current, and thus the connection is no longer torsion free, that is, $D_{[\mu}{e_{\nu]}}^a\neq 0 $.
Eq.~(\ref{tff}) relates the tetrad to the current, which is itself a bilinear function of the fermion field.
This is the same manner in which the background metric $g$ in GR is specified in terms of matter fields present in the Lagrangian
through the Einstein equations.
\par 
%-------------------------------------------------------------------------------------------------------------------------------------------------------% 
Because of (\ref{tff}), 
the second term in (\ref{holst_action})
no longer vanishes.
The steps leading up the solution to (\ref{tff})
are given in \cite{Perez:2005pm}.
The solution   denoted $\omega[e,\psi]$  carries an explicit dependence on the fermion field $\psi$ itself, and has the form
%-------------------------------------------------------------------------------------------------------------------------------------------------------% 
\begin{equation}
{\omega_{\mu}}^{ab}[e,\psi]
=
-\dfrac{\gamma}{\gamma^2+1}\left(2 {e_\mu}^{[a} J^{b]}-\gamma {{\epsilon}^{ab}}_{cd} J^c{e_\mu}^{d}\right) 
\ .\label{connection_with_torsion}  \end{equation}
%-------------------------------------------------------------------------------------------------------------------------------------------------------% 
% This solution for the connection $\omega$ may be substituted into the action (\ref{holst_action}).
The connection in (\ref{connection_with_torsion}) may be substituted back into the action (\ref{holst_action}).
The key ingredient is that %the  connection 
$\omega[e,\psi]$  contains fermion terms
that couple   back to the fermions via the covariant derivative. 
The full derivation is not given here, but can be found in 
\cite{Perez:2005pm}.
The resulting form for the total action is
%-------------------------------------------------------------------------------------------------------------------------------------------------------% 
\begin{equation}S[e,\psi]=S[e]
+ S_{f}[e,\psi] + S_{I}[e,\psi], \label{final_actn_1}\end{equation}
%-------------------------------------------------------------------------------------------------------------------------------------------------------% 
where the first two terms
are the standard second--order tetrad action of general relativity
with fermions,  
%-------------------------------------------------------------------------------------------------------------------------------------------------------% 
\begin{align}
S[e]+ S_{f}[e,\psi]=&  \int
dx \, e\, {e^{\mu}}_{a} {e^{\nu}}_{b}\ {F_{\mu\nu}}^{ab}[\omega[e]] \nonumber\\
&+ \frac{i}2
\int dx \,e \left(\overline\psi\, \gamma^a {e^\mu}_{a}  D_{\mu}[\omega[e]]-\overline{D_{\mu}[\omega[e]]\psi}\,\gamma^a{e^\mu}_a\psi\right)
\psi, \label{final_actn_2}
\end{align}
%-------------------------------------------------------------------------------------------------------------------------------------------------------% 
and the interaction term is 
%-------------------------------------------------------------------------------------------------------------------------------------------------------% 
\begin{equation}
S_{I}[e,\psi]= -
\frac{\gamma^2}{\gamma^2+1} \int dx \ e\
(\overline{\psi}\gamma_5\gamma_{a}\psi)\
(\overline{\psi}\gamma_5\gamma^{a}\psi)\ .\label{final_actn_3} 
\end{equation} 
%-------------------------------------------------------------------------------------------------------------------------------------------------------% 
The first term is the same
second--order tetrad action of general relativity given by (\ref{palatini_action}) with $\omega=\omega[e]$,
and the second term is the action of fermions coupled to gravity given in (\ref{S_f}).
The third term 
describes a four-fermion interaction mediated by a 
non-propagating torsion.  An interaction of
this form is well known: it is predicted by the Einstein-Cartan
theory. 
Crucially, this is precisely the  type of four-fermion interaction   that could forge a composite Higgs model,
as discussed in the introductory remarks above.\par 
%-------------------------------------------------------------------------------------------------------------------------------------------------------% 
From now on spacetime indices are omitted, and the total action in  (\ref{final_actn_1}) shall be
 written in the equivalent form
\cite{Rovelli:2014ssa,Perez:2005pm}
%-------------------------------------------------------------------------------------------------------------------------------------------------------% 
\begin{align} S[e,\psi]=&  S[e]+S_f[e,\psi]+S_I[e,\psi]\label{final_actn-4}\\[0.5em]
=&
\int dx\, e\,e^a\wedge e^b\wedge F^{cd}\epsilon_{abcd}\nonumber\\
&+\frac1{6}
\int dx \,  \theta^a \wedge e^b\wedge e^c\wedge e^d\epsilon_{abcd}
\nonumber\\
&+\int dx\,(\bar\psi\gamma_5\gamma_a\psi)\,(\bar\psi\gamma_5\gamma^a\psi) \ ,\label{final_actn-5}\end{align}
%-------------------------------------------------------------------------------------------------------------------------------------------------------% 
where
%-------------------------------------------------------------------------------------------------------------------------------------------------------% 
\begin{equation}
\theta^a\equiv\frac{i}2\left(\bar\psi\gamma^a D_\mu\psi -\overline{D_\mu\psi}\gamma^a\psi\right)dx^\mu \ .\label{theta_0^a}\end{equation}
%-------------------------------------------------------------------------------------------------------------------------------------------------------%
Eq.~(\ref{final_actn-4}) is called the \emph{minimal coupling} action.
\par 
%-------------------------------------------------------------------------------------------------------------------------------------------------------%
A more general action is derived in \cite{Alexandrov:2008iy}
that gives rise to the same equations of motion has the same form as 
Eq.~(\ref{final_actn-4}) but with 
%-------------------------------------------------------------------------------------------------------------------------------------------------------% 
\begin{equation}
\theta^a\equiv\frac{i}2\left(\bar\psi(1+\xi)\gamma^a D_\mu\psi -\overline{D_\mu\psi}(1+\xi^\ast)\gamma^a\psi\right)dx^\mu \ ,\label{theta^a}\end{equation}
%-------------------------------------------------------------------------------------------------------------------------------------------------------%
where $\xi$ is a complex constant.
The resulting action is called the \emph{non-minimal coupling} action.
%-------------------------------------------------------------------------------------------------------------------------------------------------------% 
\section{The Weyl anomaly and the role of  dimension-zero scalar fields}
\label{sec_weyl_anomaly}
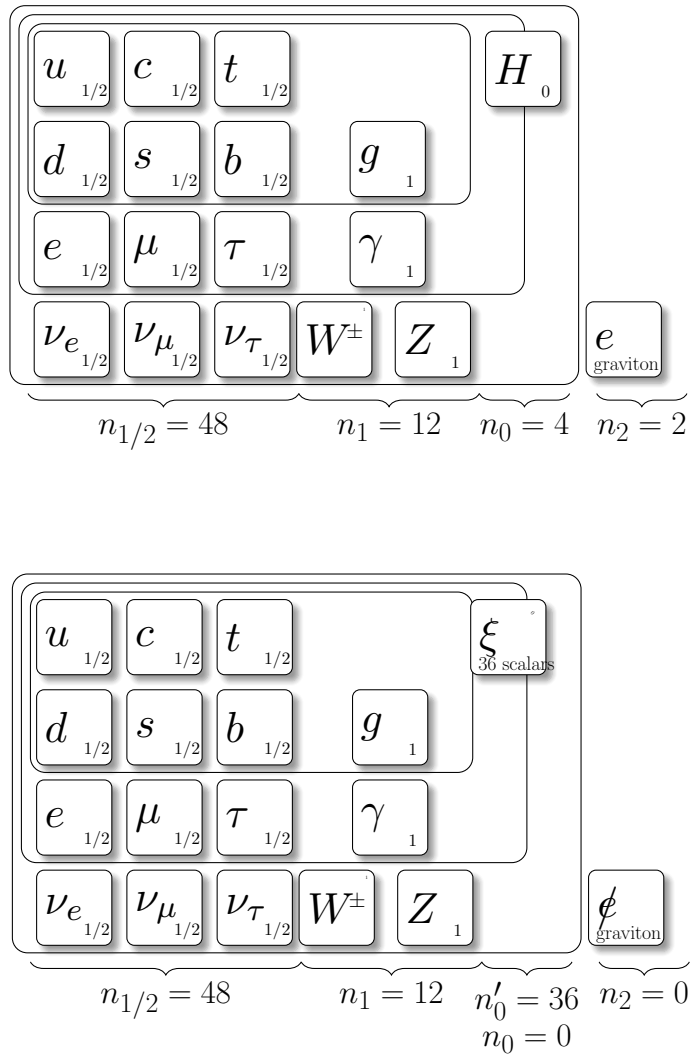
\begin{figure}
\begin{tikzpicture}[x=1.2cm, y=1.2cm]
\draw[round] (-0.5,0.5) rectangle (4.4,-1.5);
\draw[round] (-0.6,0.6) rectangle (5.0,-2.5);
\draw[round] (-0.7,0.7) rectangle (5.6,-3.5);

\node at(0, 0)   {\particl 
			{$u$}       
			{} {} {$1/2$} {} {} 
			};
\node at(0,-1)   {\particl 
			{$d$}        
			{} {} {$1/2$} {} {} 
			};
\node at(0,-2)   {\particl 
			{$e$}       
			{} {} {$1/2$} {} {} 
			};
\node at(0,-3)   {\particl 
			{$\nu_e$}   
			{} {} {$1/2$} {} {} 
			};
\node at(1, 0)   {\particl
			{$c$}        
			{} {} {$1/2$} {} {} 
			};
\node at(1,-1)   {\particl 
			{$s$}        
			{} {} {$1/2$} {} {} 
			};
\node at(1,-2)   {\particl
			{$\mu$}       
			{} {} {$1/2$} {} {} 
			};
\node at(1,-3)   {\particl
			{$\nu_\mu$}   
			{} {} {$1/2$} {} {} 
			};
\node at(2, 0)   {\particl
			{$t$}         
			{} {} {$1/2$} {} {} 
			};
\node at(2,-1)   {\particl
			{$b$}        
			{} {} {$1/2$} {} {} 
			};
\node at(2,-2)   {\particl
			{$\tau$}      
			{} {} {$1/2$} {} {} 
			};
\node at(2,-3)   {\particl
			{$\nu_\tau$}  
			{} {} {$1/2$} {} {} 
			};
\node at(3,-3)   {\particl 
			{$W^{\hspace{-.3ex}\scalebox{.5}{$\pm$}}$}
							
						{} {} {1} {} {} };
\node at(4,-3)   {\particl 
			{$Z$}       
			{} {} {$1$} {} {} }; 
\node at(3.5,-2) {\particl 
			{$\gamma$}  
			{} {} {$1$} {} {} };
\node at(3.5,-1) {\particl 
			{$g$}       
			{} {} {$1$} {} {} };
\node at(5,0)    {\particl 
			{$H$}       
			{} {} {$0$} {} {} };
\node at(6.1,-3) {\particl
			{$e$}           {{\huge graviton}}                       {}{}{}{}};

\draw [mbrace] (-0.5,-3.6) -- (2.5,-3.6)
			node[bottomlabel]
			{{\Huge $ n_{1/2} 
			=48$}}; 

\draw [mbrace] (2.5,-3.6) -- (4.5,-3.6)
			node[bottomlabel]{{\Huge $ n_{1}=12$}};
			
\draw [mbrace] (4.5,-3.6) -- (5.5,-3.6)
			node[bottomlabel]{{\Huge $ n_{0}=4$}};
\draw [mbrace] (5.8,-3.6) -- (6.8,-3.6)
			node[bottomlabel]{{\Huge $ n_{2}=2$}};

\end{tikzpicture}\,\\[4em]
\begin{tikzpicture}[x=1.2cm, y=1.2cm]
\draw[round] (-0.5,0.5) rectangle (4.4,-1.5);
\draw[round] (-0.6,0.6) rectangle (5.0,-2.5);
\draw[round] (-0.7,0.7) rectangle (5.6,-3.5);

\node at(0, 0)   {\particl 
			{$u$}       
			{} {} {$1/2$} {} {} 
			};
\node at(0,-1)   {\particl 
			{$d$}       
			{} {} {$1/2$} {} {} 
			};
\node at(0,-2)   {\particl 
			{$e$}       
			{} {} {$1/2$} {} {} 
			};
\node at(0,-3)   {\particl 
			{$\nu_e$}   
			{} {} {$1/2$} {} {} 
			};
\node at(1, 0)   {\particl
			{$c$}        
			{} {} {$1/2$} {} {} 
			};
\node at(1,-1)   {\particl 
			{$s$}      
			{} {} {$1/2$} {} {} 
			};
\node at(1,-2)   {\particl
			{$\mu$}     
			{} {} {$1/2$} {} {} 
			};
\node at(1,-3)   {\particl
			{$\nu_\mu$}   
			{} {} {$1/2$} {} {} 
			};
\node at(2, 0)   {\particl
			{$t$}         
			{} {} {$1/2$} {} {} 
			};
\node at(2,-1)   {\particl
			{$b$}        
			{} {} {$1/2$} {} {} 
			};
\node at(2,-2)   {\particl
			{$\tau$}      
			{} {} {$1/2$} {} {} 
			};
\node at(2,-3)   {\particl
			{$\nu_\tau$}  
			{} {} {$1/2$} {} {} 
			};
\node at(3,-3)   {\particl 
			{$W^{\hspace{-.3ex}\scalebox{.5}{$\pm$}}$}
							
						{} {} {1} {} {} };
\node at(4,-3)   {\particl 
			{$Z$}         
			{} {} {$1$} {} {} }; 
\node at(3.5,-2) {\particl 
			{$\gamma$}  
			{} {} {$1$} {} {} };
\node at(3.5,-1) {\particl 
			{$g$}       
			{} {} {$1$} {} {} };

\node at(5,0)    {\particl 
			{$\xi$}        {{\huge 36 scalars}}               

			{} {} {$0$} {} {} };
\node at(6.1,-3) {\particl
			{$\slashed{e}$}           {{\huge graviton}}                       {}{}{}{}};

\draw [mbrace] (-0.5,-3.6) -- (2.5,-3.6)
			node[bottomlabel]
			{{\Huge $ n_{1/2} 
			=48$}};

\draw [mbrace] (2.5,-3.6) -- (4.5,-3.6)
			node[bottomlabel]{{\Huge $ n_{1}=12$}};
			
\draw [mbrace] (4.5,-3.6) -- (5.5,-3.6)
			node[bottomlabel]{{\Huge $ n_{0}'=36$}\\[0.3em]{\Huge $ n_{0} =0$}};
\draw [mbrace] (5.8,-3.6) -- (6.8,-3.6)
			node[bottomlabel]{{\Huge $ n_{2}=0$}};

\end{tikzpicture}

\caption{
\footnotesize{
Above, 
the configuration of species of particles the standard model.
Below, a modified version of the standard model 
with the difference that 
there are now $n_0'=36$ dimension-zero scalar fields, and the $n_0=4$ dimension-one scalars and the $n_2=2$ gravitons have been removed.
}
}
\label{fig_SM_1}\end{figure}

\noindent
%-------------------------------------------------------------------------------------------------------------------------------------------------------% 
Given a conformally invariant classical theory (sometimes referred to as a  Weyl  invariant classical theory),
the corresponding quantum theory preserves this Weyl invariance if the \emph{Weyl anomaly} cancels.
The Weyl anomaly itself is given by \cite{Boyle:2021jaz}
%-------------------------------------------------------------------------------------------------------------------------------------------------------% 
\begin{equation}
\langle T^{\mu}_{\mu}\rangle= \frac1{360(4\pi)^2}\,\left(\,3\,c C^{2}-a E\,\right)+\zeta \Box R,
\label{W_anomaly_1}\end{equation}
where
%-------------------------------------------------------------------------------------------------------------------------------------------------------% 
\begin{equation}
C^{2}= R^{\alpha\beta\gamma\delta}R_{\alpha\beta\gamma\delta}-2R^{\alpha\beta}R_{\alpha\beta}+\frac{1}{3}R^{2}
\label{square_of_the_Weyl_curvature}\end{equation}
%-------------------------------------------------------------------------------------------------------------------------------------------------------% 
is the square of the Weyl curvature,  
%-------------------------------------------------------------------------------------------------------------------------------------------------------% 
\begin{equation}E=R^{\alpha\beta\gamma\delta}R_{\alpha\beta\gamma\delta}-4R^{\alpha\beta}R_{\alpha\beta}+R^{2}
\label{Gauss-Bonnet_density}\end{equation}
%-------------------------------------------------------------------------------------------------------------------------------------------------------% 
is the 
Gauss-Bonnet  density, and the parameters $a$ and $c$ in (\ref{W_anomaly_1}) are given by
\cite{Duff:1993wm,Capper:1974ed,Capper:1973mv,Capper:1974ic,Deser:1974cz,Christensen:1977jc,Brown:1976wc,Brown:1977pq,Dowker:1976zf,Duff:1977ay}
%-------------------------------------------------------------------------------------------------------------------------------------------------------% 
\begin{align}
a=&%\tfrac{1}{360(4\pi)^{2}}
n_{0}+\frac{\!11\!}{2}\,n_{1/2}+62\, n_{1}+1142\, n_{2}\label{parameter_a}  \\
c=&%\tfrac{1}{120(4\pi)^{2}}
n_{0}+\,3\, n_{1/2}+12 n_{1}\,+\,522\,n_{2}\ .  \label{parameter_c}
\end{align}
%-------------------------------------------------------------------------------------------------------------------------------------------------------% 
Here,  
$n_{0}$ is the number of dimension-one real scalar fields (spin 0), 
%described by the actions (\ref{S_2}) and (\ref{S_4}), 
$n_{1/2}$ is the number of Weyl or Majorana spinor fields (spin 1/2), $n_{1}$ is the number of real vector fields (spin 1), 
and $n_{2}$ is the number of graviton fields (spin 2).
Importantly, of the standard model particles,
the fermions account for $n_{1/2}=48$ particles.
These include six leptons:
$(e,\nu_e)$,
$(\mu,\nu_\mu)$,
$(\tau,\nu_\tau)$,
six quark flavours
$(u,d)$, $(c,s)$, $(t,b)$, each of which can have three colour quantum numbers
thus yielding 18 quarks.
Overall this makes 24 fermions.   Together with the corresponding antifermions
the total comes to 48 spin $1/2$ fermions.
The bosons that are the force carriers 
include 1 photon ($\gamma$) mediating EM interactions,
the $W^+$, $W^-$ and $Z$ bosons mediating weak interactions,
and eight gluons ($g$) that mediate strong interactions, which add up to twelve spin $1$ bosons.
The standard electroweak theory also includes four spin $0$ scalar fields
that are Higgs bosons, and  one spin $2$ graviton   yet to be discovered.
This configuration is summarized in 
in the upper chart in 
Table  \ref{fig_SM_1}.
\par 
%-------------------------------------------------------------------------------------------------------------------------------------------------------% 
\par 
%-------------------------------------------------------------------------------------------------------------------------------------------------------% 
This configuration of particles
gives rise to a 
non-zero Weyl anomaly, as easily checked by substituting the appropriate quantities of particle species in  (\ref{W_anomaly_1}).
However, with 
precisely 36 spin $0$  scalar fields that have mass dimension zero,
the Weyl anomaly vanishes.
Specifically, the new Weyl anomaly has the form (\ref{W_anomaly_1})
with (\ref{square_of_the_Weyl_curvature}) and (\ref{Gauss-Bonnet_density}),
but with $a$ and $c$ given instead of (\ref{parameter_a}) and (\ref{parameter_c}) by
%-------------------------------------------------------------------------------------------------------------------------------------------------------% 
\begin{align}
a=& 
n_{0}+\tfrac{\!11\!}{2}\,n_{1/2}+62\, n_{1}+1142\, n_{2} 
{-28n_0'}\\
c=& 
n_{0}+\,3\, n_{1/2}+12 n_{1}\,+\,522\,n_{2}  {- 8n_0'} 
\end{align}
%-------------------------------------------------------------------------------------------------------------------------------------------------------% 
where $n_0'$ is the number of spin 0 dimension 0 scalar fields in the theory.
Note that 
if $n_0'=36$,   then as a result  $a=0$,  and independently $c=0$ as well.
Consequently
the Weyl anomaly cancels with this configuration of particles.
This modified configuration of particles is summarized in the lower chart in Table \ref{fig_SM_1}.
\par 
%-------------------------------------------------------------------------------------------------------------------------------------------------------% 
The observation that 36 dimension 0 scalar fields
leads to a vanishing Weyl anomaly is one of the motives for the composite Higgs model 
proposed in this article.
Not only does the cancellation of the Weyl anomaly require precisely 36 dimension 0 scalars,
but the action with non-minimal coupling of fermions to gravity (given in (\ref{actn_dirac_ferms_1}) below)
admits precisely $36$ different coupling coefficients, 
as explained below.
\par 
The non-minimal coupling action was given in (\ref{final_actn-5}).
The piece containing the non-minimal coupling of fermions to gravity
concerning this discussion is 
%-------------------------------------------------------------------------------------------------------------------------------------------------------% 
\begin{equation}
S_f= \frac1{6}\epsilon_{abcd}\int dx\ \theta^a\wedge e^b\wedge e^c\wedge e^d \ ,\label{actn_dirac_ferms_1}
\end{equation}
%-------------------------------------------------------------------------------------------------------------------------------------------------------% 
where $\theta^a$ contributions from all of the fermions in the standard model:
%-------------------------------------------------------------------------------------------------------------------------------------------------------% 
\begin{align}
\theta^a=&\theta^a_l+\theta^a_q \ ,\label{non-minimal-coupling-2}
\\[0.5em]
\theta^a_l=&\quad\frac{i}2\sum_{A,B=e,\mu,\tau}\bar\psi_A\left(1+(\xi_l)_{AB}\right)\gamma^a D \psi_B 
-\frac{i}2\sum_{A,B=e,\mu,\tau}\overline{D \psi_A}\left(1+(\xi^\dagger_l)_{AB}\right)\gamma^a\psi_B 
\label{non-minimal-coupling-3}
\\
\theta^a_q=&\quad\frac{i}2\sum_{A,B=u,s,t}\bar\psi_A\left(1+(\xi_q)_{AB}\right)\gamma^a D \psi_B  
-\frac{i}2\sum_{A,B=u,s,t}\overline{D \psi_A}\left(1+(\xi^\dagger_q)_{AB}\right)\gamma^a\psi_B  
\label{non-minimal-coupling-4}
\end{align}
%-------------------------------------------------------------------------------------------------------------------------------------------------------% 
If it assumed that each fermion in the standard model couples to the background vielbein field differently, 
then there are $36$ different coefficients,
denoted as $\xi$, required  to couple each one of the fermions to the background tetrad. \par 
%-------------------------------------------------------------------------------------------------------------------------------------------------------% 
In our model these coupling coefficients 
become dimension 0 scalar fields in their own right.
Accordingly, the action of our model is comprises
the kinetic action of the scalar fields themselves and the 
coupling piece with non-minimal coupling of fermions to gravity
through 36 coupling coefficients, where the very coupling coefficients are scalar fields.
\par 
%-------------------------------------------------------------------------------------------------------------------------------------------------------% 
A \emph{dimension-zero} conformally coupled scalar field $\phi$ has the action
%-------------------------------------------------------------------------------------------------------------------------------------------------------% 
\begin{equation}
S_{\rm scalar}[\phi]=\frac1{2}\int dx\, \sqrt{g}\phi\Delta_4 \phi
\label{S_4-0}\end{equation}
%-------------------------------------------------------------------------------------------------------------------------------------------------------% 
where $\Delta_{4}$ is the unique conformally-invariant fourth order differential operator \cite{DeWitt:1964mxt}
%-------------------------------------------------------------------------------------------------------------------------------------------------------% 
\begin{equation}
\Delta_4 = \Box^2+2 R^{\alpha\beta}\nabla_\alpha\nabla_\beta-\frac2{3}R\Box+\frac1{3}\left(\nabla^\alpha R\right)\nabla_\alpha\ .
\label{Delta_4}
\end{equation}
%-------------------------------------------------------------------------------------------------------------------------------------------------------% 
The  action in Eq.~(\ref{S_4-0}) is invariant under the Weyl transformation
%-------------------------------------------------------------------------------------------------------------------------------------------------------% 
$
g_{\alpha\beta}(x) \to\Omega^2(x) g_{\alpha\beta}(x)$, $\phi(x) \to\Omega^0(x)\phi(x)
$.
\par
%-------------------------------------------------------------------------------------------------------------------------------------------------------% 
The total action of our model is
%-------------------------------------------------------------------------------------------------------------------------------------------------------% 
\begin{equation}S=S_{\rm scalar}[e]+S_f[e]\end{equation}
%-------------------------------------------------------------------------------------------------------------------------------------------------------% 
where $S_{\rm scalar}[e]$ is the  conformally invariant action of dimension 0 scalar fields (see (\ref{S_4-0}))
%-------------------------------------------------------------------------------------------------------------------------------------------------------% 
\begin{equation}S_{\rm scalar}=\sum_{\substack {A,B,C,D=1,2,3\\{\rm generations}}}\, \sum_{u,v=q,l}\alpha^{uv}_{ABCD}\int dx\ 
\left(\xi_u^{AB}\right)^\ast \Box^2 \xi^{CD}_v\ , \label{S_scalar}\end{equation}
%-------------------------------------------------------------------------------------------------------------------------------------------------------% 
where $\alpha^{uv}_{ABCD}$ are 
coupling constants.
$S_f[e]$  is the  covariant Dirac action with 36 constant coupling
coefficients replaced by dimension 0 scalar fields:
\begin{align}
S_f[e]=& \frac1{6}\epsilon_{abcd}\int dx\ \theta^a\wedge e^b\wedge e^c\wedge e^d
\\[1em]
\theta^a=&\theta^a_l+\theta^a_q
\\
\theta^a_l=&\quad\frac{i}2\sum_{A,B=e,\mu,\tau}\bar\psi_A\left[1+(\xi_l)_{AB}\right]\gamma^a D \psi_B
-\overline{D \psi_A}\left[1+(\xi^\dagger_l)_{AB}\right]\gamma^a\psi_B
\\
\theta^a_q=&\quad\frac{i}2\sum_{A,B=u,s,t}\bar\psi_A\left[1+(\xi_q)_{AB}\right]\gamma^a D \psi_B  
-\overline{D \psi_A}\left[1+(\xi^\dagger_q)_{AB}\right]\gamma^a\psi_B 
\end{align}
where $\xi_l$, $\xi_q$ are $3\times 3$ complex-valued fields.
\par 
%-------------------------------------------------------------------------------------------------------------------------------------------------------% 
In \cite{Miller:2022qil}
a simplified toy-model version of this action was formed, comprising  
the dominant contribution from the 3rd generation of quarks and 1 scalar field $\xi$, viz
%-------------------------------------------------------------------------------------------------------------------------------------------------------% 
\begin{equation}S=S_{\rm scalar}+S_q\end{equation}
%-------------------------------------------------------------------------------------------------------------------------------------------------------% 
where 
%-------------------------------------------------------------------------------------------------------------------------------------------------------% 
\begin{equation} {S_{\rm scalar}=\frac1{2}\int dx\ \sqrt{g}\xi\,\Delta_4\, \xi}\end{equation}
%-------------------------------------------------------------------------------------------------------------------------------------------------------% 
with $\Delta_4$ given by (\ref{Delta_4}) and 
%-------------------------------------------------------------------------------------------------------------------------------------------------------% 
\begin{equation}S_q=\int d x\ \left[\alpha\xi^\ast\Box^2\xi+\frac{i}2\bar Q
\left(1+\xi\right)
\slashed D Q
-\frac{i}2
\overline{\slashed D Q} 
\left(1+\xi^\ast\right) Q\right]\end{equation}
where %-------------------------------------------------------------------------------------------------------------------------------------------------------% 
$Q=\begin{pmatrix}t\\b
\end{pmatrix}$.
It was assumed that the $t$ and $b$ quarks couple in the same way to the background tetrad and that their masses are equal.
In this simplified model, dynamical symmetry breaking occurs in the action
with the consequent generation of a mass term ascribing mass to the quarks.
In the following section this model is generalized to the case where dynamical symmetry breaking
leads to different mass terms for the $t$ and $b$ quarks.
%-------------------------------------------------------------------------------------------------------------------------------------------------------% 
\section{Fundamental scalars in models with three generations of SM fermions}
\label{sec_fundamental_scalars_in_models_of_3_generations_of_SM_fermions}\noindent
%-------------------------------------------------------------------------------------------------------------------------------------------------------% 
\subsection{Non-minimal coupling of fermions with gravity}
\label{sec_nonminimal_coupling_of_fermions_to_gravity}\noindent
%-------------------------------------------------------------------------------------------------------------------------------------------------------% 
The starting point of the discussion in this section is the action for Dirac fermions coupled to the vielbein field as given in Eq.~(\ref{actn_dirac_ferms_1}). 
%-------------------------------------------------------------------------------------------------------------------------------------------------------% 
In turn, the action for the right - handed Weyl fermions
may be written as Eq.~(\ref{actn_dirac_ferms_1})
with ${{{\theta}}}^a \to  {{{\theta}}} ^a_R$, where
%-------------------------------------------------------------------------------------------------------------------------------------------------------% 
\begin{equation} 
{{{\theta}}}^a_R=\frac i{2}\left[\bar\psi_R\sigma^aD_\mu\psi_R-\overline{D_\mu \psi}_R\sigma^a\psi_R\right] dx^\mu \ ,\label{actn_dirac_ferms_3}\end{equation}
%-------------------------------------------------------------------------------------------------------------------------------------------------------% 
while the action for the left - handed Weyl fermions may
be written as   Eq. (\ref{actn_dirac_ferms_1}) with ${{{\theta}}}^a \to  {{{\theta}}} ^a_L$ and
%-------------------------------------------------------------------------------------------------------------------------------------------------------% 
\begin{equation} {{{\theta}}}^a_L=\frac i{2}\left[\bar\psi_L\bar \sigma^aD_\mu\psi_L-\overline{D_\mu \psi}_L
\bar \sigma^a\psi_L\right] dx^\mu \ ,\label{actn_dirac_ferms_3}\end{equation}
%-------------------------------------------------------------------------------------------------------------------------------------------------------% 
A non - minimal coupling of fermions to gravity 
is achieved 
via the substitution
\cite{Alexandrov:2008iy}
%-------------------------------------------------------------------------------------------------------------------------------------------------------% 
\begin{equation} {{{\theta}}}^a_R\to
{{{\theta}}}^a_R
+\frac i{2}\left[ \bar\psi_R \xi_R\sigma^aD_\mu\psi_R
-
\xi^\ast_R\overline{ D_\mu \psi}_R \sigma^a\psi_R\right] dx^\mu 
\label{non_min_cplng_1}\end{equation}
%-------------------------------------------------------------------------------------------------------------------------------------------------------% 
and 
%-------------------------------------------------------------------------------------------------------------------------------------------------------% 
\begin{equation} {{{\theta}}}^a_L\to
{{{\theta}}}^a_L
+\frac i{2}\left[ \bar\psi_L \xi_L\bar \sigma^aD_\mu\psi_L
-
\xi^\ast_L \overline{D_\mu \psi}_L \bar \sigma^a\psi_L\right] dx^\mu  \label{non_min_cplng_2}\end{equation}
%-------------------------------------------------------------------------------------------------------------------------------------------------------% 
with  complex - valued constants $\xi_R$ and $\xi_L$. 
Notice that in the absence of the spin connection, $D_\mu=\PD_\mu$ and the imaginary
parts of  $\xi_R$ and $\xi_L$ decouple and do not interact with
fermions. 
At the same time the real parts of these constants may be absorbed by a rescaling of the fermion fields.
\par 
%-------------------------------------------------------------------------------------------------------------------------------------------------------% 
That said, suppose that this construction is extended to the
case where  $\xi_R$ and $\xi_L$ become coordinate-dependent
fields. In this case
both the real
and imaginary parts of $\xi_R$ and $\xi_L$  interact
with the fermions of the theory,
and cannot be removed by any rescaling of the
fields. 
Further more,  let $\xi_R$ and $\xi_L$ 
be assigned flavour indices.
In the presence of $SU (3)\otimes SU (2) \otimes  U (1)$
gauge fields, there are 
$n_1 = 8 + 3 + 1 = 12$
vector fields. In accordance the number of Weyl fermions must equal
$
n_{1/2} = 4n_1 = 48$.
This number coincides with the number of Weyl fermions
in the SM with three generations: 
$n_{ 1/2} = (1_{\rm leptons} +
3_{\rm quarks} ) \times  2_{{\rm up}\,{\rm \&}\, {\rm down}} \times  2_{{\rm left}\, {\rm\&}\,{\rm right}} \times  3_{\rm generations} = 48$.
%-------------------------------------------------------------------------------------------------------------------------------------------------------% 
\subsection{Parity breaking interactions with zero dimension scalar fields}
\label{sec_parity_breaking_interactions_zero_dim_scalar_fields}\noindent
%-------------------------------------------------------------------------------------------------------------------------------------------------------% 
The theory can be extended one stage further to the case where 
$\xi_L$ and $\xi_R$ carry not only flavor indices but also 
generation indices.
Even more they can be further distinguished by assigning  
different forms for the matrices $\xi$
for the left - handed and the right - handed particles, while
$\xi$ the matrices for the quarks and leptons remain identical. 
In this approach the fermion term $\theta^a$ in (\ref{actn_dirac_ferms_1}) now reads
\begin{equation}{{{\theta}}}^a= {{{\theta}}}^a_L+ {{{\theta}}}^a_R \ ,\label{parity_brkng_zero-dim_sclr_flds}\end{equation}
%-------------------------------------------------------------------------------------------------------------------------------------------------------% 
where 
%-------------------------------------------------------------------------------------------------------------------------------------------------------% 
\begin{align} {{{\theta}}}^a_R=&
\frac i{2} \bigg[\bar\psi_R \left(1+\xi_R\right)\sigma^aD_\mu\psi_R
\nonumber\\&\qquad\qquad- (D_\mu\bar\psi_R)\left(1+
\xi^\dagger_R\right)\sigma^a\psi_R \bigg] dx^\mu \ ,\label{parity_brkng_zero-dim_sclr_flds_1}\end{align}
%-------------------------------------------------------------------------------------------------------------------------------------------------------% 
and 
%-------------------------------------------------------------------------------------------------------------------------------------------------------% 
\begin{align} {{{\theta}}}^a_L =&
\frac i{2}\bigg[ \bar\psi_L \left(1+
\xi_L\right)\bar \sigma^aD_\mu\psi_L
\nonumber\\ &\qquad\qquad-
(D_\mu\bar\psi_L)\left(1+
\xi_L^\dagger\right)\bar \sigma^a\psi_L\bigg]
dx^\mu \ .\label{parity_brkng_zero-dim_sclr_flds_2}\end{align}
%-------------------------------------------------------------------------------------------------------------------------------------------------------% 
This gives rise to
$n_0'  = 2_{\rm chiralities} \times  2_{{\rm real}\,{\rm \&}\,{\rm imaginary}\,{\rm parts}} \times 
3_{\rm generations}  \times  3_{\rm generations}  = 36$
components of the scalar
fields. This is precisely the number of scalars $n_0'=3n_1$ 
needed for the cancellation of the Weyl anomaly
\cite{Boyle:2021jaz}.
\par 
%-------------------------------------------------------------------------------------------------------------------------------------------------------% 
\subsection{Interactions that conserve parity}\label{sec_interactions_that_conserve_parity}\noindent
%-------------------------------------------------------------------------------------------------------------------------------------------------------% 
There would be the same number of scalar fields as
above if $\xi_L$ were identical to $\xi_R$, but with the matrices $\xi$ being
different for quarks and leptons. In this case the interaction
with $\xi$ does not break parity. The fermion
action is still defined as in  Eq. (\ref{actn_dirac_ferms_1}), but instead of Eq. (\ref{parity_brkng_zero-dim_sclr_flds}), 
%-------------------------------------------------------------------------------------------------------------------------------------------------------% 
\begin{equation}{{{\theta}}}^a= {{{\theta}}}^a_l+ {{{\theta}}}^a_q \ ,\label{10}\end{equation}
%-------------------------------------------------------------------------------------------------------------------------------------------------------% 
where 
%-------------------------------------------------------------------------------------------------------------------------------------------------------% 
\begin{align} {{{\theta}}}^a_q=&
\frac i{2}\bigg[ \bar\psi_q \left(1+
\xi_q\right)\gamma^aD_\mu\psi_q
\nonumber\\&\qquad\qquad 
-
(D_\mu\bar\psi_q)\left(1+
\xi_q^\dagger\right)\gamma^a\psi_q\bigg] dx^\mu \ ,\label{11}\end{align}
%-------------------------------------------------------------------------------------------------------------------------------------------------------% 
and 
%-------------------------------------------------------------------------------------------------------------------------------------------------------% 
\begin{align} {{{\theta}}}^a_l=&
\frac i{2}\bigg[ \bar\psi_q \left(1+
\xi_l\right)\gamma^aD_\mu\psi_l
\nonumber\\&\qquad\qquad
-
(D_\mu\bar\psi_l)\left(1+
\xi_l^\dagger\right)\gamma^a\psi_l\bigg] dx^\mu \ .
\label{parity_conservatn_zero-dim_sclr_flds_2}\end{align}
%-------------------------------------------------------------------------------------------------------------------------------------------------------% 
Here both $\xi_l$ and $\xi_q$ are the $3 \times  3$ complex - valued
matrices in flavor space. As before, 
$n_0'  = 2_{{\rm leps}\,{\rm \&}\,{\rm qrks}} \times  2_{{\rm real}\,{\rm \&}\,{\rm imag}\,{\rm parts}} \times 
3_{\rm generations}  \times  3_{\rm generations}  = 36$
zero dimension scalar fields.\par 
\subsection{Additional coupling constants}
The above mentioned scheme may be extended to allow different couplings of the same field $\xi_q$ with different quarks/leptons. Namely, we consider the action of the form
\begin{align}S=&
\int dx \, 
\bigg[\alpha^{uv}_{ABCD} \, 
\left(\xi_u^{AB}\right)^\ast \Box^2 \xi^{CD}_v\ 
\nonumber\\&
+\frac i{2}\bar Q_{a L}(1+\beta^{ab}_{cd L}\xi^{cd}_q)\slashed D Q_{b L}
-\frac i{2}\overline{\slashed D Q_{aL}} (1+\beta_{cd L}^{ba\ast}\xi^{cd \ast}_q )Q_{bL} 
\nonumber\\&
+\frac i{2}\bar U_{aR}(1+\beta^{ab}_{cd U}\xi^{cd}_q)\slashed D U_{bR}
-\frac i{2}\overline{\slashed D U_{aR}} (1+\beta_{cd U}^{ba\ast}\xi^{cd \ast}_q )U_{bR}
\nonumber\\&
+\frac i{2}\bar D_{aR}(1+\beta_{cd D}^{ab}\xi^{cd}_q)\slashed D D_{bR}
-\frac i{2}\overline{\slashed D D_{aR}} (1+\beta_{cd D}^{ba\ast}\xi^{cd\ast}_q )D_{bR}
\nonumber\\&
+\frac i{2}\bar L_{a L}(1+\gamma^{ab}_{cd L}\xi^{cd}_l)\slashed D L_{b L}
-\frac i{2}\overline{\slashed D L_{aL}} (1+\gamma_{cd L}^{ba\ast}\xi^{cd \ast}_l )L_{bL} 
\nonumber\\&
+\frac i{2}\bar N_{aR}(1+\gamma^{ab}_{cd N}\xi^{cd}_l)\slashed D N_{bR}
-\frac i{2}\overline{\slashed D N_{aR}} (1+\gamma_{cd N}^{ba\ast}\xi^{cd \ast}_l )N_{bR}
\nonumber\\&
+\frac i{2}\bar E_{aR}(1+\gamma_{cd E}^{ab}\xi^{cd}_l)\slashed D E_{bR}
-\frac i{2}\overline{\slashed D E_{aR}} (1+\gamma_{cd E}^{ba\ast}\xi^{cd\ast}_l )E_{bR}
\bigg]\   \label{action_1}
\end{align}
Here $Q_{aL}$ is the $SU(2)$ doublet of left - handed quarks $(u,d),(c,s),(t,b)$ (index $a$ takes values $1,2,3$). $U_{aR}$ is the $SU(2)$ singlet of right - handed quarks $u,c,t$. $D_{aR}$ is the $SU(2)$ singlet of right - handed quarks $d,s,b$. In a similar way  $L_{aL}$ is the $SU(2)$ doublet of left - handed leptons $(\nu,e),(\nu_\mu,\mu),(\nu_\tau,\tau)$. $N_{aR}$ is the $SU(2)$ singlet of right - handed neutrinos $\nu,\nu_\mu,\nu_\tau$. $E_{aR}$ is the $SU(2)$ singlet of right - handed leptons $e,\mu,\tau$. Tensors $\beta$ and $\gamma$ contain coupling constants of the field $\xi_q$ and $\xi_l$ to fermions.
%-------------------------------------------------------------------------------------------------------------------------------------------------------% 
\section{Electroweak symmetry breaking and mass generation}
\label{sec_EW_symmetry_breaking_and_mass_generation}
\noindent
\subsection{Introduction of the toy model with top and bottom quarks}
%-------------------------------------------------------------------------------------------------------------------------------------------------------% 
Consider the SM with three generations and the
fundamental scalar $\xi$ fields constructed in \S \ref{sec_interactions_that_conserve_parity}.
With variations of the vielbein and   spin connection
restrained,  define an action for the zero dimension scalar fields
as
%-------------------------------------------------------------------------------------------------------------------------------------------------------% 
\begin{equation}S_B=\alpha^{uv}_{ABCD}\int dx \, 
\left(\xi_u^{AB}\right)^\ast \Box^2 \xi^{CD}_v\ ,\label{S_B}\end{equation}
%-------------------------------------------------------------------------------------------------------------------------------------------------------% 
where $A,B,C,D=1,2,3$ are generation indices,  $u,v = q, l$ are flavor indices, and $\alpha^{uv}_{ABCD}$
are a set of coupling constants.
%-------------------------------------------------------------------------------------------------------------------------------------------------------% 
The effective four - fermion interactions arise from the 
exchange by quanta of the fields $\xi_q$ and $\xi_l$.  
The effective four - fermion interaction is non-local.\par
%-------------------------------------------------------------------------------------------------------------------------------------------------------% 
Consider now a certain sector of the theory that describes 
the dominant contributions 
from the sector corresponding to the third generation of quarks, and 
the field $\xi_q^{33}$.
Let the kinetic piece of the action given in 
(\ref{S_B}) be added to the action in (\ref{actn_dirac_ferms_1}),
with $\theta^a$ given by (\ref{10}), 
with only the sector describing the fields $\xi_q^{33}$ included.
Note that
the left- and right-handed fermion fields $\psi_L$, $\psi_R$ appearing 
in (\ref{actn_dirac_ferms_1})
in this sector
correspond to the doublets
$Q_{3L}=\begin{pmatrix}t_L\\b_L
\end{pmatrix}$
and 
$Q_{3R}=\begin{pmatrix}t_R\\b_R
\end{pmatrix}$,
where 
$t_L$, $t_R$, $b_L$, $b_R$
stand for left- and right-handed top and bottom quark fields.
The action obtained has the form
%-------------------------------------------------------------------------------------------------------------------------------------------------------% 
\begin{align}S=&
\int dx \, 
\bigg[\alpha^{qq}_{3333}
\xi_q^{33\ast} \Box^2 \xi^{33}_q
\nonumber\\&
+\frac i{2}\bar Q_{3L}\left(1+\beta_L\xi^{33}_q\right)\slashed D Q_{3L}
-\frac i{2}\overline{\slashed D Q_{3L}} \left(1+\beta_L^\ast\xi^{33\ast}_q \right)Q_{3L} 
\nonumber\\&
+\frac i{2}\bar Q_{3R}\left(1+\beta_R\xi^{33}_q\right)\slashed D Q_{3R}
-\frac i{2}\overline{\slashed D Q_{3R}} \left(1+\beta_R^\ast\xi^{33\ast}_q \right)Q_{3R} 
\bigg]\ . \label{action_3rd_generation}
\end{align}
%-------------------------------------------------------------------------------------------------------------------------------------------------------% 
Here, $\beta_L$ and $\beta_R$ are non-identical constant coupling coefficients.
Note that the covariant
derivative $D$ contains gauge fields.
This action admits global $SU(2)$ symmetry and,
given the necessary form for the gauge fields, local $SU(2)$ symmetry as well.
Next this action is modified to a different expression.
The symmetry of the left-handed fields remains, 
but that of the right-handed fields is broken by allowing
different couplings
of the top and bottom quarks to the scalar field.
The modified expression is
%-------------------------------------------------------------------------------------------------------------------------------------------------------% 
\begin{align}S=&
\int dx \, 
\bigg[\alpha^{qq}_{3333}
\xi_q^{33  \ast} \Box^2 \xi^{33}_q
\nonumber\\&
+\frac i{2}\bar Q_{3L}(1+\beta_L\xi^{33}_q)\slashed D Q_{3L}
-\frac i{2}\overline{\slashed D Q_{3L}} (1+\beta_L^\ast\xi^{33\ast}_q )Q_{3L} 
\nonumber\\&
+\frac i{2}\bar t_{R}(1+\beta_t\xi^{33}_q)\slashed D t_R
-\frac i{2}\overline{\slashed D t_R} (1+\beta_t^\ast\xi^{33\ast}_q )t_R
\nonumber\\&
+\frac i{2}\bar b_{R}(1+\beta_b\xi^{33}_q)\slashed D b_R
-\frac i{2}\overline{\slashed D b_R} (1+\beta_b^\ast\xi^{33\ast}_q )b_R
\bigg]\   \label{action_1}
\end{align}
%-------------------------------------------------------------------------------------------------------------------------------------------------------% 
where  $\beta_t$ and $\beta_b$ are non-identical constant coupling coefficients
corresponding to the couplings of the top and bottom quarks to the scalar field, respectively.
To save on notation, use the shortened symbols 
$\xi_q\equiv \xi^{33}_q$, $\alpha \equiv\alpha^{qq}_{3333}$ 
$Q  \equiv Q_{3 }$, $Q_L\equiv Q_{3L}=\begin{pmatrix}t_L\\b_L
\end{pmatrix}$, and $Q_R\equiv Q_{3R}=\begin{pmatrix}t_R\\b_R
\end{pmatrix}$ . Like this 
Eq.~(\ref{action_1}) reads
%-------------------------------------------------------------------------------------------------------------------------------------------------------% 
\begin{align}S=&
\int dx \, 
\bigg[\alpha 
\xi  ^\ast \Box^2 \xi  
\nonumber\\&
+\frac i{2}\bar Q_{ L}\left(1+\beta_L\xi  \right)\slashed D Q_{ L}
-\frac i{2}\overline{\slashed D Q_{ L}} \left(1+\beta_L^\ast\xi^{ \ast}  \right)Q_{ L} 
\nonumber\\&
+\frac i{2}\bar t_{R}\left(1+\beta_t\xi  \right)\slashed D t_R
-\frac i{2}\overline{\slashed D t_R} \left(1+\beta_t^\ast\xi^{ \ast} \right)t_R
\nonumber\\&
+\frac i{2}\bar b_{R}\left(1+\beta_b\xi \right)\slashed D b_R
-\frac i{2}\overline{\slashed D b_R} \left(1+\beta_b^\ast\xi^{ \ast}  \right)b_R
\bigg]\   \label{action_2}
\end{align}
%-------------------------------------------------------------------------------------------------------------------------------------------------------%
Let the last two terms on the right be re-written in such a way that
%-------------------------------------------------------------------------------------------------------------------------------------------------------% 
\begin{align}S=&
\int dx \, 
\bigg[\alpha
\xi  ^\ast \Box^2 \xi  
\nonumber\\&
+\frac i{2}\bar Q_{ L}\left(1+\beta_L\xi  \right)\slashed D Q_{ L}
-\frac i{2}\overline{\slashed D Q_{ L}} \left(1+\beta_L^\ast\xi^{ \ast}  \right)Q_{ L} 
\nonumber\\&
+\frac i{2}
\left(\begin{array}{cc}\bar t_{R}&0\end{array}\right)\left(1+\beta_t\xi  \right)\slashed D 
\begin{pmatrix} t_{R}\\0\end{pmatrix}
-\frac i{2}\begin{pmatrix}\overline{\slashed D t}_{R}& 0\end{pmatrix} 
\left(1+\beta_t^\ast\xi^{ \ast} \right)\begin{pmatrix} t_{R}\\0\end{pmatrix}
\nonumber\\[1em]&
+\frac i{2}\left(\begin{array}{cc}0&\bar b_{R}\end{array}\right)\left(1+\beta_b\xi \right)\slashed D 
\begin{pmatrix}0\\b_R\end{pmatrix}
-\frac i{2} \begin{pmatrix}0&\overline{\slashed D b}_{R}\end{pmatrix}  \left(1+\beta_b^\ast\xi^{ \ast}  \right)
\begin{pmatrix} 0\\b_R\end{pmatrix}
\bigg]\   \label{action_3}
\end{align}
\par 
%-------------------------------------------------------------------------------------------------------------------------------------------------------%
Define the top and bottom projection operators $\Pi_t$ and $\Pi_b$ 
%that project the doublet $Q=\begin{pmatrix} t \\b\end{pmatrix}$ to purely top and bottom flavours respectively:
%-------------------------------------------------------------------------------------------------------------------------------------------------------%
\begin{equation}
\Pi_t=\left(\begin{array}{cc}1&0\\0&0\end{array}\right)\ ,\qquad 
\Pi_b =\left(\begin{array}{cc}0&0\\0&1\end{array}\right)\ ,\label{TB_proj_ops}
\end{equation}
%-------------------------------------------------------------------------------------------------------------------------------------------------------%
such that given the doublet $Q=\begin{pmatrix} t \\b\end{pmatrix}$,
%-------------------------------------------------------------------------------------------------------------------------------------------------------% 
\begin{equation}\begin{aligned}
\Pi_t Q=&\begin{pmatrix} t \\0\end{pmatrix}\ ,\qquad 
&&\Pi_b Q=\begin{pmatrix} 0 \\b\end{pmatrix}\ ,\\[1em]
\bar Q\Pi_t =&\begin{pmatrix} \bar t &0\end{pmatrix}\ ,\qquad 
&&\bar Q\Pi_b =\begin{pmatrix} 0 &\bar b\end{pmatrix}\ ,
\end{aligned}
\label{TB_proj_ops_1}
\end{equation}
%-------------------------------------------------------------------------------------------------------------------------------------------------------%
Note the properties
%-------------------------------------------------------------------------------------------------------------------------------------------------------%
\begin{equation}\Pi_t ^2=\Pi_t  \ ,\qquad \Pi_b ^2=\Pi_b \ ,\qquad \Pi_t \Pi_b =\Pi_b \Pi_t =0\ .\label{TB_proj_ops_2}\end{equation}
\par 
%-------------------------------------------------------------------------------------------------------------------------------------------------------%
Accordingly (\ref{action_3}) may be written in terms of $\Pi_t $ and $\Pi_b $ as
%-------------------------------------------------------------------------------------------------------------------------------------------------------%
\begin{align}S=&
\int dx \, 
\bigg[\alpha
\xi  ^\ast \Box^2 \xi  
\nonumber\\&
+\frac i{2}\bar Q_{ L}\left(1+\beta_L\xi  \right)\slashed D Q_{ L}
-\frac i{2}\overline{\slashed D Q_{ L}} \left(1+\beta_L^\ast\xi^{ \ast}  \right)Q_{ L} 
\nonumber\\&
+\frac i{2}
\bar Q_R \Pi_t \left(1+\beta_t\xi  \right)\slashed D 
\Pi_t Q_R
-\frac i{2}\overline{\slashed D Q}_{R}\Pi_t 
\left(1+\beta_t^\ast\xi^{ \ast} \right)\Pi_t Q_R
\nonumber\\&
+\frac i{2}\bar Q_R \Pi_b \left(1+\beta_b\xi \right)\slashed D 
\Pi_b Q_R
-\frac i{2} \overline{\slashed D Q}_{R}\Pi_b   \left(1+\beta_b^\ast\xi^{ \ast}  \right)
\Pi_b Q_R
\bigg]\label{action_4}\end{align}
%-------------------------------------------------------------------------------------------------------------------------------------------------------%
Note that the projection matrices $\Pi_t $,$\Pi_b $ 
and $\gamma$ matrices act on different spaces
($\Pi_t $,$\Pi_b $ act on $SU(2)$ doublets in flavour space
while the $4\times 4$ $\gamma$ matrices act on four-component Dirac spinors), so  
$\Pi_t $,$\Pi_b $ matrices in can be commuted past the $\gamma$ matrices in the operators $\slashed D$ to yield
%-------------------------------------------------------------------------------------------------------------------------------------------------------%
\begin{align}S
=&
\int dx \, 
\bigg[\alpha
\xi  ^\ast \Box^2 \xi  
\nonumber\\&
+\frac i{2}\bar Q_{ L}\left(1+\beta_L\xi  \right)\slashed D Q_{ L}
-\frac i{2}\overline{\slashed D Q_{ L}} \left(1+\beta_L^\ast\xi^{ \ast}  \right)Q_{ L} 
\nonumber\\&
+\frac i{2}
\bar Q_R \left(\Pi_t ^2+\beta_t\xi \Pi_t ^2 \right)\slashed D 
Q_R
-\frac i{2}\overline{\slashed D Q}_{R}
\left(\Pi_t ^2+\beta_t^\ast\xi^{ \ast} \Pi_t ^2\right)Q_R
\nonumber\\&
+\frac i{2}\bar Q_R \left(\Pi_b ^2+\beta_b\xi \Pi_b ^2\right)\slashed D 
Q_R
-\frac i{2} \overline{\slashed D Q}_{R}  \left(\Pi_b ^2+\beta_b^\ast\xi^{ \ast} \Pi_b ^2 \right)
Q_R
\bigg]
\nonumber\\[1em]
=&
\int dx \, 
\bigg[\alpha 
\xi  ^\ast \Box^2 \xi  
\nonumber\\&
+\frac i{2}\bar Q_{ L}\left(1+\beta_L\xi  \right)\slashed D Q_{ L}
-\frac i{2}\overline{\slashed D Q_{ L}} \left(1+\beta_L^\ast\xi^{ \ast}  \right)Q_{ L} 
\nonumber\\&
+\frac i{2}
\bar Q_R \left(\Pi_t +\beta_t\xi \Pi_t  \right)\slashed D 
Q_R
-\frac i{2}\overline{\slashed D Q}_{R}
\left(\Pi_t +\beta_t^\ast\xi^{ \ast} \Pi_t \right)Q_R
\nonumber\\&
+\frac i{2}\bar Q_R \left(\Pi_b +\beta_b\xi \Pi_b \right)\slashed D 
Q_R
-\frac i{2} \overline{\slashed D Q}_{R}  \left(\Pi_b +\beta_b^\ast\xi^{ \ast} \Pi_b  \right)
Q_R
\bigg]\ ,\label{action_5}
\end{align}
%-------------------------------------------------------------------------------------------------------------------------------------------------------%
where in the 2nd step (\ref{TB_proj_ops_2}) was used.
This can be further simplified to
%-------------------------------------------------------------------------------------------------------------------------------------------------------%
\begin{align}S=&
\int dx \, 
\bigg[\alpha 
\xi  ^\ast \Box^2 \xi  
\nonumber\\&
+\frac i{2}\bar Q_{ L}\left(1+\beta_L\xi  \right)\slashed D Q_{ L}
-\frac i{2}\overline{\slashed D Q_{ L}} \left(1+\beta_L^\ast\xi^{ \ast}  \right)Q_{ L} 
\nonumber\\&
+\frac i{2}
\bar Q_R \left(\Pi_t +\Pi_b +\beta_t\xi \Pi_t  +\beta_b\xi \Pi_b \right)\slashed D 
Q_R
-\frac i{2}\overline{\slashed D Q}_{R}
\left(\Pi_t +\Pi_b +\beta_t^\ast\xi^{ \ast} \Pi_t +\beta_b^\ast\xi^{ \ast} \Pi_b \right)Q_R
\bigg]
\nonumber\\[1em]
%-------------------------------------------------------------------------------------------------------------------------------------------------------%
=&
\int dx \, 
\bigg[\alpha
\xi  ^\ast \Box^2 \xi  
\nonumber\\&
+\frac i{2}\bar Q_{ L}\left(1+\beta_L\xi  \right)\slashed D Q_{ L}
-\frac i{2}\overline{\slashed D Q_{ L}} \left(1+\beta_L^\ast\xi^{ \ast}  \right)Q_{ L} 
\nonumber\\&
+\frac i{2}
\bar Q_R \left(1+\beta_t\xi \Pi_t  +\beta_b\xi \Pi_b \right)\slashed D 
Q_R
-\frac i{2}\overline{\slashed D Q}_{R}
\left(1+\beta_t^\ast\xi^{ \ast} \Pi_t +\beta_b^\ast\xi^{ \ast} \Pi_b \right)Q_R
\bigg]
\  , \label{action_6}
\end{align}
%-------------------------------------------------------------------------------------------------------------------------------------------------------%
where in the last step the fact that $\Pi_t +\Pi_b =1$ was used.
According to the definitions in 
Eqs.~(\ref{psi_LR}) and (\ref{global_symms_light_quark_sector_2})
and the relations in (\ref{PLR_gamma}), 
%-------------------------------------------------------------------------------------------------------------------------------------------------------%
\begin{subequations}\label{DQ}\begin{eqnarray}
\slashed D Q_L=&D_\mu \gamma^\mu P_L Q=D_\mu P_R\gamma^\mu Q&=P_R\slashed DQ\ ,\label{DQ_a}\\
\slashed D Q_R=&D_\mu \gamma^\mu P_R Q=D_\mu P_L\gamma^\mu Q&=P_L\slashed DQ\ ,\label{DQ_b}
\end{eqnarray} 
%-------------------------------------------------------------------------------------------------------------------------------------------------------%
such that
%-------------------------------------------------------------------------------------------------------------------------------------------------------%
\begin{eqnarray}
\overline{\slashed D Q}_L=&\overline{P_R\slashed DQ}
=
\left(P_R\slashed DQ\right)^\dagger\gamma^0
=
\left(\slashed DQ\right)^\dagger P_R^\dagger\gamma^0
=
\left(\slashed DQ\right)^\dagger P_R\gamma^0
% =\left(\slashed DQ\right)^\dagger \gamma^0 P_L
&=
\overline{ \slashed DQ} P_L
\ ,\label{DQ_c}\\
\overline{\slashed D Q}_R=&\overline{P_L\slashed DQ}
=
\left(P_L\slashed DQ\right)^\dagger\gamma^0
=
\left(\slashed DQ\right)^\dagger P_L^\dagger\gamma^0
=
\left(\slashed DQ\right)^\dagger P_L\gamma^0
% =\left(\slashed DQ\right)^\dagger \gamma^0 P_R
&=
\overline{ \slashed DQ} P_R
\ .\label{DQ_d}
\end{eqnarray}\end{subequations}
%-------------------------------------------------------------------------------------------------------------------------------------------------------%
With the help of these relations Eq.~(\ref{action_6})  can be cast in the form
%-------------------------------------------------------------------------------------------------------------------------------------------------------%
\begin{align}S
=&
\int dx \, 
\bigg[\alpha
\xi  ^\ast \Box^2 \xi  
\nonumber\\&
+\frac i{2}\bar Q P_R\left(1+\beta_L\xi  \right)P_R\slashed D Q
-\frac i{2}\overline{\slashed D Q} P_L \left(1+\beta_L^\ast\xi^{ \ast}  \right)P_LQ
\nonumber\\&
+\frac i{2}
\bar Q P_L \left(1+\beta_t\xi \Pi_t  +\beta_b\xi \Pi_b \right) P_L\slashed D 
Q
-\frac i{2}\overline{\slashed D Q}P_{R}
\left(1+\beta_t^\ast\xi^{ \ast} \Pi_t +\beta_b^\ast\xi^{ \ast} \Pi_b \right)P_RQ 
\bigg]
\nonumber\\[1em]
%-------------------------------------------------------------------------------------------------------------------------------------------------------%
=&
\int dx \, 
\bigg[\alpha 
\xi  ^\ast \Box^2 \xi  
\nonumber\\&
+\frac i{2}\bar Q P_R^2\left(1+\beta_L\xi  \right) \slashed D Q
-\frac i{2}\overline{\slashed D Q} P_L^2 \left(1+\beta_L^\ast\xi^{ \ast}  \right) Q
\nonumber\\&
+\frac i{2}
\bar Q P_L^2 \left(1+\beta_t\xi \Pi_t  +\beta_b\xi \Pi_b \right)  \slashed D 
Q
-\frac i{2}\overline{\slashed D Q}P_{R}^2
\left(1+\beta_t^\ast\xi^{ \ast} \Pi_t +\beta_b^\ast\xi^{ \ast} \Pi_b \right) Q 
\bigg]
\  . \label{action_7}
\end{align}
%-------------------------------------------------------------------------------------------------------------------------------------------------------%
But $P_R^2=P_R$ and $P_L^2=P_L$ (see Eqs.~(\ref{PRPL_idempotent}), hence 
%-------------------------------------------------------------------------------------------------------------------------------------------------------%
\begin{align}S
=&
\int dx \, 
\bigg[\alpha
\xi  ^\ast \Box^2 \xi  
\nonumber\\&
+\frac i{2}\bar Q P_R \left(1+\beta_L\xi  \right) \slashed D Q
-\frac i{2}\overline{\slashed D Q} P_L  \left(1+\beta_L^\ast\xi^{ \ast}  \right) Q
\nonumber\\&
+\frac i{2}
\bar Q P_L \left(1+\beta_t\xi \Pi_t  +\beta_b\xi \Pi_b \right)  \slashed D 
Q
-\frac i{2}\overline{\slashed D Q}P_{R} 
\left(1+\beta_t^\ast\xi^{ \ast} \Pi_t +\beta_b^\ast\xi^{ \ast} \Pi_b \right) Q 
\bigg]
\nonumber\\[1em]
%-------------------------------------------------------------------------------------------------------------------------------------------------------%
=&
\int dx \, 
\bigg[\alpha 
\xi  ^\ast \Box^2 \xi  
\nonumber\\&
+\frac i{2}\bar Q   \left(P_R+P_L+\beta_L\xi P_R +\left[\beta_t  \Pi_t  +\beta_b  \Pi_b \right]\xi P_L\right) \slashed D Q
\nonumber\\&
-\frac i{2}\overline{\slashed D Q}   \left(P_L+P_R+\beta_L^\ast\xi^{ \ast} P_L
+\left[\beta_t^\ast  \Pi_t +\beta_b^\ast \Pi_b \right]\xi^{ \ast} P_R\right)Q
\bigg]
\nonumber\\[1em]
%-------------------------------------------------------------------------------------------------------------------------------------------------------%
=&
\int dx \, 
\bigg[\alpha 
\xi  ^\ast \Box^2 \xi  
\nonumber\\&
+\frac i{2}\bar Q   \left(1+\beta_L\xi P_R +\left[\beta_t \Pi_t  +\beta_b  \Pi_b \right]\xi P_L\right) \slashed D Q
\nonumber\\&
-\frac i{2}\overline{\slashed D Q}   \left(1+\beta_L^\ast\xi^{ \ast} P_L
+\left[\beta_t^\ast  \Pi_t +\beta_b^\ast \Pi_b \right]\xi^{ \ast}P_R\right)Q
\bigg]
\nonumber\\[1em]
%-------------------------------------------------------------------------------------------------------------------------------------------------------%
=&
\int dx \, 
\bigg[\alpha
\xi  ^\ast \Box^2 \xi  
+\frac i{2}\bar Q    \slashed D Q
+\frac i{2}\bar Q   \Gamma  \xi\slashed D Q
-\frac i{2}\overline{\slashed D Q}   Q
-\frac i{2}\overline{\slashed D Q}    \chi  \xi^\ast Q
\bigg]
\  , \label{action_8}
\end{align}
%-------------------------------------------------------------------------------------------------------------------------------------------------------%
where 
%-------------------------------------------------------------------------------------------------------------------------------------------------------%
\begin{equation}
\Gamma = \beta_L P_R +\left(\beta_t  \Pi_t  +\beta_b  \Pi_b \right)  P_L\ ,\qquad 
\chi = \beta_L^\ast P_L
+\left(\beta_t^\ast \Pi_t +\beta_b^\ast  \Pi_b \right)  P_R\ .
\label{M1_M2}
\end{equation}
%-------------------------------------------------------------------------------------------------------------------------------------------------------%
Note that in accordance with the identities in (\ref{PLR_bar}),
%-------------------------------------------------------------------------------------------------------------------------------------------------------%
\begin{equation}
\gamma^0 \chi 
= \beta_L^\ast  \gamma^0P_L
+\left(\beta_t^\ast \Pi_t +\beta_b^\ast  \Pi_b \right) \gamma^0P_R
=\left( \beta_L^\ast  P_R^\dagger
+\left(\beta_t^\ast \Pi_t +\beta_b^\ast  \Pi_b \right)  P_L^\dagger\right)\gamma^0
=\Gamma ^\dagger \gamma^0
\end{equation}
%-------------------------------------------------------------------------------------------------------------------------------------------------------%
such that 
%-------------------------------------------------------------------------------------------------------------------------------------------------------%
\begin{equation}\overline{\slashed D Q}    \chi   Q=\left(\slashed D Q\right)^\dagger   \gamma^0 \chi   Q
=
\left(\slashed D Q\right)^\dagger  \Gamma ^\dagger \gamma^0  Q
=
\left(\Gamma \slashed D Q\right)^\dagger  \gamma^0  Q
=
\left(\overline{ \Gamma \slashed D Q} \right)\  Q
\end{equation}
%-------------------------------------------------------------------------------------------------------------------------------------------------------%
and consequently (\ref{action_8}) takes the succinct form
%-------------------------------------------------------------------------------------------------------------------------------------------------------%
\begin{align}S
=&
\int dx \, 
\bigg[\alpha
\xi  ^\ast \Box^2 \xi  
+\frac i{2}\bar Q    \slashed D Q
+\frac i{2}\bar Q   \Gamma  \xi\slashed D Q
-\frac i{2}\overline{\slashed D Q}   Q
-\frac i{2}\overline{\Gamma  \slashed D Q}      \xi^\ast Q
\bigg]
\  . \label{action_9}
\end{align}
%-------------------------------------------------------------------------------------------------------------------------------------------------------%
\par 
%-------------------------------------------------------------------------------------------------------------------------------------------------------% 
\subsection{Effective action resulting from the integration over scalar fields}
\begin{figure}[htb!]
\begin{center} 
\begin{tikzpicture}
\begin{feynman}
\vertex (i1) at (0,0) ;
\vertex (v1) at (2,0) ;
\vertex (v2) at (4,0) ;
\vertex (v1a) at (2,0.01) ;
\vertex (v2a) at (4,0.01) ;
\vertex (v1b) at (2,0.02) ;
\vertex (v2b) at (4,0.02) ;
\vertex (f1) at (6,0) ;

\diagram* {
(i1) --  [fermion, edge label=\(p\)](v1),
(v2) --  [fermion,edge label=\(p\)](f1),
(v1) -- [edge label=\(k\)](v2),
(v1a)-- (v2a),
(v1b)-- (v2b),
(v1) -- [scalar, half left,edge label=\(p-k\)] (v2)
};
\end{feynman}
\end{tikzpicture}\end{center}
\caption{\footnotesize 
Feynman diagram of the fermion self-energy corresponding to Eq.~(\ref{Sigma_p}).
The incoming and outgoing lines are initial and final fermion states with momentum $p$.
The dashed line in the loop is a scalar boson propagator carrying virtual momentum $p-k$, and the thick
horizontal line corresponds to the full fermion propagator with renormalized mass $\Sigma (k)$
inclusive of 
all self-energy corrections, as related in Eq.~(\ref{Sigma_p}).
}\label{fig_1}
\end{figure}
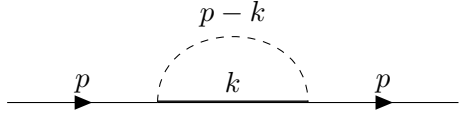

Define an effective action, $S_{\rm eff}$
as
%-------------------------------------------------------------------------------------------------------------------------------------------------------% 
\begin{equation}e^{iS_{\rm eff}}=\frac1{Z_0}\int D\xi \,D\xi ^\ast   e^{iS}\ \label{Seff}\end{equation}
%-------------------------------------------------------------------------------------------------------------------------------------------------------% 
where $S$ appearing in the integrand is the action in (\ref{action_9}).
To integrate out the scalar fields, let
%-------------------------------------------------------------------------------------------------------------------------------------------------------% 
\begin{align}
\xi'=&\xi-\dfrac{i}{2\alpha}\displaystyle\int dy  \overline{  \Gamma   \slashed D Q}\ Q \Box^{-2}(x,y)\label{xi_prime}\\[0.5em]
\xi^{\ast\prime}=&\xi^\ast+\dfrac{i}{2\alpha}\displaystyle\int dy \overline{Q} \Gamma \slashed D Q\Box^{-2}(x,y).\label{xi_ast_prime}
\end{align}
%-------------------------------------------------------------------------------------------------------------------------------------------------------%
where $\Box^{-2}(x,y)$ is the square of the inverse propagator of the boson field $\xi=\xi^{33}_q$.
In terms of the new variables $S$ has the form
%-------------------------------------------------------------------------------------------------------------------------------------------------------%
\begin{align}
S
=&
\int dx\, 
\bigg[\alpha
\left( \xi^{\ast\prime} 
-\frac{i}{2\alpha} \int dz \overline{Q} \Gamma \slashed D Q\Box^{-2}(x,z)
\right)\Box^2_x 
\left(\xi ' 
+\frac{i}{2\alpha} \int dy  \overline{  \Gamma   \slashed D Q} \ Q\Box^{-2}(x,y)
\right)
\nonumber\\
&
+\frac i{2}\bar Q   \Gamma  \xi'\slashed D Q
+\frac i{2}\bar Q   \Gamma  \left(\frac{i}{2\alpha} \int dy  \overline{  \Gamma   \slashed D Q}\ Q \Box^{-2}(x,y)\right) \slashed D Q
\nonumber\\
&
-\frac i{2}\overline{\Gamma  \slashed D Q}      \xi^{\ast\prime} Q
+\frac i{2}\overline{\Gamma  \slashed D Q} \left(\frac{i}{2\alpha} \int dy \overline{Q} \Gamma \slashed D Q\Box^{-2}(x,y) \right) Q
\nonumber\\
&
+\frac i{2}\bar Q    \slashed D Q
-\frac i{2}\overline{\slashed D Q}   Q
\bigg]
\nonumber\\[1em]
%-------------------------------------------------------------------------------------------------------------------------------------------------------%
=&
\int dx\, 
\bigg[\alpha
\xi^{\ast\prime} \Box^2_x \xi ' 
-\frac{i}{2}  \int dz \overline{Q} \Gamma \slashed D Q\Box^{-2}(x,z)\Box^2_x\xi'
+\xi^{\ast\prime}  \Box^2_x \frac{i}{2}
\int dy  \overline{  \Gamma   \slashed D Q}\  Q\Box^{-2}(x,y)
\nonumber\\
&
+
\frac1{4\alpha}
\int dz \overline{Q} \Gamma \slashed D Q\Box^{-2}(x,z)
\Box_x^2
\int dy  \overline{  \Gamma   \slashed D Q}\ Q \Box^{-2}(x,y)
\nonumber\\
&
+\frac i{2}\bar Q   \Gamma  \xi'\slashed D Q
-\frac 1{4\alpha}\bar Q   \Gamma   \int dy  \left(\overline{  \Gamma   \slashed D Q}\ Q \Box^{-2}(x,y)\right) \slashed D Q
\nonumber\\
&
-\frac i{2}\overline{\Gamma  \slashed D Q}      \xi^{\ast\prime} Q
-\frac 1{4\alpha}\overline{\Gamma  \slashed D Q}  \int dy \left( \overline{Q} \Gamma \slashed D Q\Box^{-2}(x,y) \right) Q
\nonumber\\
&
+\frac i{2}\bar Q    \slashed D Q
-\frac i{2}\overline{\slashed D Q}   Q
\bigg]
\nonumber\\[1em]
%-------------------------------------------------------------------------------------------------------------------------------------------------------%
=&
\int dx\, 
\bigg[\alpha
\xi^{\ast\prime} \Box^2_x \xi ' 
-\frac{i}{2}   \overline{Q} \Gamma \slashed D Q \xi'
+\xi^{\ast\prime}   \frac{i}{2}
\overline{  \Gamma   \slashed D Q} \ Q 
\nonumber\\
&
+
\frac1{4\alpha}
\int dy  \overline{Q} \Gamma \slashed D Q  (y)
\Box^{-2}(x,y)
\overline{  \Gamma   \slashed D Q}\ Q (x)
\nonumber\\
&
+\frac i{2}\bar Q   \Gamma  \xi'\slashed D Q
-\frac 1{4\alpha}\bar Q   \Gamma   \int dy  \left(\overline{  \Gamma   \slashed D Q} \ Q\Box^{-2}(x,y)\right) \slashed D Q
\nonumber\\
&
-\frac i{2}\overline{\Gamma  \slashed D Q}      \xi^{\ast\prime} Q
-\frac 1{4\alpha}\overline{\Gamma  \slashed D Q}  \int dy \left( \overline{Q} \Gamma \slashed D Q\Box^{-2}(x,y) \right) Q
\nonumber\\
&
+\frac i{2}\bar Q    \slashed D Q
-\frac i{2}\overline{\slashed D Q}   Q
\bigg]
\  . \label{action_10}
\end{align}
%-------------------------------------------------------------------------------------------------------------------------------------------------------%
The 2nd and 5th terms cancel,  the 3rd and 7th cancel, and the 4th and 8th terms cancel, leaving
%-------------------------------------------------------------------------------------------------------------------------------------------------------%
\begin{align}
S
=&
\int dx\, 
\bigg[\alpha
\xi^{\ast\prime} \Box^2_x \xi ' 
-\frac 1{4\alpha}\bar Q   \Gamma   \int dy  \left(\overline{  \Gamma   \slashed D Q} \ Q\Box^{-2}(x,y)\right) \slashed D Q
% \nonumber\\
% &
+\frac i{2}\bar Q    \slashed D Q
-\frac i{2}\overline{\slashed D Q}   Q
\bigg]
\  . \label{action_11}
\end{align}
%-------------------------------------------------------------------------------------------------------------------------------------------------------%
or equivalently
%-------------------------------------------------------------------------------------------------------------------------------------------------------%
\begin{align}
S
=&
\int dx\, 
\bigg[\alpha
\xi^{\ast\prime} \Box^2_x \xi ' 
+\frac i{2}\bar Q    \slashed D Q
-\frac i{2}\overline{\slashed D Q}   Q\bigg]
\nonumber\\
&\qquad -\frac 1{4\alpha}
\int dx \,  dy\,\left( \overline{ Q}   \Gamma   \slashed D Q\right)(x) \Box^{-2}(x,y) \left(\overline{  \Gamma   \slashed D Q} \ Q\right)(y)
\  . \label{action_12}
\end{align}
%-------------------------------------------------------------------------------------------------------------------------------------------------------%
By substituting (\ref{action_12}) in (\ref{Seff}) and integrating out the scalar fields
it is obtained that
%-------------------------------------------------------------------------------------------------------------------------------------------------------% 
\begin{align}e^{iS_{\rm eff}}=&\frac1{Z_0}\left({\rm Det}\left(\frac{i\Box_x}{\pi}\right)\right)^{-1}
\exp\bigg[\alpha \int dx\left(\frac i{2}\bar Q    \slashed D Q
-\frac i{2}\overline{\slashed D Q}   Q\right)\nonumber\\[0.5em]&
-\frac 1{4\alpha}
\int dx \,  dy\,\left( \overline{ Q}   \Gamma   \slashed D Q\right)(x) \Box^{-2}(x,y)  (\overline{  \Gamma   \slashed D Q} \ Q )(y)
\bigg] \ \label{Seff_1}\end{align}
%-------------------------------------------------------------------------------------------------------------------------------------------------------% 
such that the effective action is found to be
%-------------------------------------------------------------------------------------------------------------------------------------------------------% 
\begin{equation}S_{\rm eff}=
-\frac 1{4\alpha}
\int dx \,  dy\,\left( \overline{ Q}   \Gamma   \slashed D Q\right)(x) \Box^{-2}(x,y)  (\overline{  \Gamma   \slashed D Q} \ Q )(y)
\ .\label{S_4}\end{equation}
%-------------------------------------------------------------------------------------------------------------------------------------------------------% 
Here $\Box^{-2}(x,y)$ is the square of the inverse propagator of the boson field $\xi^{33}_q$.
The one-loop contribution to the two-point Green function is straightforwardly read off Eq.~(\ref{S_4}).
The form of the self - energy of the third generation quarks
then follows.
% , as written in  
% Eq.~(\ref{Sigma_p}) below.
The corresponding Feynman diagram is shown in Fig.~\ref{fig_1}.
%-------------------------------------------------------------------------------------------------------------------------------------------------------%
\par 
%-------------------------------------------------------------------------------------------------------------------------------------------------------%
\section{Schwinger-Dyson approach for calculating fermion mass}
\label{sec_SD_for_calculating_mf}
%-------------------------------------------------------------------------------------------------------------------------------------------------------%
\subsection{Schwinger - Dyson equation in rainbow approximation}
Analogous to the method used for the toy model in \cite{Miller:2022qil},
at this point a Schwinger-Dyson equation is assembled to express the inverse of the full propagator, $D^{-1}(p)$
in terms of the self-energy function $\Sigma(p)$,
% in the form of a simplified ansatz, 
through which  $\Sigma(p)$ can be determined.
The self - energy function,
$\Sigma (p)$
of the third generation quarks through leading order,
has the form
%-------------------------------------------------------------------------------------------------------------------------------------------------------% 
\begin{equation}
\Sigma(p)=\frac 1{\alpha }\int \dbar k\,\Gamma  \gamma k \frac{i  }{\gamma k- \Sigma(k)} \Gamma  \gamma k
\frac1{(p-k)^4}\ ,
\ \label{Sigma_p}\end{equation}
%-------------------------------------------------------------------------------------------------------------------------------------------------------% 
where 
the standard notation
$\int\dbar k\left(\dots\right)=(2\pi)^{-4}\int d^4k\left(\dots\right)$ has been  used.
Note that this is equivalent to expression (19) in \cite{Miller:2022qil} but with $\Gamma $ set to $1$.
Now apply a Wick rotation (see Appendix \ref{sec_wick_rotatns} for details) to obtain 
%-------------------------------------------------------------------------------------------------------------------------------------------------------% 
\begin{equation}
i\Sigma(p)
=\frac {1}{\alpha }\int \dbar k_E\, \Gamma  \gamma_Ek_E \frac{1}{\left(\gamma_E k_E-i\Sigma(k)\right)} \Gamma  \gamma_Ek_E 
\frac1{(p_E-k_E)^4}
\ .\label{Sigma_p_wick3A}\end{equation}
%-------------------------------------------------------------------------------------------------------------------------------------------------------%
Since all expressions from now on are in terms of Euclidean-space variables 
the subscript `$E$' can be  dropped.\par
%-------------------------------------------------------------------------------------------------------------------------------------------------------% 
% In \cite{Miller:2022qil} (see E.(22) therein) the Schwinger-Dyson equation was written for the toy model
% in which there was just one fermion propagator in 
%-------------------------------------------------------------------------------------------------------------------------------------------------------% 
The self-energy function contains combinations of $\left\{\gamma_\mu\right\}$ matrices  and $\gamma_5$,
and in accordance is a $4\times4$ matrix on the space of four-spinors.
It also contains combinations of the $2\times 2$ projection matrices $\Pi_1 ,\Pi_2 $
making it a function on the space of $SU(2)$ doublets.
(In all $\Sigma(p)$ is an $8\times 8$ matrix.)
So, $\Sigma(p)$ can be expressed in terms of a linear combination of a basis of $4\times 4$ matrices
on the space of Dirac spinors and a basis of $2\times 2$ matrices on the space of $SU(2)$ doublets.
A basis of $2\times2$ matrices is
%-------------------------------------------------------------------------------------------------------------------------------------------------------% 
\begin{equation}
\Pi_1:=\Pi_t=\begin{pmatrix}1&0\\0&0\end{pmatrix}\ ,\quad
\Pi_2:=\begin{pmatrix}0&1\\0&0\end{pmatrix}\ ,\quad 
\Pi_3:= \begin{pmatrix}0&0\\1&0\end{pmatrix}\ ,\quad 
\Pi_4:=\Pi_b=\begin{pmatrix}0&0\\0&1\end{pmatrix}\ .
\label{Pis}
\end{equation}
%-------------------------------------------------------------------------------------------------------------------------------------------------------% 
A basis of $4\times 4$ matrices may be constructed from the $\gamma$ matrices
as the set $\left\{\mathbf 1,\gamma_\mu,\gamma_5,\gamma_\mu\gamma_5,\gamma_{\mu\nu}\right\}$,
where their precise definitions can be found in Appendix~\ref{sec_dirac_matrices}.
%-------------------------------------------------------------------------------------------------------------------------------------------------------%
\begin{equation}\begin{array}{llllllllll}
\Sigma(p)=& & a_1(p) \Pi_1&+&a_2(p) \Pi_2&+&a_3(p) \Pi_3&+&a_4(p) \Pi_4 \\[0.5em]
&+&b_1(p) \gamma_5 \Pi_1&+&b_2(p)\gamma_5  \Pi_2&+&b_3(p) \gamma_5 \Pi_3&+&b_4(p)\gamma_5  \Pi_4
\\[0.5em]
&+&c^\mu_1(p) \gamma_\mu \Pi_1&+&c^\mu_2(p)\gamma_\mu  \Pi_2&+&c^\mu_3(p) \gamma_\mu \Pi_3&+&c^\mu_4(p)\gamma_\mu  \Pi_4
\\[0.5em]
&+&d^\mu_1(p) \gamma_\mu \gamma_5 \Pi_1&+&d^\mu_2(p)\gamma_\mu  \gamma_5 \Pi_2&+&d^\mu_3(p) \gamma_\mu \gamma_5 \Pi_3&+&d^\mu_4(p)\gamma_\mu  \gamma_5  \Pi_4
\\[0.5em]
&+&f^{\mu\nu}_1(p) \gamma_{\mu\nu}  \Pi_1&+&f^{\mu\nu}_2(p)\gamma_{\mu\nu}   \Pi_2&+&f^{\mu\nu}_3(p) \gamma_{\mu\nu}  \Pi_3&+&f^{\mu\nu}_4(p)\gamma_{\mu\nu}   \Pi_4
\end{array}\end{equation}
%-------------------------------------------------------------------------------------------------------------------------------------------------------%
Certain symmetry properties of $\Sigma(p)$ means that a handful of these terms can be dropped.
$\Sigma(p)$ depends on the absolute value of the momentum $p_\mu$, so  the $d^{\mu\nu}(p)$
could only have symmetric combinations of $p^\mu p^\nu$ or $p\delta^{\mu\nu}$, 
which contract on $\Sigma_{\mu\nu}$ to give zero.
Since (\ref{Sigma_p_wick3A}) only contains $\Pi_1 $ and $\Pi_2 $ 
terms containing $\Pi_3$ and $\Pi_4$ shall be disregarded.
Hence $\Sigma(p)$ has the form
%-------------------------------------------------------------------------------------------------------------------------------------------------------%
\begin{equation}\begin{array}{llllllllll}
\Sigma(p)=& & a_1(p) \Pi_1&+&a_2(p) \Pi_2&+&a_3(p) \Pi_3&+&a_4(p) \Pi_4 \\[0.5em]
&+&b_1(p) \gamma_5 \Pi_1&+&b_2(p)\gamma_5  \Pi_2&+&b_3(p) \gamma_5 \Pi_3&+&b_4(p)\gamma_5  \Pi_4
\\[0.5em]
&+&c^\mu_1(p) \gamma_\mu \Pi_1&+&c^\mu_2(p)\gamma_\mu  \Pi_2&+&c^\mu_3(p) \gamma_\mu \Pi_3&+&c^\mu_4(p)\gamma_\mu  \Pi_4
\\[0.5em]
&+&d^\mu_1(p) \gamma_\mu \gamma_5 \Pi_1&+&d^\mu_2(p)\gamma_\mu  \gamma_5 \Pi_2&+&d^\mu_3(p) \gamma_\mu \gamma_5 \Pi_3&+&d^\mu_4(p)\gamma_\mu  \gamma_5  \Pi_4
\end{array}\end{equation}
%-------------------------------------------------------------------------------------------------------------------------------------------------------%
In light of this structure for $\Sigma(p)$ 
a suitable ansatz for the Schwinger-Dyson equation is
%-------------------------------------------------------------------------------------------------------------------------------------------------------%
\begin{equation}
D^{-1}(p)=\gamma  p-i\Sigma(p)
\label{SD_eqn_0}
\end{equation}
%-------------------------------------------------------------------------------------------------------------------------------------------------------%
where 
%-------------------------------------------------------------------------------------------------------------------------------------------------------%
\begin{align}
D^{-1}(p)=&\nonumber\\[-1.4em]
&\begin{array}{rlllllllll}
&\big[A_1(p^2)\Pi_1 &+&A_2(p^2)\Pi_2&+&A_3(p^2)\Pi_3 &+&A_4(p^2)\Pi_4 \big] \gamma  p 
\\[1em]
-i&\big[B_1(p^2) \Pi_1 &+&B_2(p^2) \Pi_2&+& B_3(p^2) \Pi_3&+&B_4(p^2) \Pi_4 \big]
\\[1em]
+&
\big[C_1(p^2)\Pi_1 &+&C_2(p^2)\Pi_2&+& C_3(p^2) \Pi_3&+&C_4(p^2) \Pi_4\big] \gamma  p  \gamma_5
\\[1em]
-i&\big[D_1(p^2)\Pi_1 &+&D_2(p^2) \Pi_2  &+& D_3(p^2) \Pi_3&+&D_4(p^2) \Pi_4\big]\gamma_5
\end{array}\nonumber\\[1em]
=&\hat A\gamma p-i\hat B\otimes \mathbf 1_{4\times 4}+\hat C\gamma p\gamma_5-i\hat D\gamma_5\ .
\label{SD_eqn}\end{align}
%-------------------------------------------------------------------------------------------------------------------------------------------------------%
where $\hat A=\begin{pmatrix}A_1&A_2\\A_3&A_4\end{pmatrix}$ and similarly for $\hat B$, $\hat C$ and $\hat D$.
%-------------------------------------------------------------------------------------------------------------------------------------------------------%
Note the identity 
%-------------------------------------------------------------------------------------------------------------------------------------------------------%
\begin{align}
&\left(\hat A\gamma  p-i\hat  B+\hat  C   \gamma  p\gamma_5-i\hat D \gamma_5\right)
\left(\hat A \gamma  p+i\hat B  +\hat C   \gamma  p\gamma_5-i\hat D \gamma_5\right)
\nonumber\\[1em]
%-------------------------------------------------------------------------------------------------------------------------------------------------------%
=&\left(\hat A \gamma  p-i\hat B  \right)\left(\hat A \gamma  p+i\hat B  \right)
+\left(\hat C   \gamma  p\gamma_5-i\hat D \gamma_5\right)\left(\hat C   \gamma  p\gamma_5-i\hat D \gamma_5\right)
\nonumber\\&
+\left(\hat A \gamma  p-i\hat B  \right)\left(\hat C   \gamma  p\gamma_5-i\hat D \gamma_5\right)
+\left(\hat C   \gamma  p\gamma_5-i\hat D \gamma_5\right)\left(\hat A \gamma  p+i\hat B  \right)
\nonumber\\[1em]
%-------------------------------------------------------------------------------------------------------------------------------------------------------%
=&\hat A ^2(\gamma  p)^2+\hat B  ^2
+i[\hat A,\hat B]\gamma p
+\hat C^2\gamma  p\gamma_5\gamma  p\gamma_5-\hat D ^2\gamma_5^2
-i[\hat C,\hat D]\gamma p\gamma_5^2
\nonumber\\
&
+[\hat A,\hat C](\gamma p)^2\gamma_5
-i[\hat B,\hat C]\gamma p\gamma_5
-i[\hat A,\hat D]\gamma p\gamma_5
-[\hat B,\hat D]\gamma_5
\nonumber\\[1em]
%-------------------------------------------------------------------------------------------------------------------------------------------------------%
=&\hat A ^2(\gamma  p)^2+\hat B  ^2
+i[\hat A,\hat B]\gamma p
-\hat C^2(\gamma  p)^2\gamma_5^2-\hat D ^2\gamma_5^2
-i[\hat C,\hat D]\gamma p\gamma_5^2
\nonumber\\
&
+[\hat A,\hat C](\gamma p)^2\gamma_5
-i[\hat B,\hat C]\gamma p\gamma_5
-i[\hat A,\hat D]\gamma p\gamma_5
-[\hat B,\hat D]\gamma_5
\nonumber\\[1em]
%-------------------------------------------------------------------------------------------------------------------------------------------------------%
=&\left(\hat A^2-\hat C^2\right)p^2+\hat B^2-\hat D^2
\nonumber\\
&+i\left([\hat A,\hat B]-[\hat C,\hat D]\right)\gamma p
+[\hat A,\hat C]p^2\gamma_5
-i\left([\hat B,\hat C]+[\hat A,\hat D]\right)\gamma p\gamma_5
-[\hat B,\hat D]\gamma_5
\ .
\label{SD_eqn_-2}
\end{align}
%-------------------------------------------------------------------------------------------------------------------------------------------------------%
At this stage we seek solutions for which 
the second line of (\ref{SD_eqn_-2}) vanishes.
This is the case provided the commutator of any pair of non-identical matrices from
$\{\hat A,\hat B,\hat C,\hat D\}$ vanish, namely
\begin{equation}
[\hat A,\hat B]=0\ ,\qquad 
[\hat A,\hat C]=0\ ,\qquad
[\hat A,\hat D]=0\,\qquad
[\hat B,\hat C]=0\,\qquad
[\hat B,\hat D]=0\,\qquad
[\hat C,\hat D]=0\ .\label{vanishing_commutators}
\end{equation}
After imposing (\ref{vanishing_commutators}) then
%-------------------------------------------------------------------------------------------------------------------------------------------------------%
\begin{equation}
\left(\hat A\gamma  p-i\hat  B+\hat  C   \gamma  p\gamma_5-i\hat D \gamma_5\right)
\left(\hat A \gamma  p+i\hat B  +\hat C   \gamma  p\gamma_5-i\hat D \gamma_5\right)
%-------------------------------------------------------------------------------------------------------------------------------------------------------%
=(\hat A^2-\hat C^2t)p^2+\hat B^2-\hat D^2
\label{SD_eqn_-1}
\end{equation}
%-------------------------------------------------------------------------------------------------------------------------------------------------------%
and the inverse of (\ref{SD_eqn_0}) is found to be
%-------------------------------------------------------------------------------------------------------------------------------------------------------%
\begin{equation}
\frac1{\gamma p-i\Sigma(p)}
=
\left(\hat A \gamma  p+i\hat B  +\hat C   \gamma  p\gamma_5-i\hat D \gamma_5\right)\cdot \frac{1}{ (\hat A^2-\hat C^2 )p^2+\hat B^2-\hat D^2}\ .
\label{propagator}
\end{equation}
%-------------------------------------------------------------------------------------------------------------------------------------------------------%
where $ \dfrac1{  (\hat A^2-\hat C^2 )p^2+\hat B^2-\hat D^2 }$ on the right stands for the inverse of the matrix
$ (\hat A^2-\hat C^2 )p^2+\hat B^2-\hat D^2$.\par
%-------------------------------------------------------------------------------------------------------------------------------------------------------%
Now (\ref{propagator})   is substituted inside (\ref{Sigma_p_wick3A}) to obtain
%-------------------------------------------------------------------------------------------------------------------------------------------------------% 
\begin{equation}
i\Sigma(p)
= \frac {1}{\alpha }\int \dbar k 
\Gamma  \gamma k 
\left(\hat A \gamma  k+i\hat B  +\hat C   \gamma  k\gamma_5-i\hat D \gamma_5\right)\ \frac1{ (\hat A^2-\hat C^2 )k^2+\hat B^2-\hat D^2}
\Gamma  \gamma k  
\frac1{(p -k )^4}
\ .\label{schwinger_dyson_1}\end{equation}
%-------------------------------------------------------------------------------------------------------------------------------------------------------%
Note that the matrix $\Gamma $  defined in (\ref{M1_M2})
($\Gamma=\beta_LP_R+(\beta_t\Pi_1+\beta_b\Pi_4)P_L$
where $\Pi_t$, $\Pi_b$ have been replaced with $\Pi_1$,$\Pi_4$
in accordance with (\ref{Pis}))
can be anti-commuted past $\gamma$ matrices in accordance with (\ref{PLR_gamma}).
Let
%-------------------------------------------------------------------------------------------------------------------------------------------------------%
\begin{equation}
\tilde \Gamma = \beta_L  P_L
+\left(\beta_t  \Pi_1 +\beta_b   \Pi_4 \right)  P_R\ .\label{schwinger_dyson_2}
\end{equation}
%-------------------------------------------------------------------------------------------------------------------------------------------------------%
Then the following relations are derived from (\ref{PLR_gamma}) and (\ref{PLR_gamma_5}):
%-------------------------------------------------------------------------------------------------------------------------------------------------------% 
{\small \begin{equation}
\Gamma \gamma^\mu =\gamma^\mu \tilde \Gamma \ ,\quad \gamma^\mu \Gamma  =\tilde \Gamma  \gamma^\mu\ ,\quad \tilde \Gamma \gamma^\mu =\gamma^\mu \Gamma \ ,\quad \gamma^\mu \tilde \Gamma  =\Gamma  \gamma^\mu\ ,\quad 
\Gamma \gamma_5=\gamma_5\Gamma \ ,\quad \tilde \Gamma \gamma_5=\gamma_5\tilde \Gamma \ .
\label{schwinger_dyson_3}
\end{equation}}\noindent
%-------------------------------------------------------------------------------------------------------------------------------------------------------%
By using (\ref{schwinger_dyson_3}) and the anticommutation relation in (\ref{gamma5_gamma_anticommutation}), then Eq.~(\ref{schwinger_dyson_1}) becomes
%-------------------------------------------------------------------------------------------------------------------------------------------------------% 
\begin{align}
i\Sigma(p)
=&\frac {1}{\alpha }\int \dbar k 
\Gamma  \gamma k 
\left(\hat A \gamma  k+i\hat B  +\hat C   \gamma  k\gamma_5-i\hat D \gamma_5\right) \frac1{(\hat A^2-\hat C^2)k^2+\hat B^2-\hat D^2}
\gamma k \tilde \Gamma    
\frac1{(p -k )^4}
\nonumber\\[0.5em]
=&\frac {1}{\alpha }\int \dbar k 
\Gamma  \gamma k 
\left(\hat A \gamma  k+i\hat B  +\hat C   \gamma  k\gamma_5-i\hat D \gamma_5\right) \gamma k   \frac1{ (\hat A^2-\hat C^2 )k^2+\hat B^2-\hat D^2}
\tilde \Gamma    
\frac1{(p -k )^4}
\nonumber\\[0.5em]
=&\frac {1}{\alpha }\int \dbar k 
\Gamma  \gamma k
\left(\hat Ak^2 +i\hat B\gamma k   -\hat C   k^2\gamma_5+i\hat D \gamma k\gamma_5\right)   \frac1{ (\hat A^2-\hat C^2)k^2+\hat B^2-\hat D^2}
\tilde \Gamma    
\frac1{(p -k )^4}
\nonumber\\[0.5em]
=&\frac {1}{\alpha }\int \dbar k 
\Gamma  
\left(\hat A  k^2 \gamma k +i\hat B k^2   - \hat C k^2 \gamma k \gamma_5+i \hat Dk^2 \gamma_5\right) \frac1{ (\hat A^2-\hat C^2 )k^2+\hat B^2-\hat D^2}
\tilde \Gamma    
\frac1{(p -k )^4}
\nonumber\\[0.5em]
=&\ \ \ \frac {1}{\alpha } \Gamma  \int \dbar k (\hat A+\hat C\gamma_5)
\frac1{ (\hat A^2-\hat C^2 )k^2+\hat B^2-\hat D^2}
\Gamma   
\frac{k^2\gamma k}{(p -k )^4}
\nonumber\\&
+
\frac {i}{\alpha }\Gamma   \int \dbar k ( \hat B+\hat D \gamma_5)
\frac1{ (\hat A^2-\hat C^2 )k^2+\hat B^2-\hat D^2}
\tilde \Gamma    
\frac{k^2}{(p -k )^4}
\ .\label{schwinger_dyson_4}\end{align}\par 
%-------------------------------------------------------------------------------------------------------------------------------------------------------%
Let (\ref{schwinger_dyson_4}) be written as
%-------------------------------------------------------------------------------------------------------------------------------------------------------% 
\begin{align}
i\Sigma(p)
=&\frac {1}{\alpha } \Gamma     
\Sigma_1(p)
\Gamma   
+
\frac {i}{\alpha }\Gamma   
\Sigma_2(p)\tilde \Gamma    
\ ,\label{schwinger_dyson_5}\end{align}
%-------------------------------------------------------------------------------------------------------------------------------------------------------%
where 
%-------------------------------------------------------------------------------------------------------------------------------------------------------% 
\begin{align}
\Sigma_1(p)
=&  \int \dbar k \ (\hat A+\hat C\gamma_5)\cdot 
\frac{1}{ (\hat A^2-\hat C^2 )k^2+\hat B^2-\hat D^2}\cdot 
\frac{k^2\gamma k}{(p -k )^4}
\ ,\label{schwinger_dyson_6}\\[0.5em]
\Sigma_2(p)
=&  \int \dbar k \ ( \hat B+\hat D \gamma_5)\cdot 
\frac{1}{ (\hat A^2-\hat C^2 )k^2+\hat B^2-\hat D^2}\cdot 
\frac{k^2}{(p -k )^4}
\ .\label{schwinger_dyson_7}\end{align} 
%-------------------------------------------------------------------------------------------------------------------------------------------------------%
$\Sigma_1(p)$ needs to be in a form in which $\gamma p$ appears explicitly such that
$\hat A(p^2)$ and $\hat C(p^2)$ can be easily read off (\ref{SD_eqn_0})   by comparing coefficients.
Multiply  $\Sigma_1$ by $\gamma p$ to obtain 
%-------------------------------------------------------------------------------------------------------------------------------------------------------% 
\begin{align}
\gamma p \Sigma_1(p)
%-------------------------------------------------------------------------------------------------------------------------------------------------------% 
=&  -\frac { 1}{2  } \int \dbar k \ (\hat A+\hat C\gamma_5)\cdot
\frac{k^2   }
{ (\hat A^2-\hat C^2 )k^2+\hat B^2-\hat D^2}  
\bigg(
\frac{1 }{(p-k)^2 }
-
\frac{ p^2+k^2   }{ (p-k)^4}
\bigg)\ .
%-------------------------------------------------------------------------------------------------------------------------------------------------------% 
\label{SD_eqn_3}
\end{align}
%-------------------------------------------------------------------------------------------------------------------------------------------------------% 
Now multiply (\ref{SD_eqn_3}) by $\gamma p$, bearing in mind that 
by (\ref{wick_rot_3A})  $(\gamma p)^2=p^2$, to obtain
%-------------------------------------------------------------------------------------------------------------------------------------------------------% 
\begin{align}
p^2 \Sigma_1(p)=&\frac {-1}{2  }\gamma p
\int \dbar k \
(\hat A+\hat C\gamma_5)\cdot
\frac{k^2    }
{ (\hat A ^2-\hat C  ^2 )k^2+\hat B ^2-\hat D^2} \cdot  
\left(
\frac1{ (p-k)^2 }
-
\frac{  p^2+k^2     }{(p-k)^4}
\right)
%-------------------------------------------------------------------------------------------------------------------------------------------------------% 
\ .\label{SD_eqn_4}
\end{align}
% -------------------------------------------------------------------------------------------------------------------------------------------------------% 
In this form the $\gamma p$ term is explicit and  
(\ref{schwinger_dyson_5}) reads
%-------------------------------------------------------------------------------------------------------------------------------------------------------% 
\begin{align}
i\Sigma(p)
=& \frac {-1}{2\alpha p^2 }\Gamma 
\gamma p
\int \dbar k  \    (\hat A+\hat C\gamma_5)
\cdot\frac{k^2 }
{ (\hat A ^2-\hat C  ^2 )k^2+\hat B ^2-\hat D^2}\cdot 
\left(
\frac1{ (p-k)^2 }
-
\frac{  p^2+k^2     }{(p-k)^4}
\right)  \Gamma   \nonumber\\&
+
\frac {i}{\alpha }\Gamma  
\int \dbar k \  ( \hat B+\hat D \gamma_5)\cdot
\frac{1}{ (\hat A^2-\hat C^2 )k^2+\hat B^2-\hat D^2}\cdot
\frac{k^2}{(p -k )^4} \tilde \Gamma    
\ .\label{schwinger_dyson_8}\end{align}\par 
%-------------------------------------------------------------------------------------------------------------------------------------------------------%
Now substitute (\ref{schwinger_dyson_8}) back into (\ref{SD_eqn_0}) to obtain
%-------------------------------------------------------------------------------------------------------------------------------------------------------%
\begin{align}
&\hat A\gamma p-i\hat B +\hat C\gamma p\gamma_5-i\hat D\gamma_5
\nonumber\\
=&
\gamma p +
\frac {1}{2\alpha p^2 }\Gamma 
\gamma p
\int \dbar k \   (\hat A+\hat C\gamma_5)\cdot 
\frac{k^2 }
{ (\hat A ^2-\hat C  ^2 )k^2+\hat B ^2-\hat D^2}\cdot
\left(
\frac1{ (p-k)^2 }
-
\frac{  p^2+k^2     }{(p-k)^4}
\right)
\Gamma  
\nonumber\\&
-
\frac {i}{\alpha }\Gamma  
\int \dbar k \ ( \hat B-\hat D \gamma_5)\cdot 
\frac1{ (\hat A^2+\hat C^2 )k^2+\hat B^2-\hat D^2}\cdot 
\frac{k^2}{(p -k )^4} \tilde \Gamma    
\nonumber\\[1em]
=&
\gamma p +
\frac {1}{2\alpha p^2 }\Gamma 
\int \dbar k \  (\hat A-\hat C\gamma_5)\cdot 
\frac{k^2     }
{ (\hat A ^2-\hat C  ^2 )k^2+\hat B ^2-\hat D^2} \cdot 
\left(
\frac1{ (p-k)^2 }
-
\frac{  p^2+k^2     }{(p-k)^4}
\right)
\tilde \Gamma  \gamma p
\nonumber\\&
-
\frac {i}{\alpha }\Gamma  
\int \dbar k\  ( \hat B-\hat D \gamma_5)\cdot 
\frac1{ (\hat A^2-\hat C^2 )k^2+\hat B^2-\hat D^2}\cdot
\frac{k^2}{(p -k )^4} \tilde \Gamma \ .  
\label{schwinger_dyson_9}
\end{align}
%-------------------------------------------------------------------------------------------------------------------------------------------------------%
The coefficients of the (linearly-independent) terms $1$, $\gamma p$, $\gamma_5$, $\gamma_5\gamma p$ may be equated to end up with the relations
%-------------------------------------------------------------------------------------------------------------------------------------------------------%
\begin{subequations}\label{sub_schwinger_dyson_10}\begin{align}
\hat A(p^2)
=&1
+
\frac {1}{2\alpha p^2 }\Gamma  
\int \dbar k \ \hat A 
\frac{k^2\,     }
{ (\hat A ^2-\hat C  ^2 )k^2+\hat B ^2-\hat D^2}\cdot 
\left(
\frac1{ (p-k)^2 }
-
\frac{  p^2+k^2     }{(p-k)^4}
\right)
\tilde \Gamma  \ ,
\label{schwinger_dyson_10}\\[1em]
\hat B(p^2)
=&
\frac { 1}{\alpha }\Gamma    
\int \dbar k \ \hat B 
\frac1{ (\hat A^2-\hat C^2 )k^2+\hat B^2-\hat D^2}\cdot 
\frac{k^2}{(p -k )^4} \tilde \Gamma    \ ,
\label{schwinger_dyson_11}\\[1em]
\hat C(p^2)
=& 
\frac {1}{2\alpha p^2 }\Gamma  
\int \dbar k \ \hat C  
\frac{k^2\,     }
{ (\hat A ^2-\hat C  ^2 )k^2+\hat B ^2-\hat D^2}
\cdot 
\left(
\frac1{ (p-k)^2 }
-
\frac{  p^2+k^2     }{(p-k)^4}
\right)
\tilde \Gamma  \ ,
\label{schwinger_dyson_12}\\[1em]
\hat D(p^2)
=&-
\frac {1}{\alpha }\Gamma    
\int \dbar k \ \hat D
\frac1{ (\hat A^2-\hat C^2 )k^2+\hat B^2-\hat D^2}\cdot 
\frac{k^2}{(p -k )^4} \tilde \Gamma    \ .
\label{schwinger_dyson_13}
\end{align}\end{subequations}
%-------------------------------------------------------------------------------------------------------------------------------------------------------%
\subsection{Calculation of integrals over angular degrees of freedom}
Write Eqs.~(\ref{sub_schwinger_dyson_10})  in terms of
four-dimensional polar coordinates:
%-------------------------------------------------------------------------------------------------------------------------------------------------------%
{\small \begin{subequations}\label{schwinger_dyson_14}\begin{align}
\hat A(p^2)
=&1
+
\frac {1}{2\alpha p^2 }\frac1{(2\pi)^4}\Gamma  
\int^\infty_0 dk\int^\pi_0d\theta_1\int^\pi_0d\theta_2\int^{2\pi}_0d\theta_3 k^3\sin^2\theta_1\sin\theta_2
\nonumber\\&
\ k^2\hat A 
\cdot \frac{1}
{ (\hat A ^2-\hat C  ^2 )k^2+\hat B ^2-\hat D^2}  
\left(
\frac1{ p^2+k^2-2pk\cos\theta_1 }
-
\frac{  p^2+k^2     }{(p^2+k^2-2pk\cos\theta_1 )^2}
\right)
\tilde \Gamma 
\nonumber\\
%-------------------------------------------------------------------------------------------------------------------------------------------------------% 
=&1
+
\frac {\pi}{\alpha p^2 }\frac1{(2\pi)^4}\Gamma  
\int^\infty_0 dk^2\int^\pi_0d\theta_1 \sin^2\theta_1 
\nonumber\\&
\ k^4\hat A 
\frac{1}
{ (\hat A ^2-\hat C  ^2 )k^2+\hat B ^2-\hat D^2}
\cdot 
\left(
\frac1{ p^2+k^2-2pk\cos\theta_1 }
-
\frac{  p^2+k^2     }{(p^2+k^2-2pk\cos\theta_1 )^2}
\right)
\tilde \Gamma \ ,
\label{schwinger_dyson_14a}\\[1em]
%-------------------------------------------------------------------------------------------------------------------------------------------------------% 
\hat B(p^2)
=&
\frac {1}{\alpha }\frac1{(2\pi)^4}\Gamma    
\int^\infty_0 dk\int^\pi_0d\theta_1\int^\pi_0d\theta_2\int^{2\pi}_0d\theta_3 k^3\sin^2\theta_1\sin\theta_2
\nonumber\\&
k^2 \hat B 
\frac1{ (\hat A^2-\hat C^2 )k^2+\hat B^2-\hat D^2}\cdot 
\frac1{ (p^2+k^2-2pk\cos\theta_1)^2 }\tilde \Gamma   
\nonumber\\
%-------------------------------------------------------------------------------------------------------------------------------------------------------% 
=&
\frac {2\pi}{\alpha }\frac1{(2\pi)^4}\Gamma    
\int^\infty_0 dk^2\int^\pi_0d\theta_1 \sin^2\theta_1 
\nonumber\\&
k^4 \hat B 
\frac1{ (\hat A^2-\hat C^2 )k^2+\hat B^2-\hat D^2}\cdot 
\frac1{ (p^2+k^2-2pk\cos\theta_1)^2 }\tilde \Gamma   
\ ,
\label{schwinger_dyson_14b}
\\[1em]
%-------------------------------------------------------------------------------------------------------------------------------------------------------% 
\hat C(p^2)
=& 
\frac {1}{2\alpha p^2 }\frac1{(2\pi)^4}\Gamma  
\int^\infty_0 dk\int^\pi_0d\theta_1\int^\pi_0d\theta_2\int^{2\pi}_0d\theta_3 k^3\sin^2\theta_1\sin\theta_2
\nonumber\\& \ k^2\hat C 
\frac{1}
{ (\hat A ^2-\hat C  ^2 )k^2+\hat B ^2-\hat D^2}
\cdot
\left(
\frac1{ p^2+k^2-2pk\cos\theta_1 }
-
\frac{  p^2+k^2     }{(p^2+k^2-2pk\cos\theta_1 )^2}
\right)
\tilde \Gamma 
\nonumber\\
%-------------------------------------------------------------------------------------------------------------------------------------------------------% 
=&
\frac {\pi}{\alpha p^2 }\frac1{(2\pi)^4}\Gamma  
\int^\infty_0 dk^2\int^\pi_0d\theta_1 \sin^2\theta_1 
\nonumber\\&
\ k^4\hat C 
\frac{1}
{ (\hat A ^2-\hat C  ^2 )k^2+\hat B ^2-\hat D^2}\cdot 
\left(
\frac1{ p^2+k^2-2pk\cos\theta_1 }
-
\frac{  p^2+k^2     }{(p^2+k^2-2pk\cos\theta_1 )^2}
\right)
\tilde \Gamma \ .
\label{schwinger_dyson_14c}
\\[1em]
%-------------------------------------------------------------------------------------------------------------------------------------------------------% 
\hat D(p^2)
=&-
\frac {1}{\alpha }\frac1{(2\pi)^4}\Gamma    
\int^\infty_0 dk\int^\pi_0d\theta_1\int^\pi_0d\theta_2\int^{2\pi}_0d\theta_3 k^3\sin^2\theta_1\sin\theta_2
\nonumber\\&
k^2 \hat D
\frac1{ (\hat A^2-\hat C^2 )k^2+\hat B^2-\hat D^2}\cdot
\frac1{ (p^2+k^2-2pk\cos\theta_1)^2 }\tilde \Gamma   
\nonumber\\
=&-
\frac {2\pi}{\alpha }\frac1{(2\pi)^4}\Gamma    
\int^\infty_0 dk^2\int^\pi_0d\theta_1 \sin^2\theta_1 
\nonumber\\&
k^4 \hat D 
\frac1{ (\hat A^2-\hat C^2 )k^2+\hat B^2-\hat D^2}\cdot 
\frac1{ (p^2+k^2-2pk\cos\theta_1)^2 }\tilde \Gamma   
\ .
\label{schwinger_dyson_14d}
\end{align}\end{subequations}}\noindent
%-------------------------------------------------------------------------------------------------------------------------------------------------------% 
% The corresponding relation for $\hat C$ is similar to relation (\ref{schwinger_dyson_14a}) for $\hat A$,
% and 
% The corresponding relation for $\hat D$ is similar to relation (\ref{schwinger_dyson_14b}) for $\hat B$.
The $\theta_1$ integrals have the forms
%-------------------------------------------------------------------------------------------------------------------------------------------------------% 
\begin{align}
I_{1}=&
\int^\pi_0 d\theta_1\frac{\sin^2\theta_1}
{a-b\,\cos\theta_1 }
\ ,\label{angular_integral_1}\\
I_{ 2}=&
\int^\pi_0 d\theta_1\frac{\sin^2\theta_1}
{(a-b\,\cos\theta_1)^2}
\ ,\label{angular_integral_2}
\end{align}
%-------------------------------------------------------------------------------------------------------------------------------------------------------% 
where $a= p^2+k^2 >0$ and $b=2pk$,
such that 
Eqs.~(\ref{schwinger_dyson_14}) 
read
%-------------------------------------------------------------------------------------------------------------------------------------------------------%
\begin{subequations}\label{schwinger_dyson_15}\begin{align}
\hat A(p^2)
=&1
+
\frac {\pi}{\alpha p^2 }\frac1{(2\pi)^4}\Gamma  
\int^\infty_0 dk^2
\ k^4\hat A \frac{1}
{ (\hat A ^2-\hat C  ^2 )k^2+\hat B ^2-\hat D^2}\cdot 
\left(I_1-(p^2+k^2)I_2\right) 
\tilde \Gamma \ ,
\label{schwinger_dyson_15a}\\[1em]
\hat B(p^2)
=&
\frac {2\pi}{\alpha }\frac1{(2\pi)^4}\Gamma    
\int^\infty_0 dk^2
k^4 \hat B  \frac{1}
{ (\hat A ^2-\hat C  ^2 )k^2+\hat B ^2-\hat D^2}\cdot  
I_2 \tilde \Gamma   
\ ,
\label{schwinger_dyson_15b}\\[1em]
\hat C(p^2)
=& 
\frac {\pi}{\alpha p^2 }\frac1{(2\pi)^4}\Gamma  
\int^\infty_0 dk^2
\ k^4\hat C  \frac{1}
{ (\hat A ^2-\hat C  ^2 )k^2+\hat B ^2-\hat D^2}\cdot
\left(I_1-(p^2+k^2)I_2\right) 
\tilde \Gamma \ ,
\label{schwinger_dyson_15c}\\[1em]
\hat D(p^2)
=&-
\frac {2\pi}{\alpha }\frac1{(2\pi)^4}\Gamma    
\int^\infty_0 dk^2  
k^4 \hat D  \frac{1}
{ (\hat A ^2-\hat C  ^2 )k^2+\hat B ^2-\hat D^2} 
I_2 \tilde \Gamma   
\ .
\label{schwinger_dyson_15d}
\end{align}\end{subequations}
%-------------------------------------------------------------------------------------------------------------------------------------------------------% 
The values of the integrals $I_{ 1}$ and $I_{ 2}$ are calculated in 
Appendix \S \ref{sec_angular_integral}.
Their final forms are given in Eqs.~(\ref{I1_eps_expansn})~and~(\ref{I2_eps_expansn}). 
After substituting them in Eqs.~(\ref{schwinger_dyson_15})  the resulting expressions are
%-------------------------------------------------------------------------------------------------------------------------------------------------------% 
\begin{subequations}\label{schwinger_dyson_16}\begin{align}
\hat A(p^2)
=&1
+
\frac {1}{16\pi^2\alpha p^2 }\Gamma  
\int^\infty_0 dk^2
\ k^4\hat A  \frac{1}
{ (\hat A ^2-\hat C  ^2 )k^2+\hat B ^2-\hat D^2} 
\nonumber\\&
\left[
\frac{p^2}{k^2(p^2-k^2)}\theta(k-p)+\frac{k^2}{p^2(k^2-p^2)}\theta(p-k)
\right] 
\tilde \Gamma \ ,
\label{schwinger_dyson_16a}\\[1em]
\hat B(p^2)
=&
\frac {1}{16\pi^2\alpha  }\Gamma    
\int^\infty_0 dk^2  
k^4 \hat B  \frac{1}
{ (\hat A ^2-\hat C  ^2 )k^2+\hat B ^2-\hat D^2} 
\nonumber\\&
\left[
\frac1{k^2(k^2-p^2)}\theta(k-p)
+
\frac1{p^2(p^2-k^2)}\theta(p-k)
\right]
\tilde \Gamma   
\ ,
\label{schwinger_dyson_16b}\\[1em]
\hat C(p^2)
=& 
\frac {1}{16\pi^2\alpha p^2 }\Gamma  
\int^\infty_0 dk^2
\ k^4\hat C \ \frac{1}
{ (\hat A ^2-\hat C  ^2 )k^2+\hat B ^2-\hat D^2} 
\nonumber\\&
\left[
\frac{p^2}{k^2(p^2-k^2)}\theta(k-p)+\frac{k^2}{p^2(k^2-p^2)}\theta(p-k)
\right] 
\tilde \Gamma \ ,
\label{schwinger_dyson_16c}\\[1em]
\hat D(p^2)
=&
-\frac {1}{16\pi^2\alpha  }\Gamma    
\int^\infty_0 dk^2  
k^4 \hat D  \frac{1}
{ (\hat A ^2-\hat C  ^2 )k^2+\hat B ^2-\hat D^2} 
\nonumber\\&
\left[
\frac1{k^2(k^2-p^2)}\theta(k-p)
+
\frac1{p^2(p^2-k^2)}\theta(p-k)
\right] \tilde \Gamma   
\ ,
\label{schwinger_dyson_16d}
\end{align}\end{subequations}
%-------------------------------------------------------------------------------------------------------------------------------------------------------% 
Clearly the integrands in both expressions diverge at $k=p$.
However, 
the integration limits will be adjusted
to take values that agree with current experimental data, such that this
singular point will lie outside of the integration range.
This is 
elucidated in the next paragraph. 
The next step is to 
drop the 
contributions to 
the loop integrals in Eqs.~(\ref{schwinger_dyson_16})
from the region  $k<p$ while retaining 
just the pieces 
from the  region $k>p$.
This too is justified in the following paragraph.
To aid the discussion below
it is useful to expand the remaining terms in $A$ and $B$ in powers of $p^2$ 
to yield
%-------------------------------------------------------------------------------------------------------------------------------------------------------% 
\begin{subequations}\label{schwinger_dyson_17}\begin{align}
\hat A(p^2)
=&1
+
\frac {1}{16\pi^2\alpha   }\Gamma  
\int^\infty_0 \!\!\!dk^2
\  \hat A  \frac{1}
{ (\hat A ^2-\hat C  ^2 )k^2+\hat B ^2-\hat D^2}  
% \nonumber\\&
\left[
-1
-\frac{p^2 }{k^{2}}
-\frac{p^4 k^4}{k^{4}}\right] 
\tilde \Gamma +O(p^6)\ ,
\label{schwinger_dyson_17a}\\[1em]
\hat B(p^2)
=&
\frac {1}{16\pi^2\alpha  }\Gamma    
\int^\infty_0 dk^2  
\hat B  \frac{1}
{ (\hat A ^2-\hat C  ^2 )k^2+\hat B ^2-\hat D^2}  
% \nonumber\\&
\left[1
+
\frac{p^2   }{k^2}
+
\frac{p^4 }{k^{4}}
\right]
\tilde \Gamma   +O(p^6)
\ ,
\label{schwinger_dyson_17b}\\[1em]
\hat C(p^2)
=&
\frac {1}{16\pi^2\alpha   }\Gamma  
\int^\infty_0 dk^2
\  \hat C  \frac{1}
{ (\hat A ^2+\hat C  ^2 )k^2+\hat B ^2-\hat D^2} 
% \nonumber\\&
\left[
-1
-\frac{p^2 }{k^{2}}
-\frac{p^4 k^4}{k^{4}}\right]
\tilde  \Gamma +O(p^6) \ ,
\label{schwinger_dyson_17c}\\[1em]
\hat D(p^2)
=&-
\frac {1}{16\pi^2\alpha  }\Gamma    
\int^\infty_0 dk^2  
\hat D  \frac{1}
{ (\hat A ^2-\hat C  ^2 )k^2+\hat B ^2-\hat D^2} 
% \nonumber\\&
\left[1
+
\frac{p^2   }{k^2}
+
\frac{p^4 }{k^{4}}
\right] \tilde \Gamma   +O(p^6)
\ .
\label{schwinger_dyson_17d}
\end{align}\end{subequations}
%-------------------------------------------------------------------------------------------------------------------------------------------------------%
\subsection{The leading order approximation at $\Lambda \gg \lambda$.}
The leading order in $p^2$ contributions to $A$ and $B$ are straightforwardly read off
Eqs.~(\ref{schwinger_dyson_17}). By  
inserting a cut-off at the ultra-violet and at the infra-red end of the spectrum,
these LO contributions have the forms
%-------------------------------------------------------------------------------------------------------------------------------------------------------% 
\begin{subequations}\label{schwinger_dyson_18}\begin{align}
\hat A_0 
=&1
-
\frac {1}{16\pi^2\alpha   }\Gamma  
\int^{\Lambda^2}_{\lambda^2}  dk^2
\hat A_0  \frac{1}
{ (\hat A_0 ^2-\hat C_0  ^2 )k^2+\hat B_0 ^2-\hat D_0^2} 
\tilde \Gamma \ ,
\label{schwinger_dyson_18a}\\[1em]
\hat B_0 
=&
\frac {1}{16\pi^2\alpha  }\Gamma    
\int^{\Lambda^2}_{\lambda^2}  dk^2  
\hat B_0  \frac{1}
{ (\hat A_0 ^2-\hat C_0  ^2 )k^2+\hat B_0 ^2-\hat D_0^2}  
\tilde \Gamma  
\ ,
\label{schwinger_dyson_18b}\\[1em]
\hat C_0 
=&-
\frac {1}{16\pi^2\alpha   }\Gamma  
\int^{\Lambda^2}_{\lambda^2}  dk^2
\hat C_0  \frac{1}
{ (\hat A_0 ^2-\hat C_0  ^2 )k^2+\hat B_0 ^2-\hat D_0^2} 
\tilde  \Gamma \ ,
\label{schwinger_dyson_18c}\\[1em]
\hat D_0 
=&-
\frac {1}{16\pi^2\alpha  }\Gamma    
\int^{\Lambda^2}_{\lambda^2}  dk^2  
\hat D_0  \frac{1}
{ (\hat A_0 ^2-\hat C_0  ^2 )k^2+\hat B_0 ^2-\hat D_0^2}  
\tilde \Gamma  
\ .
\label{schwinger_dyson_18d}
\end{align}\end{subequations}
%-------------------------------------------------------------------------------------------------------------------------------------------------------%
Let 
%-------------------------------------------------------------------------------------------------------------------------------------------------------%
\begin{equation}\hat B_0 ^2-\hat D_0^2=M^2(\hat A_0 ^2-\hat C_0  ^2 ) .\label{M}\end{equation}
%-------------------------------------------------------------------------------------------------------------------------------------------------------%
Then Eqs.~(\ref{schwinger_dyson_18}) reduce to
%-------------------------------------------------------------------------------------------------------------------------------------------------------% 
\begin{subequations}\label{schwinger_dyson_19_0}\begin{align}
\hat A_0 
=&1
-
\frac {1}{16\pi^2\alpha   }\Gamma \hat A_0 
\int^{\Lambda^2}_{\lambda^2}  dk^2
\left[(k^2+M^2)(\hat A_0 ^2-\hat C_0  ^2)\right]^{-1} 
\tilde \Gamma \ ,
\label{schwinger_dyson_19_0a}\\[1em]
\hat B_0 
=&
\frac {1}{16\pi^2\alpha  }\Gamma \hat B_0   
\int^{\Lambda^2}_{\lambda^2}  dk^2  
\left[(k^2+M^2)(\hat A_0 ^2-\hat C_0  ^2) \right]^{-1} 
\tilde \Gamma  
\ ,
\label{schwinger_dyson_19_0b}\\[1em]
\hat C_0 
=& -
\frac {1}{16\pi^2\alpha  }\Gamma \hat C_0 
\int^{\Lambda^2}_{\lambda^2}  dk^2
\left[(k^2+M^2)(\hat A_0 ^2-\hat C_0  ^2) \right]^{-1}  
\tilde \Gamma \ ,
\label{schwinger_dyson_19_0c}\\[1em]
\hat D_0 
=&-
\frac {1}{16\pi^2\alpha  }\Gamma \hat D_0  
\int^{\Lambda^2}_{\lambda^2}  dk^2  
\left[(k^2+M^2)(\hat A_0 ^2-\hat C_0  ^2) \right]^{-1} 
\tilde \Gamma  
\ .\label{schwinger_dyson_19_0d}
\end{align}\end{subequations}
%-------------------------------------------------------------------------------------------------------------------------------------------------------%
Note that $\left[(k^2+M^2)(\hat A_0 ^2-\hat C_0  ^2) \right]^{-1} =(\hat A_0 ^2-\hat C_0  ^2)^{-1}(k^2+M^2)^{-1}$.
Hence Eqs.~(\ref{schwinger_dyson_19_0}) can be written as
%-------------------------------------------------------------------------------------------------------------------------------------------------------% 
\begin{subequations}\label{schwinger_dyson_19}\begin{align}
\hat A_0 
=&1
-
\frac {1}{16\pi^2\alpha   }\Gamma \hat A_0\frac{1}{\hat A_0 ^2-\hat C_0  ^2} 
\int^{\Lambda^2}_{\lambda^2}  dk^2
\frac{1}
{ k^2+M^2 }  
\tilde \Gamma \ ,
\label{schwinger_dyson_19a}\\[1em]
\hat B_0 
=&
\frac {1}{16\pi^2\alpha  }\Gamma \hat B_0 \frac{1}{\hat A_0 ^2-\hat C_0  ^2}   
\int^{\Lambda^2}_{\lambda^2}  dk^2  
\frac{1}
{  k^2+M^2 } 
\tilde \Gamma  
\ ,
\label{schwinger_dyson_19b}\\[1em]
\hat C_0 
=& -
\frac {1}{16\pi^2\alpha  }\Gamma \hat C_0 \frac{1}{\hat A_0 ^2-\hat C_0  ^2}   
\int^{\Lambda^2}_{\lambda^2}  dk^2
\frac{1}
{  k^2+M^2  }  
\tilde \Gamma \ ,
\label{schwinger_dyson_19c}\\[1em]
\hat D_0 
=&-
\frac {1}{16\pi^2\alpha  }\Gamma \hat D_0 \frac{1}{\hat A_0 ^2-\hat C_0  ^2}     
\int^{\Lambda^2}_{\lambda^2}  dk^2  
\frac{1}
{  k^2+M^2 }  
\tilde \Gamma  
\ .\label{schwinger_dyson_19d}
\end{align}\end{subequations}\par 
%-------------------------------------------------------------------------------------------------------------------------------------------------------%
From (\ref{schwinger_dyson_19b}),
%-------------------------------------------------------------------------------------------------------------------------------------------------------%
\begin{equation}
\hat B_0^{-1}\Gamma^{-1}\hat B_0=
\frac {1}{16\pi^2\alpha  }  \frac{1}{\hat A_0 ^2-\hat C_0  ^2}   
\int^{\Lambda^2}_{\lambda^2}  dk^2  
k^4   \frac{1}
{  k^2+M^2 } 
\tilde \Gamma  
\ .\label{schwinger_dyson_20}
\end{equation}
%-------------------------------------------------------------------------------------------------------------------------------------------------------%
Therefore, (\ref{schwinger_dyson_20}) may be substituted in Eqs.~(\ref{schwinger_dyson_19}) to yield
%-------------------------------------------------------------------------------------------------------------------------------------------------------% 
\begin{subequations}\label{schwinger_dyson_21}\begin{align}
\hat A_0 
=&1
-
\Gamma \hat A_0\hat B_0^{-1}\Gamma^{-1}\hat B_0\ ,
\label{schwinger_dyson_21a}\\[0.5em]
\hat B_0=&\Gamma\hat B_0\hat B_0^{-1}\Gamma^{-1}\hat B_0\ ,
\label{schwinger_dyson_21b}\\[0.5em]
\hat C_0=&-\Gamma\hat C_0\hat B_0^{-1}\Gamma^{-1}\hat B_0\ ,
\label{schwinger_dyson_21c}\\[0.5em]
\hat D_0=& -\Gamma\hat D_0\hat B_0^{-1}\Gamma^{-1}\hat B_0\ .
\label{schwinger_dyson_21d}
\end{align}\end{subequations}
%-------------------------------------------------------------------------------------------------------------------------------------------------------%
The commutation relations (\ref{vanishing_commutators}) are assumed to hold at each order of $p^2$,
hence they apply also for $\{\hat A_0,\hat B_0,\hat C_0,\hat D_0\}$.
And if $[M,N]=0$ for matrices $M,N\in\mathds C^{n\times n}$ ,
then it can be shown that $[M,N^{-1}]=[M^{-1},N]=[M^{-1},N^{-1}]=0$.
From this $\hat A_0$, $\hat B_0^{-1}$ can be swapped in Eqs.~(\ref{schwinger_dyson_21}),
as well as $\hat C_0$, $\hat B_0^{-1}$ 
and $\hat D_0$, $\hat B_0^{-1}$.
Hence Eqs.~(\ref{schwinger_dyson_21}) can be reduced to
%-------------------------------------------------------------------------------------------------------------------------------------------------------% 
\begin{subequations}\label{schwinger_dyson_22}\begin{align}
\hat A_0 
=&1
-
\Gamma \hat A_0\hat B_0^{-1}\Gamma^{-1}\hat B_0\ ,
\label{schwinger_dyson_21a}\\[0.5em]
\hat B_0=&\Gamma\hat B_0\hat B_0^{-1}\Gamma^{-1}\hat B_0\ ,
\label{schwinger_dyson_21b}\\[0.5em]
\hat C_0=&-\Gamma\hat C_0\hat B_0^{-1}\Gamma^{-1}\hat B_0\ ,
\label{schwinger_dyson_21c}\\[0.5em]
\hat D_0=& -\Gamma\hat D_0\hat B_0^{-1}\Gamma^{-1}\hat B_0\ .
\label{schwinger_dyson_21d}
\end{align}\end{subequations}
%-------------------------------------------------------------------------------------------------------------------------------------------------------%
The simplest solution is when all of the 
$\{\hat A_0,\hat B_0,\hat C_0,\hat D_0\}$ are diagonal.
Then, given that $\Gamma$, $\tilde \Gamma$ are diagonal as well,
Eqs.~(\ref{schwinger_dyson_22}) reduce to
%-------------------------------------------------------------------------------------------------------------------------------------------------------% 
\begin{subequations}\label{schwinger_dyson_23}\begin{align}
\hat A_0 
=&1
-
\frac {1}{16\pi^2\alpha   }\Gamma \hat A_0\frac{1}{\hat A_0 ^2 } 
\int^{\Lambda^2}_{\lambda^2}  dk^2
\frac{1}
{ k^2+M^2 }  
\tilde \Gamma \ ,
\label{schwinger_dyson_23a}\\[1em]
\mathds 1_{2\times 2}
=&
\frac {1}{16\pi^2\alpha  }\Gamma   \frac{1}{\hat A_0 ^2 }   
\int^{\Lambda^2}_{\lambda^2}  dk^2  
\frac{1}
{  k^2+M^2 } 
\tilde \Gamma  
\ ,
\label{schwinger_dyson_23b}\\[1em]
\hat C_0=&0\ ,
\label{schwinger_dyson_23c}\\[1em]
\hat D_0=&0\ .
\label{schwinger_dyson_23d}
\end{align}\end{subequations} 
%-------------------------------------------------------------------------------------------------------------------------------------------------------%
In relation to this solution Eq.~(\ref{M}) 
implies
%-------------------------------------------------------------------------------------------------------------------------------------------------------%
\begin{equation}\hat B_0=M\hat A_0\ .\label{M-1}\end{equation}
%-------------------------------------------------------------------------------------------------------------------------------------------------------%
Eqs.~(\ref{schwinger_dyson_23a}) and (\ref{schwinger_dyson_23b})
are $2\times 2$ matrix equations, with completely diagonal matrices on both sides.
Importantly, by constraining 
$\hat A_0$ and $\hat B_0$ to be diagonal, then consistent with 
(\ref{M})
the matrix $M^2$ is forced to be diagonal.
In light of this let it be written as
%-------------------------------------------------------------------------------------------------------------------------------------------------------%
\begin{equation}
M =\begin{pmatrix}m_t &0\\0&m_b \end{pmatrix}\ .\label{mass_matrix}
\end{equation}
%-------------------------------------------------------------------------------------------------------------------------------------------------------%
$M$ will be referred to as the \emph{mass matrix}.
Then 
%-------------------------------------------------------------------------------------------------------------------------------------------------------%
\begin{equation}
\left(k^2+M^2\right)^{-1}=\begin{pmatrix}\frac1{k^2+m_t^2} &0\\0&\frac1{k^2+m_b^2} \end{pmatrix}\ .\label{mass_matrix-1}
\end{equation}
%-------------------------------------------------------------------------------------------------------------------------------------------------------%
According to the definitions in (\ref{M1_M2}) and (\ref{schwinger_dyson_2}),
%-------------------------------------------------------------------------------------------------------------------------------------------------------% 
\begin{align}
\Gamma=&\begin{pmatrix} \beta_L P_R+\beta_t P_L  &0\\0& \beta_LP_R+\beta_b P_L  \end{pmatrix}\ ,\label{Gamma_explicit}\\[1em]
\tilde \Gamma=&\begin{pmatrix} \beta_L P_L+\beta_t P_R  &0\\0& \beta_LP_L+\beta_b P_R  \end{pmatrix}\ .\label{Gamma_tilde_explicit}
\end{align}
%-------------------------------------------------------------------------------------------------------------------------------------------------------%
Let $\hat A_0$, $\hat B_0$ be written in a notation more fitting with these relations:
%-------------------------------------------------------------------------------------------------------------------------------------------------------%
\begin{equation}
\hat A_0= \begin{pmatrix}A_t&0\\0&A_b \end{pmatrix}\ ,\label{A0_diagonal}
\end{equation}
%-------------------------------------------------------------------------------------------------------------------------------------------------------%
and 
%-------------------------------------------------------------------------------------------------------------------------------------------------------%
\begin{equation}
\hat B_0= \begin{pmatrix}B_t&0\\0&B_b \end{pmatrix}=
\begin{pmatrix}m_t A_t&0\\0&m_b A_b \end{pmatrix}
\ ,\label{B0_diagonal}
\end{equation}
%-------------------------------------------------------------------------------------------------------------------------------------------------------%
where the last equality follows directly from Eqs.~(\ref{M-1}), (\ref{mass_matrix}) and (\ref{A0_diagonal}).
Altogether 
the two diagonal elements of the matrix equation in 
Eq.~(\ref{schwinger_dyson_23a}) are
%-------------------------------------------------------------------------------------------------------------------------------------------------------% 
\begin{subequations}\label{schwinger_dyson_24}\begin{align}
A_t 
% =&1
% -
% \frac {1}{16\pi^2\alpha   }(\beta_L P_R+\beta_t P_L)A_t \frac{1}{ A_t^2 } 
% \int^{\Lambda^2}_{\lambda^2}  dk^2
%     \frac{1}
% { k^2+m_t^2 }  
% (\beta_L P_L+\beta_t P_R)\nonumber\\[0.5em]
=&
1
-
\frac {1}{16\pi^2\alpha   }\beta_L\beta_t A_t\frac{1}{ A_t^2 } 
\int^{\Lambda^2}_{\lambda^2}  dk^2
\frac{1}
{ k^2+m_t^2 }  
\ ,
\label{schwinger_dyson_24a}\\[1em]
A_b 
% =&1
% -
% \frac {1}{16\pi^2\alpha   }(\beta_L P_R+\beta_b P_L)A_b \frac{1}{ A_b^2} 
% \int^{\Lambda^2}_{\lambda^2}  dk^2
%     \frac{1}
% { k^2+m_b^2 }  
% (\beta_L P_L+\beta_b P_R)\nonumber\\[0.5em]
=&
1
-
\frac {1}{16\pi^2\alpha   }\beta_L\beta_b A_b\frac{1}{ A_b^2 } 
\int^{\Lambda^2}_{\lambda^2}  dk^2
\frac{1}
{ k^2+m_b^2 }  
\ ,
\label{schwinger_dyson_24b}
\end{align}\end{subequations} 
%-------------------------------------------------------------------------------------------------------------------------------------------------------%
while
the two diagonal elements of the matrix equation in 
Eq.~(\ref{schwinger_dyson_23b}) are
%-------------------------------------------------------------------------------------------------------------------------------------------------------% 
\begin{subequations}\label{schwinger_dyson_25}\begin{align}
1 
% =& 
% \frac {1}{16\pi^2\alpha   }(\beta_L P_R+\beta_t P_L) \frac{1}{ A_t^2 } 
% \int^{\Lambda^2}_{\lambda^2}  dk^2
%     \frac{1}
% { k^2+m_t^2 }  
% (\beta_L P_L+\beta_t P_R)\nonumber\\[0.5em]
=&
\frac {1}{16\pi^2\alpha   }\beta_L\beta_t \frac{1}{ A_t^2 } 
\int^{\Lambda^2}_{\lambda^2}  dk^2
\frac{1}
{ k^2+m_t^2 }  
\ ,
\label{schwinger_dyson_25a}\\[1em]
1
% =& 
% \frac {1}{16\pi^2\alpha   }(\beta_L P_R+\beta_b P_L) \frac{1}{ A_b^2} 
% \int^{\Lambda^2}_{\lambda^2}  dk^2
%     \frac{1}
% { k^2+m_b^2 }  
% (\beta_L P_L+\beta_b P_R)\nonumber\\[0.5em]
=& 
\frac {1}{16\pi^2\alpha   }\beta_L\beta_b \frac{1}{ A_b^2 } 
\int^{\Lambda^2}_{\lambda^2}  dk^2
\frac{1}
{ k^2+m_b^2 }  
\ .
\label{schwinger_dyson_25b} 
\end{align}\end{subequations} \par 
%-------------------------------------------------------------------------------------------------------------------------------------------------------%
Eq.~(\ref{schwinger_dyson_25a}) may be substituted into (\ref{schwinger_dyson_24a}),
to obtain
%-------------------------------------------------------------------------------------------------------------------------------------------------------%
\begin{equation}A_t=1-A_t\qquad\Rightarrow\qquad A_t=\frac1{2}\ .\label{soln_At}\end{equation}
%-------------------------------------------------------------------------------------------------------------------------------------------------------%
Now substitute (\ref{soln_At}) in (\ref{schwinger_dyson_25a}) to find
%-------------------------------------------------------------------------------------------------------------------------------------------------------%
\begin{equation}1=\frac1{4\pi^2\alpha_t}\ln\left(\frac{\Lambda^2+m_t^2}{\lambda^2+m_t^2}\right)\label{gapt}
\ ,\end{equation}
%-------------------------------------------------------------------------------------------------------------------------------------------------------%
where
%-------------------------------------------------------------------------------------------------------------------------------------------------------%
\begin{equation}
\alpha_t=\frac{\alpha}{\beta_L\beta_t}\ .\label{alpha_t}
\end{equation}
%-------------------------------------------------------------------------------------------------------------------------------------------------------%
Equally Eq.~(\ref{schwinger_dyson_25b}) may be substituted into (\ref{schwinger_dyson_24b}) 
to obtain 
%-------------------------------------------------------------------------------------------------------------------------------------------------------%
\begin{equation}A_b=1-A_b\qquad\Rightarrow\qquad A_b=\frac1{2}\ ,\label{soln_Ab}\end{equation}
%-------------------------------------------------------------------------------------------------------------------------------------------------------%
which may be substituted  in (\ref{schwinger_dyson_25b}) to yield
%-------------------------------------------------------------------------------------------------------------------------------------------------------%
\begin{equation}1=\frac1{4\pi^2\alpha_b}\ln\left(\frac{\Lambda^2+m_b^2}{\lambda^2+m_b^2}\right)\label{gapb}
\ ,\end{equation}
%-------------------------------------------------------------------------------------------------------------------------------------------------------%
where
%-------------------------------------------------------------------------------------------------------------------------------------------------------%
\begin{equation}
\alpha_b=\frac{\alpha}{\beta_L\beta_b}\ .\label{alpha_b}
\end{equation}
We assume that the ultraviolet cutoff is at the scale of Plank mass while the infrared scale is $1$ TeV. 
Then the critical value of the coupling constant $\alpha_t$ is given by 
%-------------------------------------------------------------------------------------------------------------------------------------------------------%
\begin{equation}\alpha^{(c)} = \frac{1}{4\pi^2}\ln\left( \frac{\Lambda^2 }{\lambda^2}\right) \sim 1.866405176474873\ .
\label{alpha_c}
\end{equation}
%-------------------------------------------------------------------------------------------------------------------------------------------------------%
At $\alpha_t < \alpha^{(c)}$ the top quark mass is generated dynamically, with the value
%-------------------------------------------------------------------------------------------------------------------------------------------------------% 
\begin{equation}
m_t=\Lambda e^{-2\pi^2\alpha_t} \sqrt{1-e^{4\pi^2 (\alpha_t - \alpha^{(c)})}}
\ .\label{B_LO_3}\end{equation}
%-------------------------------------------------------------------------------------------------------------------------------------------------------% 
while for $\alpha_b < \alpha^{(c)}$ the $b$ quark mass is generated dynamically, with the value
%-------------------------------------------------------------------------------------------------------------------------------------------------------% 
\begin{equation}
m_b=\Lambda e^{-2\pi^2\alpha_b} \sqrt{1-e^{4\pi^2 (\alpha_b - \alpha^{(c)})}}
\ .\label{B_LO_3_b}\end{equation}
%-------------------------------------------------------------------------------------------------------------------------------------------------------% 
It follows from (\ref{B_LO_3})  
that the generated top mass varies from $0$ at $\alpha_t = \alpha^{(c)}$ to a value that approaches $\Lambda$ 
at strong coupling $1/\alpha_t \gg 1$. The top quark mass, which is approximately  $175$ GeV, is generated for 
%-------------------------------------------------------------------------------------------------------------------------------------------------------%
\begin{equation}
\alpha_t = \frac{1}{4 \pi^2} \, \log\, \frac{\Lambda^2 + m_t^2}{\lambda^2 + m_t^2} = 1.865641077603585\ .
\label{alpha_t}
\end{equation}
%-------------------------------------------------------------------------------------------------------------------------------------------------------%
% $\alpha_t = 1.865641077603585$. 
In order to generate the observable bottom quark mass $m_b=4.18$ GeV we need
%-------------------------------------------------------------------------------------------------------------------------------------------------------%
\begin{equation}
\alpha_b = \frac{1}{4 \pi^2} \, \log\, \frac{\Lambda^2 + m_b^2}{\lambda^2 + m_b^2} = 1.866404733897677\ .
\label{alpha_b}
\end{equation}
%-------------------------------------------------------------------------------------------------------------------------------------------------------%
Without loss of generality we can set $\beta_L=1/2$. Then the original coupling constants to be set up are $\alpha, \beta_b, \beta_t$. The following choice of their values leads to the observable values of top quark and bottom quark masses:
%-------------------------------------------------------------------------------------------------------------------------------------------------------%
$$
\beta_L = \alpha = 1/2; \quad \beta_b =0.53578946829590691109\ ,\quad  \beta_t = 0.53600878111265624799\ ,
$$  
%-------------------------------------------------------------------------------------------------------------------------------------------------------%
At $\beta_L = \alpha = 1/2$ the critical values of $\beta_b, \beta_t$ corresponding to nearly vanishing values of masses are
%-------------------------------------------------------------------------------------------------------------------------------------------------------%
$$
\beta^{(c)}_L = \alpha^{(c)} = 1/2; \quad \beta^{(c)}_b = \beta^{(c)}_t = 0.53578934124514456055\ .
$$ 
%-------------------------------------------------------------------------------------------------------------------------------------------------------%
It is immediately clear that 
the above adjustment of the couplings is 
a certain kind of fine tuning, given that 
the values of the couplings that provide physical values of the quark masses are very close to their critical values. 
In fact, the value of  $\beta_b$ differs from the value of $\beta^{(c)}_b$ by an order of $10^{-7}$. 
\par
\begin{figure}[htb!]
% \centering
\includegraphics[width=9cm]{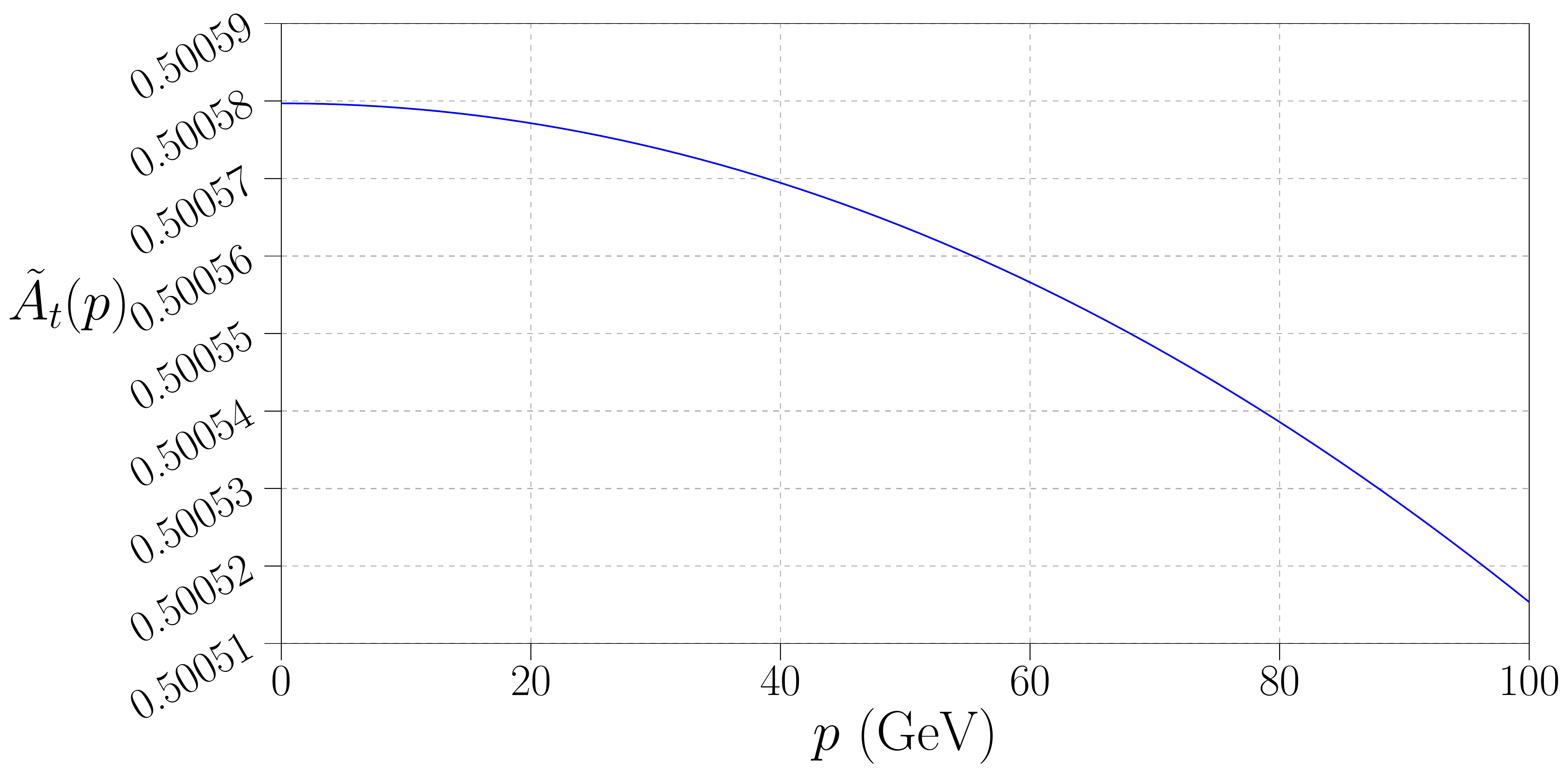}
\caption{\footnotesize 
The function $\tilde A_t(p)$ found by evaluating the integral in Eq.~(\ref{schwinger_dyson_15a}) 
for the values $\Lambda=10^{19}\,{\rm Gev}$ (the Planck mass),
$\lambda=1\,{\rm TeV}$,
$m_t=175\,{\rm GeV}$ and $\alpha_t=1.86564$.
The values of $\tilde A_t(p)$ are very close to $A_0=\frac1{2}$ as expected.
}
\label{fig_At_approx}
\end{figure}

\begin{figure}[htb!]
% \centering
\includegraphics[width=9cm]{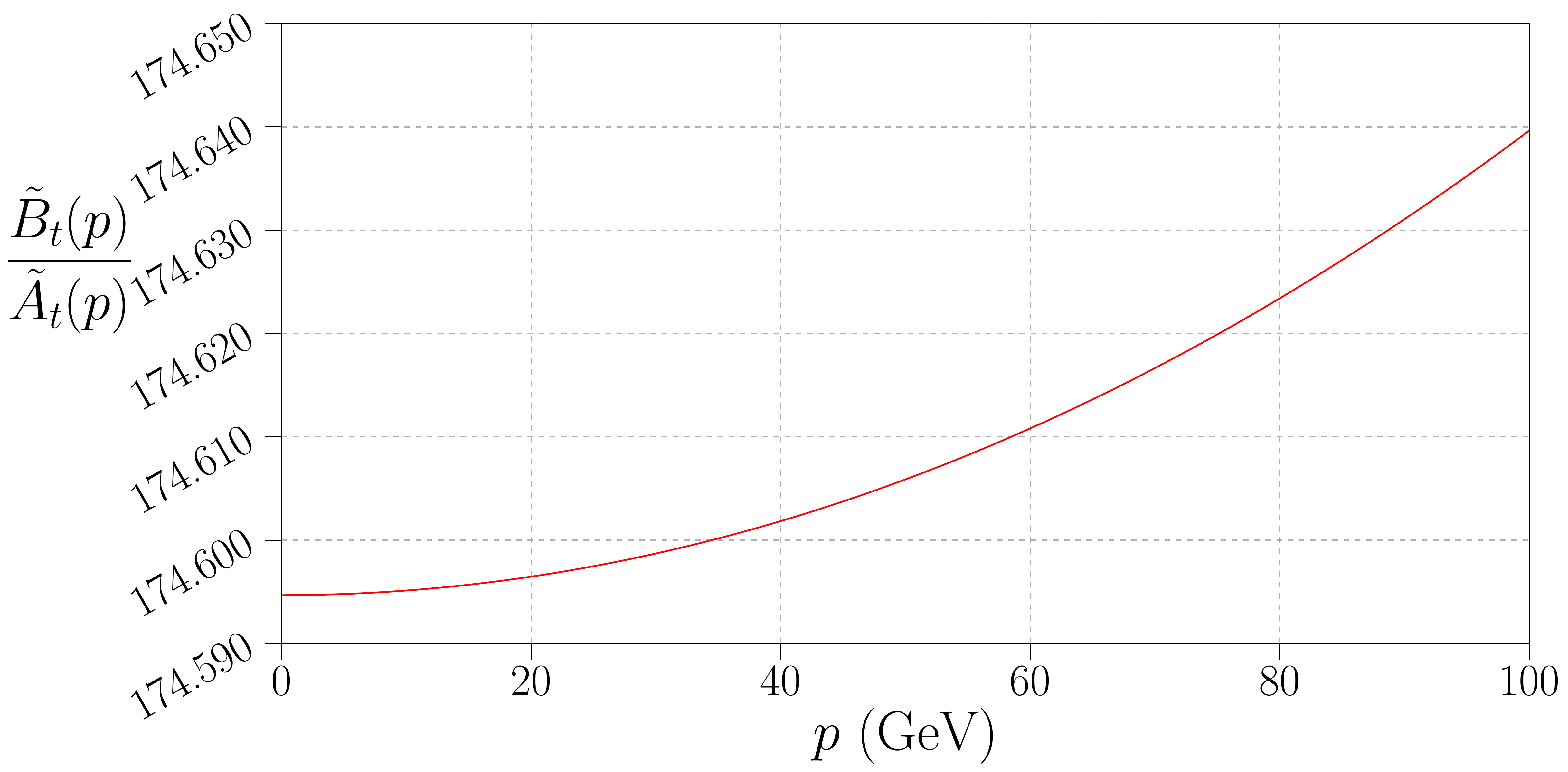}
\caption{\footnotesize 
The function $\tilde B_t(p)/\tilde A_t(p)$ found by evaluating the integral in
(\ref{schwinger_dyson_15b}) with the same numerical parameters as Fig.~\ref{fig_At_approx}.
The values of $\tilde B_t(p)/\tilde A_t(p)$ are very close to $m_t=175$ GeV, as expected.
}
\label{fig_Bt_approx}
\end{figure}

\begin{figure}[htb!]
% \centering
\includegraphics[width=9cm]{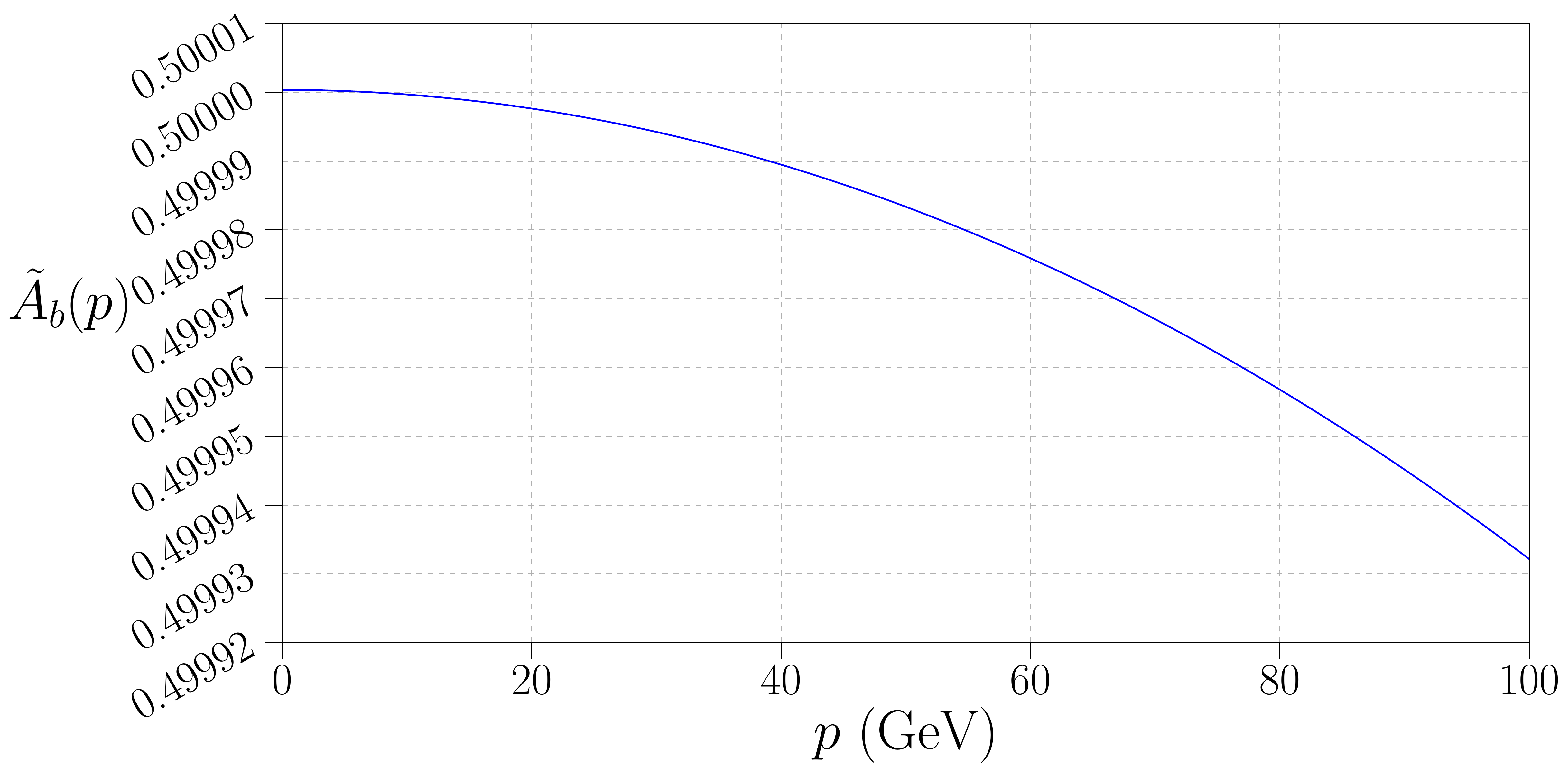}
\caption{\footnotesize 
The function $\tilde A_t(p)$ found by evaluating the integral in Eq.~(\ref{schwinger_dyson_15a}) 
for the values $\Lambda=10^{19}\,{\rm Gev}$ (the Planck mass),
$\lambda=1\,{\rm TeV}$,
$m_b=4.18\,{\rm GeV}$ and $\alpha_b=1.866404733897677$.
The values of $\tilde A_b(p)$ are very close to $A_{b0}=\frac1{2}$ as expected.
}
\label{fig_Ab_approx}
\end{figure}

\begin{figure}[htb!]
% \centering
\includegraphics[width=9cm]{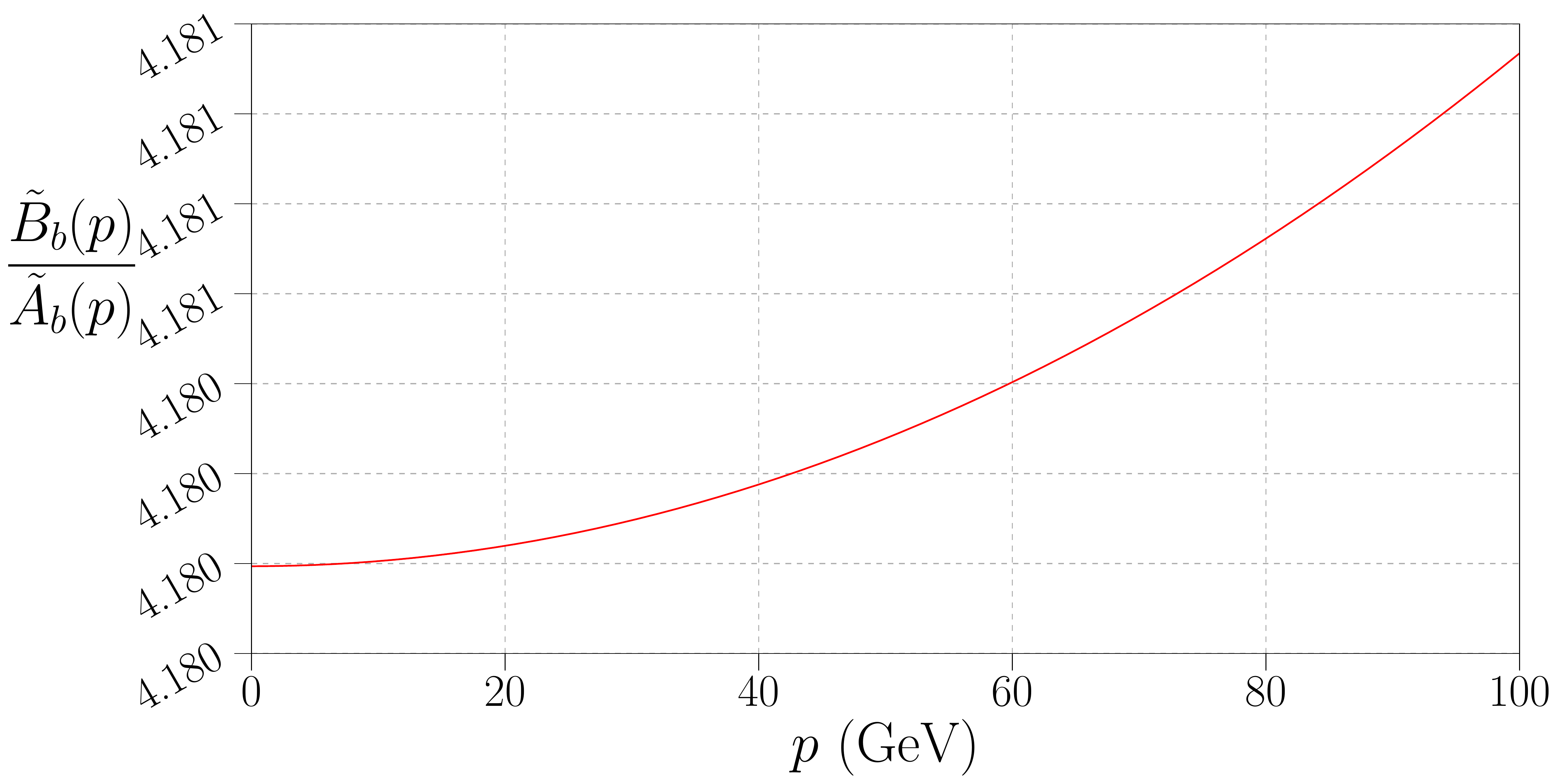}
\caption{\footnotesize 
The function $\tilde B_b(p)/\tilde A_b(p)$ found by evaluating the integral in
(\ref{schwinger_dyson_15b}) with the same numerical parameters as Fig.~\ref{fig_Ab_approx}.
The values of $\tilde B_b(p)/\tilde A_b(p)$ are very close to $m_b=4.18$ GeV, as expected.
}
\label{fig_Bb_approx}
\end{figure}

 
%-------------------------------------------------------------------------------------------------------------------------------------------------------%
As a consistency check,
take the integrals in (\ref{schwinger_dyson_16a}) and (\ref{schwinger_dyson_16b})
but with $C=D=0$,
and with the cutoffs $k^2=\lambda^2$ at the lower end  and $k^2=\Lambda^2$ at the upper end of the spectrum,
where recall that $\lambda\gg p $ is assumed  such that only the $k>p$ contribution survives.
Then 
substitute for $A$ and $B$ inside the integrands, the leading-order values found in (\ref{soln_At}) and (\ref{B_LO_3}).
Select the values $\Lambda=10^{19}\,{\rm Gev}$ (the Planck mass),
$\lambda=1\,{\rm TeV}$,
$m_t=175\,{\rm GeV}$ and $\alpha_t=1.865641077603585$ found in (\ref{alpha_t}),  then evaluate the integrals over $k^2$.
In this regime, the integrals
as functions of $p$ labeled as
$\tilde A_t(p)$ and $\tilde B_t(p)$,
have the forms
%-------------------------------------------------------------------------------------------------------------------------------------------------------%
\begin{align}
\tilde A_t(p)=&1+\frac1{16\pi^2\alpha_t}\int^{\Lambda^2}_{\lambda^2}dk^2\frac{A_{t0} k^4}{A_{t0}^2 k^2+B_{t0}^2}\frac{1}{(p^2-k^2)}\ ,
\label{tilde_A_t}\\
\tilde B_t(p)=& \frac1{16\pi^2\alpha_t}\int^{\Lambda^2}_{\lambda^2}dk^2\frac{B_{t0} k^2}{A_{t0}^2 k^2+B_{t0}^2}\frac{1}{(k^2-p^2)}\ ,
\label{tilde_B_t}
\end{align}
%-------------------------------------------------------------------------------------------------------------------------------------------------------%
where in Eqs.~(\ref{tilde_A_t}) and (\ref{tilde_B_t}),
$A_{t0}=\frac1{2}$ and
$B_{t0}=\frac{m_t}{2}=87.5$ GeV should be inserted.
The values
for $\tilde A_t(p)$ and $\tilde B_t(p)/\tilde A_t(p)$ as functions of $p$ are shown in the plots in Figs~\ref{fig_At_approx} and Fig.~\ref{fig_Bt_approx}.
The values of $\tilde A_t(p)$ are very close to $A_{t0}=\frac1{2}$ as expected, while
the values of $\tilde B_t(p)/\tilde A_t(p)$ are very close to $m_t=175$ GeV, also as expected.
%-------------------------------------------------------------------------------------------------------------------------------------------------------%
%-------------------------------------------------------------------------------------------------------------------------------------------------------%
%-------------------------------------------------------------------------------------------------------------------------------------------------------%
Now we generate analogous plots for $A_b$ and $B_b$.
We
take the integrals in (\ref{schwinger_dyson_16a}) and (\ref{schwinger_dyson_16b})
with $C=D=0$,
and cutoffs $k^2=\lambda^2$ at the lower end  and $k^2=\Lambda^2$ at the upper end of the spectrum, but this time we 
substitute for $A$ and $B$, the leading-order values found in (\ref{soln_Ab}) and (\ref{B_LO_3_b}),
with the same  values for $\Lambda$ and
$\lambda $,
$m_b=4.18\,{\rm GeV}$ and $\alpha_b=1.866404733897677$ as found in (\ref{alpha_b}), 
then evaluate the integrals over $k^2$.
In this regime, the integrals
as functions of $p$ labeled as
$\tilde A_b(p)$ and $\tilde B_b(p)$,
have the forms
%-------------------------------------------------------------------------------------------------------------------------------------------------------%
\begin{align}
\tilde A_b(p)=&1+\frac1{16\pi^2\alpha_b}\int^{\Lambda^2}_{\lambda^2}dk^2\frac{A_{b0} k^4}{A_{b0}^2 k^2+B_{b0}^2}\frac{1}{(p^2-k^2)}\ ,
\label{tilde_A_b}\\
\tilde B_b(p)=& \frac1{16\pi^2\alpha_b}\int^{\Lambda^2}_{\lambda^2}dk^2\frac{B_{b0} k^2}{A_{b0}^2 k^2+B_{b0}^2}\frac{1}{(k^2-p^2)}\ ,
\label{tilde_B_b}
\end{align}
%-------------------------------------------------------------------------------------------------------------------------------------------------------%
where in Eqs.~(\ref{tilde_A_t}) and (\ref{tilde_B_t}),
$A_{b0}=\frac1{2}$ and
$B_{b0}=\frac{m_b}{2}=2.09$ GeV should be inserted.
The values
for $\tilde A_b(p)$ and $\tilde B_b(p)/\tilde A_b(p)$ as functions of $p$ are shown in the plots in Figs~\ref{fig_Ab_approx} and Fig.~\ref{fig_Bb_approx}.
The values of $\tilde A_b(p)$ are very close to $A_{b0}=\frac1{2}$ as expected, while
the values of $\tilde B_b(p)/\tilde A_b(p)$ are very close to $m_b=4.18$ GeV, also as expected.
%-------------------------------------------------------------------------------------------------------------------------------------------------------%
%-------------------------------------------------------------------------------------------------------------------------------------------------------%
%-------------------------------------------------------------------------------------------------------------------------------------------------------%
\section{Conclusions and Discussion}\label{sec_conclusions} \noindent
%-------------------------------------------------------------------------------------------------------------------------------------------------------%
In this paper we have extended our previous  model \cite{Miller:2022qil}, which 
contained a single fermion, namely the top quark, to that comprising both top and bottom quarks with distinct masses.
In particular 
we have inserted couplings into the action in a way that allows for distinguishable masses between the top and bottom quarks. 
This form of the action leads to 
mass-gap equations  bearing solutions corresponding to non-equal masses of top and bottom quarks. 
By suitably adjusting the couplings, our model yields predicted masses for the top and bottom quarks  
that match experimental observations. In the particular case
where the ultraviolet cutoff is set to the Plank mass, while the infrared cutoff of the theory is around 1 TeV,
the gap equations can be brought to the simple form of (\ref{gapt}) and (\ref{gapb}). 
Then the above adjustment of the couplings is a form of fine tuning,
because the critical values of the coupling constants are very close to those specific values that give rise to physical masses. 
This indicates that the two cutoffs should in reality be close to each other, i.e. the UV completion of the proposed theory should enter the game
at an energy much lower than the Plank mass. 
In this situation the approximation giving rise to Eqs. (\ref{gapt}) and (\ref{gapb}) does not work, 
and we should deal directly with the more complicated form of the gap equations, namely Eqs. (\ref{schwinger_dyson_16}). Nonetheless, our solution to
these equations at extreme values of $\Lambda$, give a reasonably good description of the phenomena.
\par 
%-------------------------------------------------------------------------------------------------------------------------------------------------------%
Concerning future prospects for the development of our proposed theory, there are several questions yet to be answered, as detailed below.

\begin{enumerate}
\item 
The first question concerns the accuracy of our calculations. We use the truncated Schwinger - Dyson equations to calculate the quark masses. 
We have seen that the values of the coupling constants needed  to generate the observable values of quark masses (at least, at $\Lambda \sim m_P$)
are of the order of $0.5$. Therefore, our model cannot be used in the weak coupling regime. In this case, any application of the truncated Schwinger - Dyson equations will be limited. Nevertheless  it is generally accepted that even at non - small values of couplings, these equations in the rainbow approximation give a reasonable approximation to certain physical quantities. Still this should  be checked using a more refined scheme of calculations. Feasibly  lattice simulations could be used to check this hypothesis. 

\item
In this paper we concentrate  on the calculation of the top and bottom quark masses in the toy model constructed here.
In actual fact this model predicts appearance of not just the standard model composite Higgs but also
several scalar excitations in addition. The former Higgs boson mass should equal $125$ GeV as observed.
At the same time the masses of the remaining composite scalar bosons should be larger than  present experimental bounds. 
Those experimental bounds also include values of the decay constants of the scalar bosons.
Thus, it is necessary to calculate these decay constants as well as the extra Higgs boson masses. For this purpose, the 
Bethe - Salpeter equations for the corresponding excitations should be solved. 
This is beyond the scope of this paper but will be addressed in a follow-up paper.

\item
The theory put forward here inclusive of  fundamental scalar fields admits two types of divergences: ultraviolet and infrared. 
Correspondingly, we insert  two cutoffs. The physical interpretation of these cut-offs 
is that the  theory is an approximation to a more complicated one, in which the presence of  zero dimension scalar bosons is suppressed at  energies below $\lambda$, while at the energies above $\Lambda$ extra excitations are present. 
Since our model originates from Riemann - Cartan gravity coupled to SM fermions  non - minimally,
 we assume that the UV completion of the proposed model has to be an extension of  Riemann - Cartan gravity. 
In the same extended model,  the vielbein and spin connection have extra indices, 
and as such become matrices in flavor space. 
The construction of such a theory is an important task, that while beyond the scope of this work, will be 
addressed in our next paper.
\end{enumerate}

The authors are grateful to G.E.Volovik for useful discussions. M.A.Z. is indebted to V.A.Miransky for numerous discussions in the past on the Schwinger - Dyson equations technique.     

\appendix
%-------------------------------------------------------------------------------------------------------------------------------------------------------%
\section{Conventions for Dirac matrices and chiral 
%projection 
operators}\label{sec_dirac_matrices}\noindent 
%-------------------------------------------------------------------------------------------------------------------------------------------------------%
\subsection{Dirac matrices}\noindent
%-------------------------------------------------------------------------------------------------------------------------------------------------------%
Throughout these notes the representation of \cite{Peskin:1995ev} (see p.41) is used for the Dirac matrices,
often called the
\emph{Chiral representation}.
The Dirac matrices matrices have the form
%-------------------------------------------------------------------------------------------------------------------------------------------------------%
% Eq
%-------------------------------------------------------------------------------------------------------------------------------------------------------%
\begin{equation}
\gamma^0 =
\begin{pmatrix}0&1\\ 1&0\end{pmatrix},
\qquad
\gamma^i =\begin{pmatrix}0&\sigma^i\\-\sigma^i&0\end{pmatrix},
\label{gammas_chiral}
\end{equation}
%-------------------------------------------------------------------------------------------------------------------------------------------------------%
($i=1,2,3$) where
$\sigma^i$ are the Pauli spin matrices, 
%-------------------------------------------------------------------------------------------------------------------------------------------------------%
% Eq
%-------------------------------------------------------------------------------------------------------------------------------------------------------%
\begin{equation}
\begingroup % keep the change local
\setlength\arraycolsep{6pt}
\renewcommand*{\arraystretch}{0.8}
\sigma^1=
\begin{pmatrix}
0&1\\1&0
\end{pmatrix},\quad\quad\sigma^2=\begin{pmatrix}0&-i\\ i&0\end{pmatrix},\quad\quad\sigma^3=\begin{pmatrix}1&0\\ 0&-1\end{pmatrix},
\endgroup
\label{pauli_spin_matrices}
\end{equation}
%-------------------------------------------------------------------------------------------------------------------------------------------------------%
which have the property of all of them being Hermitian: $\sigma^{i\dagger}=\sigma^i$.
Note the identities
%-------------------------------------------------------------------------------------------------------------------------------------------------------%
\begin{equation}
\left(\sigma^i\right)^2= 1,\qquad 
\left\{\sigma^i,\sigma^j\right\}=2\delta^{ij} ,\qquad 
\sigma^i\sigma^j= i\epsilon^{ij k}\sigma_k +\delta^{ij}\sigma_i\sigma_j,\qquad 
\left[\sigma^i,\sigma^j\right] =2i\epsilon^{ijk}\sigma_k,\label{pauli_matrix_relations} \end{equation}
%-------------------------------------------------------------------------------------------------------------------------------------------------------%
from which the Dirac matrices are straightforwardly shown to satisfy the Clifford algebra 
%-------------------------------------------------------------------------------------------------------------------------------------------------------%
\begin{equation}
\left\{\gamma^\mu,\gamma^\nu\right\}=2\eta^{\mu\nu}
\label{clifford_algebra}
\end{equation}
%-------------------------------------------------------------------------------------------------------------------------------------------------------%
where $\eta^{\mu\nu}$ is the Minkowski metric, which in the convention  used here is assumed to be ``mostly positive'':
%-------------------------------------------------------------------------------------------------------------------------------------------------------%
\begin{equation}
\eta^{\mu\nu}={\rm diag}\left(1,-1,-1,-1\right)\ .
\label{minkowski_metric}
\end{equation}
%-------------------------------------------------------------------------------------------------------------------------------------------------------%
% \par 
%-------------------------------------------------------------------------------------------------------------------------------------------------------%
It follows from the representation in (\ref{gammas_chiral}) and the Hermitian property of the $\sigma^i$ that 
%-------------------------------------------------------------------------------------------------------------------------------------------------------%
\begin{equation}
\gamma^{0\dagger}=\gamma^0\ ,\qquad 
\gamma^{i\dagger}=-\gamma^i\ ,
\label{hc_gammas}
\end{equation}
%-------------------------------------------------------------------------------------------------------------------------------------------------------%
which equally  can be expressed as
%-------------------------------------------------------------------------------------------------------------------------------------------------------%
\begin{equation}
\gamma^{\mu\dagger}=\gamma^0\gamma^\mu\gamma^0\ .
\label{hc_gammas_1}
\end{equation}
%-------------------------------------------------------------------------------------------------------------------------------------------------------%
\par 
%-------------------------------------------------------------------------------------------------------------------------------------------------------%
The matrix 
%that is 
usually referred to as $\gamma_5$ is defined as
%-------------------------------------------------------------------------------------------------------------------------------------------------------%
\begin{equation}
\gamma_5=i\gamma^0\gamma^1\gamma^2\gamma^3\ ,\label{gamma5}
\end{equation}
%-------------------------------------------------------------------------------------------------------------------------------------------------------%
or equivalently, taking into account the effect of the anticommutation relation in (\ref{clifford_algebra})
on re-ordering of $\gamma$ matrices,
%-------------------------------------------------------------------------------------------------------------------------------------------------------%
\begin{equation}
\gamma_5=\frac{i}{4!}\epsilon_{\mu\nu\rho\sigma}\gamma^\mu\gamma^\nu\gamma^\rho\gamma^\sigma\ ,\label{gamma5_1}
\end{equation}
%-------------------------------------------------------------------------------------------------------------------------------------------------------%
where 
$\epsilon_{\mu\nu\rho\sigma}$ is the totally-antisymmetric
Levi-Civita symbol with four indices.
Note that 
indices are raised and lowered by contracting with the Minkowski metric,
with the effect that
$\epsilon_{0123}=-\epsilon^{0123}$.
In the convention used here,
\begin{equation}\gamma_5=\begin{pmatrix}-\mathds 1_{2\times 2}&0\\0&\mathds 1_{2\times 2}\end{pmatrix}\ .\label{gamma5_explicit}\end{equation}
It is easily verified by using (\ref{clifford_algebra}), that
%-------------------------------------------------------------------------------------------------------------------------------------------------------%
\begin{equation}\left(\gamma_5\right)^2=1\ .\label{gamma_5_squared}\end{equation}
%-------------------------------------------------------------------------------------------------------------------------------------------------------%
From the relation in (\ref{hc_gammas}) or (\ref{hc_gammas_1})
for obtaining the Hermitian conjugate of the $\gamma^\mu$,
%-------------------------------------------------------------------------------------------------------------------------------------------------------%
\begin{equation}
\gamma_5^\dagger=\gamma_5\ ,\label{gamma5_HC}
\end{equation}
%-------------------------------------------------------------------------------------------------------------------------------------------------------%
and from the Clifford algebra it is easily verified that 
%-------------------------------------------------------------------------------------------------------------------------------------------------------%
\begin{equation}
\left\{\gamma_5,\gamma^\mu\right\}=0\ .\label{gamma5_gamma_anticommutation}
\end{equation}
%-------------------------------------------------------------------------------------------------------------------------------------------------------%
Note that these identities for $\gamma_5$ are derived 
independent of the explicit form of $\gamma_5$, it is the nonetheless useful to write down it's explicit form in the 
representation used here:
%-------------------------------------------------------------------------------------------------------------------------------------------------------%
\subsection{Chiral projection operators}\noindent
%-------------------------------------------------------------------------------------------------------------------------------------------------------%
% \footnote{Unless stated otherwise, we use the convention of
% Ref.\ \cite{Bjorken_1964}.}
%-------------------------------------------------------------------------------------------------------------------------------------------------------%
The chiral projection operators are
$4\times 4$ matrices defined as
%-------------------------------------------------------------------------------------------------------------------------------------------------------%
\begin{equation}
P_R=\frac{1}{2}(1+\gamma_5)=P_R^\dagger,\quad
P_L=\frac{1}{2}(1-\gamma_5)=P_L^\dagger,
\label{chiral_projection_operators}
\end{equation}
%-------------------------------------------------------------------------------------------------------------------------------------------------------%
where the indices $R$ and $L$ refer to right-handed and 
left-handed.
% , respectively, as will become more clear below.
In the convention of these notes, according to (\ref{gamma5_explicit}),
%-------------------------------------------------------------------------------------------------------------------------------------------------------%
\begin{equation}
P_R=\begin{pmatrix}0&0\\0&\mathds 1_{2\times 2}\end{pmatrix},\quad
P_L=\begin{pmatrix}\mathds 1_{2\times 2}&0\\0&0\end{pmatrix}.
\label{chiral_projection_operators_explicit}
\end{equation}
%-------------------------------------------------------------------------------------------------------------------------------------------------------%

$P_R$ and $P_L$ satisfy a completeness relation,
%-------------------------------------------------------------------------------------------------------------------------------------------------------%
\begin{equation}
\label{PRPL_completeness}
P_R+P_L=1,
\end{equation}
%-------------------------------------------------------------------------------------------------------------------------------------------------------%
are idempotent, i.e.,
%-------------------------------------------------------------------------------------------------------------------------------------------------------%
\begin{equation}
\label{PRPL_idempotent}
P_R^2=P_R,\quad P_L^2=P_L,
\end{equation}
%-------------------------------------------------------------------------------------------------------------------------------------------------------%
and respect the orthogonality relations
%-------------------------------------------------------------------------------------------------------------------------------------------------------%
\begin{equation}
\label{PRPL_orthogonality}
P_R P_L=P_L P_R=0.
\end{equation}
%-------------------------------------------------------------------------------------------------------------------------------------------------------%
The combined properties of Eqs.\ (\ref{PRPL_completeness}) --
(\ref{PRPL_orthogonality}) guarantee that $P_R$ and $P_L$ are
indeed projection operators which project from the Dirac field variable $\psi$ to
its chiral components $\psi_R$ and $\psi_L$, defined as
%-------------------------------------------------------------------------------------------------------------------------------------------------------%
\begin{equation}
\label{psi_LR}
\psi_R\equiv P_R \psi,\quad 
\psi_L\equiv P_L \psi.
\end{equation}
%-------------------------------------------------------------------------------------------------------------------------------------------------------%
\par 
%-------------------------------------------------------------------------------------------------------------------------------------------------------%
For the goal of analyzing the symmetry of an action with
respect to independent global transformations of the left- and right-handed
fields, the identities below are required.
% In order to decompose the 16 quadratic forms $\left\{1,\gamma^\mu,\gamma_5,\gamma^\mu\gamma_5,\gamma^{[\mu,\nu]}\right\} $ into their respective
% projections to right- and left-handed fields,  make use of 
%-------------------------------------------------------------------------------------------------------------------------------------------------------%
% \cite{Gasser_1989}
\begin{equation}
\label{psibar_Gamma_psi}
\bar\psi \Gamma_i \psi=\left \{\begin{array}{lcl}
\bar\psi _R\Gamma_1 \psi_R+\bar\psi _L\Gamma_1 \psi_L&\mbox{for}&
\Gamma_1\in\{\gamma^\mu,\gamma^\mu\gamma_5\}\\[1em]
\bar\psi _R\Gamma_2 \psi_L +\bar\psi _L\Gamma_2 \psi_R&\mbox{for}& \Gamma_2
\in\{1,\gamma_5,\gamma^{[\mu,\nu]}\}
\end{array}
\right.,
\end{equation}
%-------------------------------------------------------------------------------------------------------------------------------------------------------%
where 
%-------------------------------------------------------------------------------------------------------------------------------------------------------%
\begin{equation}\bar \psi_R=\bar \psi P_L,\quad \bar \psi_L=\bar \psi P_R\ .\label{global_symms_light_quark_sector_2}\end{equation}
%-------------------------------------------------------------------------------------------------------------------------------------------------------%
% $\bar\psi _R=\bar\psi P_L$ and $\bar\psi _L=\bar\psi P_R.$
Equation (\ref{psibar_Gamma_psi}) is easily proven by inserting the completeness
relation of Eq.\ (\ref{PRPL_completeness}) both to the left and the 
right of $\Gamma_i$:
%-------------------------------------------------------------------------------------------------------------------------------------------------------%
\begin{equation}
\bar\psi \Gamma_i q=\bar\psi (P_R+P_L)\Gamma_i(P_R+P_L)\psi,
\end{equation}
%-------------------------------------------------------------------------------------------------------------------------------------------------------%
and by noting $\{\Gamma_1,\gamma_5\}=0$ and $[\Gamma_2,\gamma_5]=0$.
Together with the orthogonality relations of Eq.\ 
(\ref{PRPL_orthogonality}) it is then obtained
%-------------------------------------------------------------------------------------------------------------------------------------------------------%
\begin{equation}\label{global_symms_light_quark_sector_0}
P_R\Gamma_1 P_R=\Gamma_1P_L P_R=0,\qquad 
P_R\Gamma_1 P_L=P_R^2\Gamma_1=P_R\Gamma_1,\quad
P_L\Gamma_1 P_R=P_L^2\Gamma_1=P_L\Gamma_1,
\end{equation}
%-------------------------------------------------------------------------------------------------------------------------------------------------------%
and similarly
%-------------------------------------------------------------------------------------------------------------------------------------------------------%
\begin{equation}P_L \Gamma_1 P_L=0,\quad 
P_R \Gamma_2 P_L=0,\quad
P_L \Gamma_2 P_R=0.
\end{equation}
%-------------------------------------------------------------------------------------------------------------------------------------------------------%
% It should be stressed that the validity of Eq.\ (\ref{psibar_Gamma_psi}) is general
% and does not refer to ``massless'' quark fields.
\par 
%-------------------------------------------------------------------------------------------------------------------------------------------------------%
Due to (\ref{gamma5_gamma_anticommutation}),
%-------------------------------------------------------------------------------------------------------------------------------------------------------%
\begin{equation}
\gamma^\mu P_L= P_R\gamma^\mu\ ,\qquad \gamma^\mu P_R= P_L\gamma^\mu\ .
\label{PLR_gamma}
\end{equation}
%-------------------------------------------------------------------------------------------------------------------------------------------------------%
By (\ref{gamma_5_squared}) and (\ref{chiral_projection_operators}),
%-------------------------------------------------------------------------------------------------------------------------------------------------------%
\begin{equation}
\gamma_5 P_L= P_L\gamma_5\ ,\qquad \gamma_5 P_R= P_R\gamma_5\ .
\label{PLR_gamma_5}
\end{equation}
%-------------------------------------------------------------------------------------------------------------------------------------------------------%
And it follows from (\ref{chiral_projection_operators}) and (\ref{gamma5_gamma_anticommutation}) 
that 
%-------------------------------------------------------------------------------------------------------------------------------------------------------%
\begin{subequations}\label{PLR_bar}\begin{eqnarray}\overline{P_L}\equiv& P_L^\dagger \gamma^0=P_L\gamma^0&=\gamma^0 P_R\label{PLbar}\\
\overline{P_R}\equiv& P_R^\dagger \gamma^0=P_R\gamma^0&=\gamma^0 P_L\label{PRbar} \end{eqnarray}\end{subequations}
%-------------------------------------------------------------------------------------------------------------------------------------------------------%
\section{Wick rotations}\label{sec_wick_rotatns}\noindent 
%-------------------------------------------------------------------------------------------------------------------------------------------------------%
In Minkowski space the $\gamma$ matrices satisfy the Clifford algebra given by 
%-------------------------------------------------------------------------------------------------------------------------------------------------------%
\begin{equation}\{\gamma^\mu,\gamma^\nu\}=2\eta^{\mu\nu}\ ,\qquad (\mu,\nu=0,1,2,3)\ ,\label{wick_rot_1}\end{equation}
%-------------------------------------------------------------------------------------------------------------------------------------------------------%
where $\eta^{\mu\nu}={\rm diag}(1,-1,-1,-1)$.
A wick rotation comprises a transformation
of the $\gamma$ matrices  
to new matrices, $\gamma_E$ defined by
%-------------------------------------------------------------------------------------------------------------------------------------------------------%
\begin{equation}\gamma^4_{E}=\gamma^0
\ ,\qquad \gamma^i_E =-i \gamma^i\ ,\qquad (i=1,2,3)\label{wick_rot_2A}\end{equation}
%-------------------------------------------------------------------------------------------------------------------------------------------------------%
such that 
%-------------------------------------------------------------------------------------------------------------------------------------------------------%
\begin{equation}\{\gamma_{E}^\mu,\gamma_{E}^\nu\}=2\delta^{\mu\nu}\ ,\qquad (\mu,\nu=1,2,3,4).\label{wick_rot_3A}\end{equation}
%-------------------------------------------------------------------------------------------------------------------------------------------------------%
\par 
%-------------------------------------------------------------------------------------------------------------------------------------------------------%
Four-vector components get transformed under a Wick rotation to new components. For example given components $k^\mu$
of the four vector $k$ in Minkowski space,
%-------------------------------------------------------------------------------------------------------------------------------------------------------%
\begin{equation}k^4_{E}=ik^0
\ ,\qquad k^i_{E} =k^i\ ,\qquad (i=1,2,3)\ .\label{wick_rot_4}\end{equation}
%-------------------------------------------------------------------------------------------------------------------------------------------------------%
The implication is that 
%-------------------------------------------------------------------------------------------------------------------------------------------------------%
\begin{equation}k^2=\left(k^0\right)^2-\left(k^i\right)^2=
-\left(k_{E}^4\right)^2
-\left(k_{E}^1\right)^2-\left(k_{E}^2\right)^2-\left(k_{E}^3\right)^2
\equiv-k_{E}^2\ .
\label{wick_rot_5}\end{equation}
%-------------------------------------------------------------------------------------------------------------------------------------------------------%
\begin{equation}
\gamma k=
\gamma^0 k^0-\gamma^i k_i
=
-i\gamma^4_{{E}}k^4_{{E}}
-i\gamma^1_{{E}}k^1_{{E}}-i\gamma^2_{{E}}k^2_{{E}}-i\gamma^3_{{E}}k^3_{{E}}\equiv -i\gamma_{{E}}k_{{E}}\ .
\label{wick_rot_6A}\end{equation}
%-------------------------------------------------------------------------------------------------------------------------------------------------------%
\section{Angular integral}\label{sec_angular_integral}\noindent
%-------------------------------------------------------------------------------------------------------------------------------------------------------%
The integrals  in (\ref{angular_integral_1}) and (\ref{angular_integral_2}) are solved in this appendix,
whose forms are
%-------------------------------------------------------------------------------------------------------------------------------------------------------%
\begin{equation}\begin{aligned}
I_{ 1} =&\int^\pi_0d\theta\,\frac{\sin^2\theta}{\left(a-b\cos\theta\right) }\ ,\\  
I_{ 2} =&\int^\pi_0d\theta\,\frac{\sin^2\theta}{\left(a-b\cos\theta\right)^2}\ ,
\end{aligned}\label{angular_integral_1}\end{equation}
%-------------------------------------------------------------------------------------------------------------------------------------------------------%
where
%-------------------------------------------------------------------------------------------------------------------------------------------------------%
\begin{equation}
a=p^2+k^2\ ,\qquad b=2  p  k\ .\label{angular_integral_1A}
\end{equation}
%-------------------------------------------------------------------------------------------------------------------------------------------------------%
They both have the same structure, namely
%-------------------------------------------------------------------------------------------------------------------------------------------------------%
\begin{equation}
I_{ n} = \int^\pi_0d\theta\,\frac{\sin^2\theta}{\left(a-b\cos\theta\right)^n}\ ,\label{angular_integral_1B}
\end{equation}
%-------------------------------------------------------------------------------------------------------------------------------------------------------%
and are evaluated in the same way.\par
%-------------------------------------------------------------------------------------------------------------------------------------------------------%
First write it as an integral from $0$ to $2\pi$ so that it can be converted into a contour integral along a
closed circular path in the complex plane.
Since $\cos\theta=\cos(2\pi-\theta)$ and $\sin^2\theta=\sin^2(2\pi-\theta)$
then $I_{ n} $ can equally be written as
%-------------------------------------------------------------------------------------------------------------------------------------------------------%
\begin{align}
I_{ n} 
=&\int^{\pi}_{0}d\theta\,\frac{\sin^2(2\pi-\theta)}{\left(a-b\cos(2\pi-\theta)\right)^n}
%\nonumber\\
% =&-\int^{\pi}_{2\pi}d\xi\,\frac{\sin^2\xi}{\left(a-b\cos\xi\right)^n}
\nonumber\\
=&\int^{2\pi}_{\pi}d\xi\,\frac{\sin^2\xi}{\left(a-b\cos\xi\right)^n}\ ,\label{angular_integral_2}
\end{align}
%-------------------------------------------------------------------------------------------------------------------------------------------------------%
such that
%-------------------------------------------------------------------------------------------------------------------------------------------------------%
\begin{equation}
I_{ n} =\frac1 {2}\int^{2\pi}_0d\theta\,\frac{\sin^2\theta}{\left(a-b\cos\theta\right)^n}\ .\label{angular_integral_3}
\end{equation}
%-------------------------------------------------------------------------------------------------------------------------------------------------------%
Now write this as a closed integral over the variable $z=e^{i\theta}$ in the complex plane,
with $\sin\theta=(-i/2)(z-z^{-1})$ and
$\cos\theta=(1/2)(z+z^{-1}) $:
%-------------------------------------------------------------------------------------------------------------------------------------------------------%
\begin{align}
I_{ n}
=&
\frac i{8}\oint dz\,\frac{2^nz^{n-3}\left(z^2-1\right)^2}
{\left(2a z-b z^2-b \right)^n}
\ .\label{angular_integral_4}
\end{align}
%-------------------------------------------------------------------------------------------------------------------------------------------------------%
Write the denominator of the integrand as
%-------------------------------------------------------------------------------------------------------------------------------------------------------%
\begin{equation}
\left(2a z-b z^2-b \right) ^n
=
(-b)^n\left( z-z_1\right)^n\left( z-z_2\right)^n\ ,
\label{angular_integral_5}
\end{equation}
%-------------------------------------------------------------------------------------------------------------------------------------------------------%
where 
%-------------------------------------------------------------------------------------------------------------------------------------------------------%
\begin{equation}
z_1 =\frac{a+\sqrt{ a^2-b^2}}{b}\ ,
\qquad
z_2=\frac{a-\sqrt{ a^2-b^2}}{b}\ ,
\label{angular_integral_6}
\end{equation}
%-------------------------------------------------------------------------------------------------------------------------------------------------------%
to bring the integral into the form
%-------------------------------------------------------------------------------------------------------------------------------------------------------%
\begin{equation}
I_{ n} =2^{n-3}i \oint dz\,\frac{z^{n-3}\left(z^2-1\right)^2}
{(-b)^n\left( z-z_1\right)^n\left( z-z_2\right)^n}
\ .\label{angular_integral_7}
\end{equation}
%-------------------------------------------------------------------------------------------------------------------------------------------------------%
Now the integral has the familiar form $\displaystyle \oint dz\,\dfrac{f(z)}{(z-z_1)^n(z-z_2)^m}$
where the contour is the unit circle
with $f(z)$ analytic and continuous at $z_1$ and $z_2$,
and it can be solved  by the standard method of summing over the residues of $f(z)$.
\par 
%-------------------------------------------------------------------------------------------------------------------------------------------------------%
% First the $n=4$ is solved.
Note carefully the location of the poles. From (\ref{angular_integral_1A})
it follows that 
%-------------------------------------------------------------------------------------------------------------------------------------------------------%
\begin{equation}a-b=p^2+k^2 -2pk=(p-k)^2 >0\ , \label{angular_integral_7A}\end{equation}
%-------------------------------------------------------------------------------------------------------------------------------------------------------%
and there it is always true that
%-------------------------------------------------------------------------------------------------------------------------------------------------------%
\begin{equation}a>b  \ .\label{angular_integral_7B} \end{equation}
%-------------------------------------------------------------------------------------------------------------------------------------------------------%
Accordingly 
%-------------------------------------------------------------------------------------------------------------------------------------------------------%
\begin{align}
z_1 =&
\frac k{p} \theta(k-p)
+\frac p{k}
\theta(p-k)
\ ,
\\
z_2 =& 
\frac p{k} \theta(k-p)
+\frac k{p}
\theta(p-k)
\ .
\end{align} 
%-------------------------------------------------------------------------------------------------------------------------------------------------------%
In conclusion
$|z_2|<1$ while $|z_1|>1$,
so the pole at $z_1$ is discounted because it is not inside the unit circle.
This leaves two poles in the expression 
(\ref{angular_integral_7}): one at $z=0$ and one at $z=z_2$.
\par 
%-------------------------------------------------------------------------------------------------------------------------------------------------------%
For $n=2$ the integral in  (\ref{angular_integral_7}), by the Cauchy formula, has the form
%-------------------------------------------------------------------------------------------------------------------------------------------------------%
\begin{align}
I_{ 2} =&
\frac i{2}(2\pi i)\frac1{b^2}\left[
\left( 
\frac{\left(z^2-1\right)^2}{ \left( z-z_1\right)^2\left( z-z_2\right)^2}\right)\bigg|_{z=0}
+
\frac d{dz}\left( 
\frac{\left(z^2-1\right)^2}{ z\left( z-z_1\right)^2  }\right)\bigg|_{z=z_2}
\right]
\nonumber\\
=&
- \frac\pi{(2pk)^2}\left[ 
\frac1{ \left( z_1 z_2\right)^2}
-\frac{\left(z_2^2-1\right)^2}{z_2^2 (z_2-z_1)^2}-\frac{2 \left(z_2^2-1\right)^2}{z_2 (z_2-z_1)^3}+\frac{4 \left(z_2^2-1\right)}{(z_2-z_1)^2}
\right]
\ ,\label{angular_integral_10}
\end{align}
%-------------------------------------------------------------------------------------------------------------------------------------------------------%
while for $n=1$ it is
%-------------------------------------------------------------------------------------------------------------------------------------------------------%
\begin{align}
I_{ 1} =&
\frac i{4}(2\pi i)\frac1{(-b)}\left[
\frac d{dz}
\left( 
\frac{\left(z^2-1\right)^2}{ \left( z-z_1\right) \left( z-z_2\right) }\right)\bigg|_{z=0}
+
\left( 
\frac{\left(z^2-1\right)^2}{ z^2\left( z-z_1\right)^2 }\right)\bigg|_{z=z_2}
\right]
\nonumber\\
=&
\frac\pi{4pk }\left[ 
\frac{{z_1}+{z_2}}{{z_1}^2 {z_2}^2}
+
\frac{\left({z_2}^2-1\right)^2}{{z_2}^2 ({z_2}-{z_1})}
\right]
\ .\label{angular_integral_10}
\end{align}
%-------------------------------------------------------------------------------------------------------------------------------------------------------%
Finally
substitute the explicit forms of $z_1$ and $z_2$ to obtain
%-------------------------------------------------------------------------------------------------------------------------------------------------------%
\begin{align}
I_1=& 
\frac{\pi }{2 k^2}
\theta(k-p)
+
\frac{\pi }{2 p^2}
\theta(p-k) \ ,
\label{I1_eps_expansn}\end{align}
%-------------------------------------------------------------------------------------------------------------------------------------------------------%
and 
%-------------------------------------------------------------------------------------------------------------------------------------------------------%
\begin{align}
I_2=&-\frac{\pi  }{2k^2(p^2-k^2)} \theta(k-p)
-\frac\pi{ 2p^2(k^2-p^2)}
\theta(p-k)\ .
\label{I2_eps_expansn}\end{align} 
%-------------------------------------------------------------------------------------------------------------------------------------------------------%
% \bibliographystyle{unsrt}
% \bibliography{/home/jeremy/Dropbox/latex/references/references}
%-------------------------------------------------------------------------------------------------------------------------------------------------------%

\end{document}